\documentclass[11pt]{article}
\usepackage[utf8]{inputenc}
\usepackage[T1]{fontenc}
\usepackage{amsmath}
\usepackage{amssymb}
\usepackage{graphicx}
\usepackage{booktabs}
\usepackage{longtable}
\usepackage[margin=1in]{geometry}
\usepackage{xcolor}
\usepackage{caption}
\usepackage{authblk}
\usepackage[hidelinks]{hyperref}

\title{Perceived vertical and eye level as one orientation process: a closed-form account of the Li--Matin combination rules for egocentric space}
\author{A. Y. Shavit}
\affil{Hunter College and the Graduate Center, City University of New York, New York, NY, USA\\ \texttt{as1127@hunter.cuny.edu}}
\date{}

\begin{document}
\maketitle

\begin{abstract}
Li and Matin (2005a, 2005b) reported three regularities of the induced visually-perceived vertical (VPV), the direction judged upright. For small tilts, a peripheral line shifts it nearly linearly with tilt, two lines combine nearly linearly, and mirror-image tilts cancel. Each line votes at twice its angle, weighted by its strength. The votes' vector sum $c_1$ is the first circular moment of the doubled angles; half its angle, $\phi=\tfrac12\arg c_1$, is the induced vertical, a stimulus axis, not the judged upright. This is the \emph{orientation order-parameter model} (PLUMB). We prove that $c_1$ is, up to a constant factor, the only additive summary of axes (lines without arrowheads) that scales with line strength and turns twice as far as the field rotates. With Li and Matin's own length gain and no per-configuration fit, the model matches their multiline coefficients in-sample. Their data rule out complete summation but not simple averaging for long lines. But a line's effect is fourfold (period $90^\circ$): it reverses past $45^\circ$ (Li and Matin, 2005a; Vingerhoets et al., 2009), and near-horizontal lines shift VPV as near-vertical ones do, which $c_1$ forbids. In a two-harmonic fit to Li and Matin's data, $c_1$ carries an estimated 2--4\% of a near-vertical line's slope and a fourfold term the rest. To first order in tilt, for near-vertical segments, the readout equals the committee rule Shavit, Li, and Matin (2026) fit to eight line-array experiments, with its drives read as concentrations. If the half-fields' signals are read as sum and difference, perceived eye level (VPEL) is the difference, recovering Shavit, Li, and Matin's (2013) reversed integration rules; a fourfold term gives the same signals, so this does not single out $c_1$. Re-analysed data (30 observers) show the predicted signs and nulls, an internal check only; the cross-field gain is fitted.
\end{abstract}

\section{The regularities to be explained}
The visually-perceived vertical is biased by the orientation content of the visual field. Li and Matin (2005a, 2005b), extending Matin and Li (1995), characterized the induction produced by roll-tilted peripheral lines in two reports that year. One covered the single- and two-line orientation and length functions (\emph{Vision Research}, 2005b); the other dissected the rod-and-frame effect into 1-, 2-, 3-, and 4-line components (\emph{Perception}, 2005a).

They found: \textbf{(i)} the induced VPV shifts approximately \emph{linearly} with a single line's orientation; \textbf{(ii)} the influences of several lines combine \emph{linearly and sub-additively} --- ``the whole is less than the sum of its parts''; and \textbf{(iii)} a bilaterally \emph{symmetric} pair leaves the vertical unmoved --- symmetric tilts cancel. They further established that it is the \emph{retinal} orientation of the inducer that drives the effect. All three were measured within about $\pm22.5^\circ$ of a cardinal axis. Li and Matin (2005a) tilted horizontal as well as vertical lines, and the horizontal lines shift VPV in the same rotational sense as the vertical ones, which the first moment forbids. Over the full circle a line's effect is \emph{fourfold} (period $90^\circ$ in line tilt) and reverses past $45^\circ$ (\S5.3).

Matin and Li (1999) had earlier established the parallel rules for the perceived eye level (VPEL): the same lines, integrated with the opposite sign --- a contrast those authors state against their own rule for the vertical (Matin \& Li, 1994). \S5.5 recovers those rules from the per-hemifield orientation signals, read as a sum and a difference. We ask which computation on the stimulus produces a \emph{linear combination of orientations} --- a quantity ill-defined in general, since orientation is periodic on the half-circle.

We give a minimal answer: a single \emph{order parameter} --- the axis the field's orientations vote for, weighted by how strongly they agree. It reproduces the rules in closed form, needs no per-configuration fitted parameters in the angles, and coincides with quantities already central to image analysis and to population coding. \S\S2--5.4 develop the perceived vertical (VPV). \S5.5 then shows that the perceived eye level (VPEL) --- the \emph{difference} channel that Figure~1 previews --- is the \emph{antisymmetric} readout of the \emph{same} per-hemifield orientation signals, which a fourfold term gives equally. One stimulus representation thus supports both egocentric references: after an added per-hemifield decomposition, its symmetric (sum) readout gives the vertical and its antisymmetric (difference) readout gives eye level. That decomposition is a further structural assumption (\S5.5), not a consequence of \S2.

\paragraph{The claims in full.} The abstract compresses several claims that the body states with their conditions. The induced vertical is a stimulus-side axis, not yet the observer's setting. Orientation is a \emph{director}, an axis with no arrowhead ($\theta\equiv\theta+\pi$), and $c_1=\sum_j A_j e^{i2\gamma_j}$ counts each line at twice its angle $\gamma_j$, weighted by its salience $A_j$ (its strength or weight, set by contrast and length). The readout $\phi=\tfrac12\arg c_1$ is equivalently the principal axis of the orientation structure tensor or an orientation population vector (\S4).

We prove uniqueness in two steps (Appendix~\ref{app:unique}). For a summary that transforms under a rotation character, director symmetry alone rules out the odd rotational orders, so doubling is the lowest nonzero order available. Linear pooling and positive homogeneity (scaling every salience by a positive factor scales $c_1$ by the same factor), with a calibration that reads a vertical line as vertical, then force $c_1$ within it, up to a positive scale. When the field rotates, $c_1$ turns by twice the angle, so it is a \emph{spin-2}, rotation-covariant quantity; the axis $\phi$ it encodes turns by the angle itself. Its coherence $R=|c_1|/c_0$, the resultant's length as a fraction of the votes' total, sets how sharply the axis is defined. Once a perturbation is measured relative to the field's total effective salience, sensitivity scales inversely with coherence, so a weakly coherent field gives a poorly conditioned axis (Appendix~\ref{app:sensitivity}). The length-gain function is Li and Matin's own Fig.~5, a saturating function of total inducing length with three constants, adopted descriptively rather than as a mechanism (\S5.2). The equal-length orthogonal and square configurations have a first moment of exactly zero, so there is no angle to fit. The next moment, $c_2$, the same vote sum at four times the angle, is the \emph{fourfold} term (period $90^\circ$ in line tilt). Harmonics are counted in the doubled angle throughout, so $c_1$ is the first harmonic and $c_2$ the second, which in the raw line angle are the second and fourth. The fourfold term carries the orthogonal and square configurations' residual induction through its coupling to the upright reference, body-referenced or visual-cardinal, which the model does not separate (\S5.6), and the length term alone does not fix its magnitude. The phase readout is scale-invariant and a weighted angular average, so $n$ equal inducers combine \emph{sub-additively}, and their slope approaches the averaging limit $1/n$ as the length gain saturates. Complete summation, which Li and Matin's multiline data refute, is $k_1=1$: a multiline response equal to the sum of its single-line responses.

The decisive tests are parameter-free in the angles they predict, though not in the weights. They are the short-line regime, the $|\cos2\theta|$ coherence law, and moment-matched stimuli that share total orientation energy $c_0$ and first moment $c_1$ but differ in second moment $c_2$, read in setting precision, or in the mean once rotated off the cardinal axes (\S6). The first-order match with the committee rule (a drive-weighted average of tilts against a body drive; a drive is read here as a von Mises concentration, \S5.6), its two conditions and the size of the third-order gap are in \S5.6. In the trial-level data, the magnitude of the cross-field combination, sub-additive in its point estimates, is an estimated comparator gain. Nominally it is the gain of the stage comparing the two hemifields (\S5.5), but it is read from the data rather than derived by the architecture, and it is not yet shown to arise at that stage specifically. The first moment is not the dominant term of single-line induction. Li and Matin's (2005a) near-horizontal lines shift VPV the same way as their near-vertical lines, which $c_1$ forbids, and over the full circle a line acts fourfold. In their data $c_1$ carries an estimated $2$--$4\%$ of a near-vertical line's small-tilt slope, a point estimate from slopes printed without standard errors (\S5.3). The survivors, consistent with $c_1$ rather than specific to it, are the doubled-angle combination structure of R2 and R3 and the VPEL sum/difference readout (\S5.5). A fourfold term produces all three as well, and no published data separate $c_1$ from it. Because the account is image-computable, it links induced perceived vertical and eye level to classical descriptors of image orientation.

\begin{figure}[!htbp]\centering
\includegraphics[width=\linewidth]{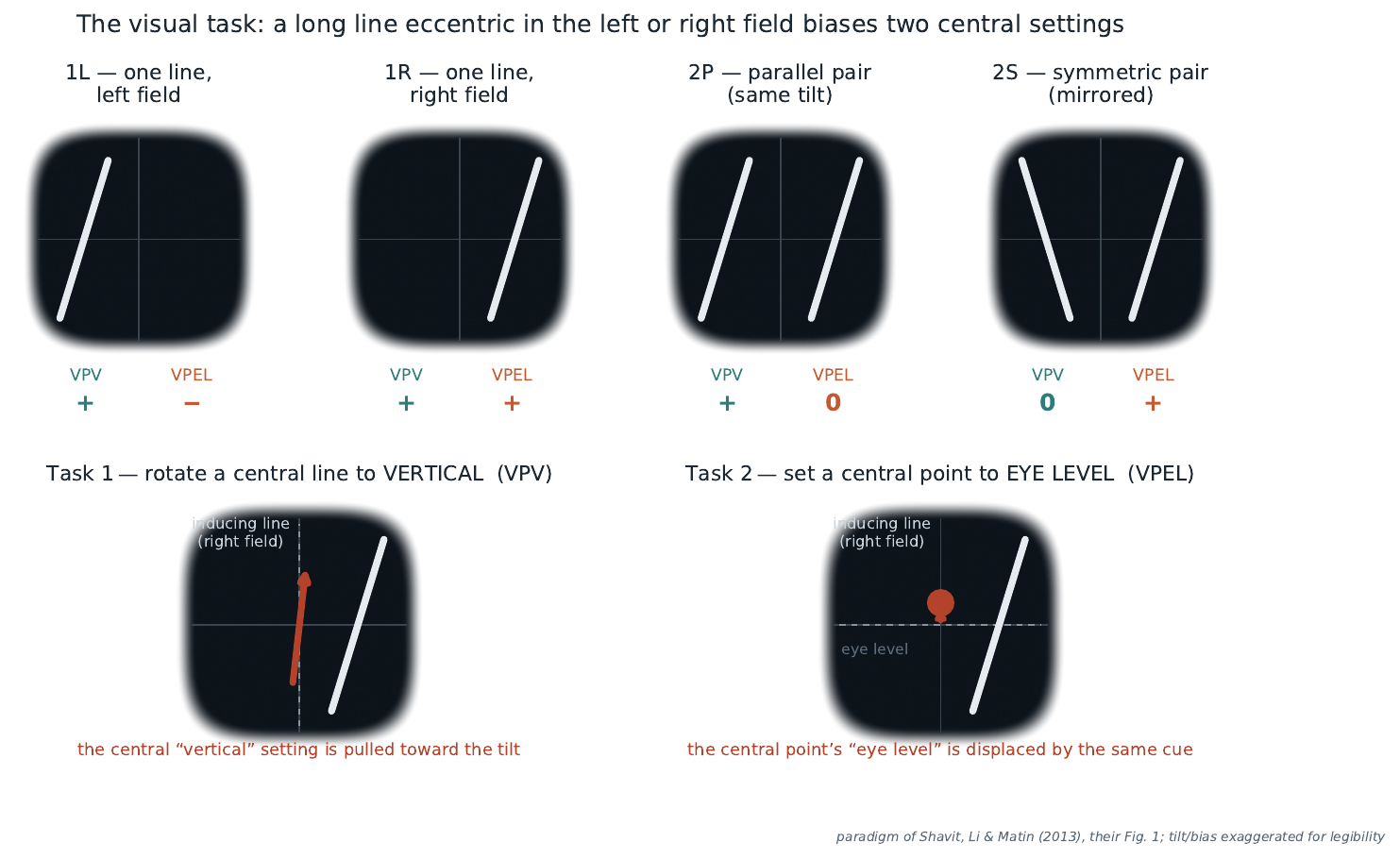}
\caption{The visual task (after Shavit, Li, \& Matin 2013, Fig.~1). A long roll-tilted line, eccentric in the left or right visual field and seen in darkness, biases two central settings: the perceived vertical (VPV, rotate a central line to upright) and the perceived eye level (VPEL, set a central point to eye level). The four inducers are one left-field line (1L), one right-field line (1R), a parallel pair (2P, same tilt), and a bilaterally symmetric pair (2S, mirrored). L/R names the field, not the tilt --- both single lines are tilted the same way. Observed effects are tagged VPV/VPEL below each, with the signs of Table~\ref{tab:readouts}: $+$ a shift with the sign of the line's tilt, $-$ the opposite sign, $0$ none. Tilt and bias are exaggerated for legibility.}
\label{fig:1}
\end{figure}

\begin{figure}[!htbp]\centering
\includegraphics[width=\linewidth]{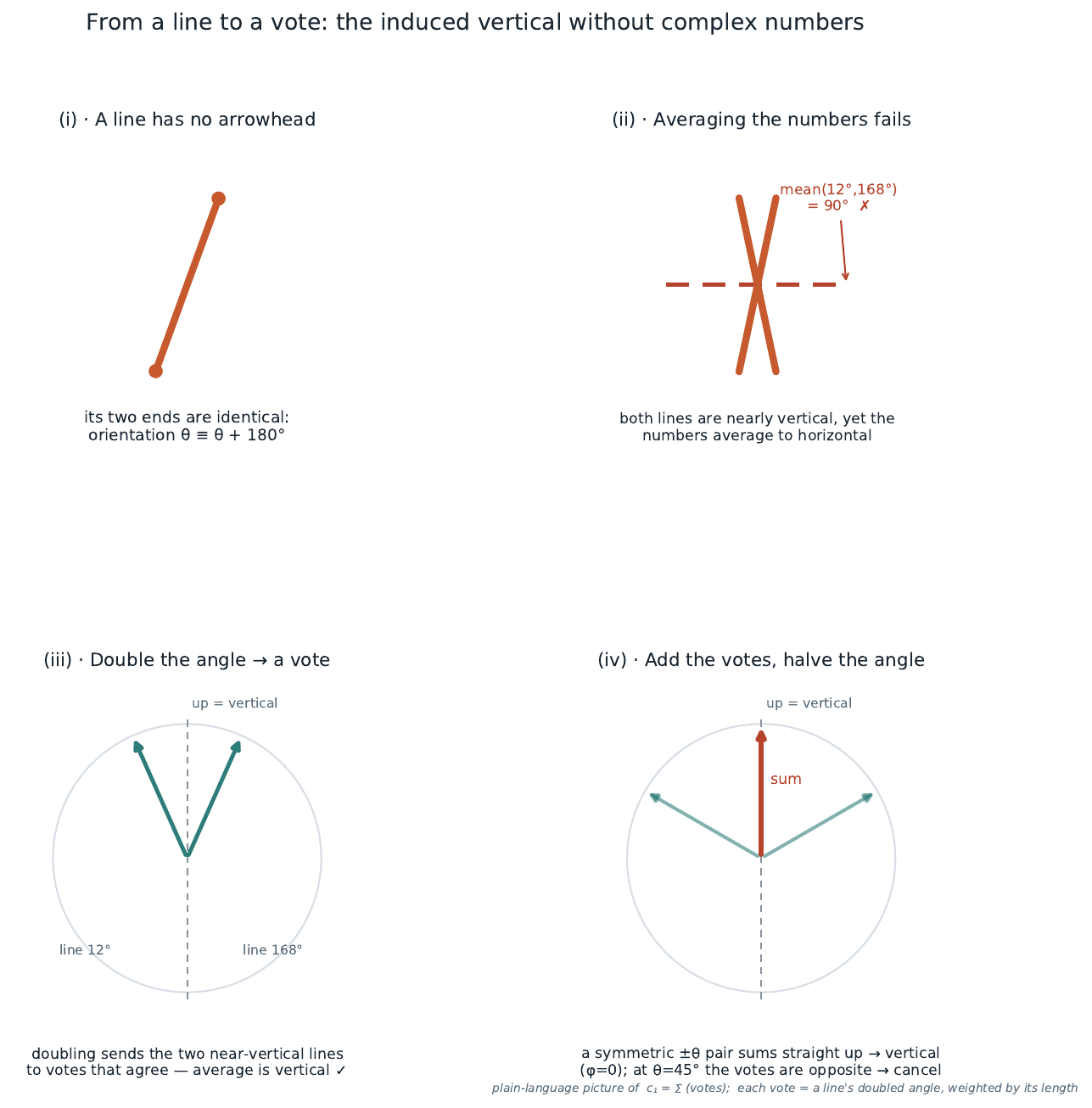}
\caption{From a line to a vote --- the induced vertical without complex numbers. (i) A line has no arrowhead, so orientation is $\theta \equiv \theta+180^\circ$. (ii) Averaging the angle numbers fails: two nearly-vertical lines at $12^\circ$ and $168^\circ$ average to $90^\circ$ (horizontal). (iii) Taking twice each line's angle turns it into a ``vote'' arrow; the two near-vertical lines now agree, both pointing up. (iv) The votes add as vectors --- drawn here from a common origin, so their sum is the arrow up the middle --- and that sum's angle is halved to give the induced vertical; a symmetric $\pm\theta$ pair sums straight up ($\phi=0$) and cancels entirely at $\theta=45^\circ$.}
\label{fig:2}
\end{figure}

\paragraph{Box 1 --- an intuition for the order parameter} \emph{(for readers less at home with the complex notation; the first-moment construction of \S\S2--4 can be read as the single picture in Figure~\ref{fig:2}).}
\begin{itemize}
\item \textbf{(i) A line's orientation has no direction.} A line points both ways at once --- ``10-o'clock--4-o'clock'' is the same line as ``4--10'' ($\theta\equiv\theta+180^\circ$), with no way to say which end is the ``tip.'' So you cannot average orientations by averaging their angle \emph{numbers}. A line at $1^\circ$ and a line at $179^\circ$ are nearly identical, yet the numbers average to $90^\circ$ --- perpendicular to both.
\item \textbf{(ii) Take twice the angle.} To average them anyway, first give each line a direction. Rotate a pointer to \emph{twice} the line's roll tilt and it becomes a \textbf{vector} --- a line that now \emph{does} have a direction. The two ends land on the same spot, so the missing tip no longer matters and ordinary vector-averaging works. You \emph{halve the answer back} at the end.
\item \textbf{(iii) Each line casts a vote.} That vector is the line's vote --- a direction at \emph{twice} its roll tilt, with a length set by how strong the line is (its contrast/length). The votes \textbf{add as vectors}. The resultant's \emph{direction}, halved back, is the induced vertical ($\phi$); its \emph{length} as a fraction of the votes' total is how strongly the field insists --- the order parameter $R$, which runs from $0$ for a disagreeing cross of lines at $\pm45^\circ$ ($90^\circ$ apart) to $1$ for perfect alignment. For mirror-symmetric lines the lateral components cancel; the whole resultant vanishes only when the two doubled-angle votes are antiparallel, at the orthogonal $\pm45^\circ$ pair. Everything that follows is this vote-sum; the symbol $e^{i2\gamma}$ in \S2 is just ``a unit vector pointing at twice the roll tilt $\gamma$.''
\end{itemize}

\section{The model}
Represent the stimulus by an orientation-energy distribution over the half-circle,
\[
E(\theta)=\sum_j A_j\,\kappa(\theta-\gamma_j),\qquad \theta\in[0,\pi),
\]
with inducer $j$ at retinal orientation $\gamma_j$, nonnegative weight $A_j\geq0$, and $\kappa$ an even, $\pi$-periodic tuning profile peaked at $0$. The weight $A_j$ is the inducer's \emph{effective} salience (contrast/length/energy) \emph{after} any early local interactions such as cross-orientation normalisation. Nonnegative weights keep $E$ an energy, the structure tensor positive semidefinite, and the normalised moment a convex orientation average. PLUMB is agnostic about that early stage. It does not assume the local orientation channels are independent --- only that whatever effective salience survives them is then pooled \emph{linearly in the doubled-angle domain} to form the global axis. Define the \textbf{induced vertical} $\phi$ --- the model's \emph{asserted axis}, the phase angle (\emph{argument}) of $c_1$, and a \emph{stimulus-side} quantity rather than the observer's setting --- and the \textbf{first-moment coherence} as
\begin{equation}
\boxed{\;\phi=\tfrac12\arg c_1,\qquad R=\frac{|c_1|}{c_0},\qquad c_1=\int_0^\pi E(\theta)e^{i2\theta}\,d\theta=\hat\kappa_1\sum_j A_j e^{i2\gamma_j},\;}
\end{equation}
where $c_0=\hat\kappa_0\sum_j A_j$ and $\hat\kappa_1>0$ is the tuning's first coefficient (it cancels in $\phi$ and, up to the constant $\hat\kappa_1/\hat\kappa_0$, sets the scale of $R$). \textbf{We fix that scale by convention: $\hat\kappa_1=\hat\kappa_0$}, so that a single line carries full coherence $R=1$. \textbf{That is an idealisation of the tuning, not a rescaling of it}, and the distinction matters. The ratio $\hat\kappa_1/\hat\kappa_0$ is unchanged by $\kappa\mapsto c\kappa$, so no choice of overall scale can set it; and for a nonnegative profile $|\hat\kappa_1|\leq\hat\kappa_0$, with equality only in the point-mass limit --- for a von Mises tuning $e^{s\cos2u}$ the ratio is $I_1(s)/I_0(s)$, which is $0.89$ at concentration $s=5$ (the concentration is the tuning's sharpness: larger $s$ is a narrower profile) and reaches $1$ only as $s\to\infty$. Setting $\hat\kappa_1=\hat\kappa_0$ is therefore the infinitely-sharp-tuning limit, and what it does operationally is divide the tuning factor out of $R$ so that $R$ measures the inducers' agreement alone. Every downstream use of $R$ --- the coherence law of R4 below, the cue-combination weight of \S5.6, the analysis code --- already assumes this; stating it here once removes the only place a reader could otherwise read $R=1$ ``for perfect agreement'' as inconsistent with $R=\hat\kappa_1/\hat\kappa_0$ for a single line.

\textbf{The stimulus axis and the measured setting differ.} $\phi$ is where the stimulus's orientation votes point, and it tracks a single line with \emph{unit} slope. The setting an observer actually makes, $\widehat v$, combines $\phi$ with the body's own upright and tracks that same line with a gain $k_V\approx0.12$--$0.15$, measured by Shavit et al.\ (2013) in two single-line conditions (\S5.6) --- roughly a seventh of $\phi$'s unit slope, so an observer follows the asserted axis only faintly. Every unit-slope statement below is a statement about $\phi$, not about a measured setting.

\textbf{Two readings of $R$ coincide only under this convention, and differ otherwise.} $R_T=|c_1|/c_0$ is whatever a structure-tensor computation on real image or stimulus data actually returns; $r=|\sum_j A_je^{i2\gamma_j}|/\sum_j A_j$ is the intrinsic, unit-normalised vote coherence, exactly in $[0,1]$ by construction. The two are related by $R_T=(\hat\kappa_1/\hat\kappa_0)\,r$, so they are the same number only under the convention above. The bound just stated settles which way they differ: since $\hat\kappa_1/\hat\kappa_0\leq1$, the structure-tensor reading always sits \emph{at or below} the vote coherence, and $R_T\in[0,1]$ automatically for nonnegative $E$ --- it is never out of range, it is attenuated. A reader who computes $R_T$ from image statistics, where the tuning is whatever the filter bank supplies, should therefore expect even an isolated line to return $R_T=\hat\kappa_1/\hat\kappa_0<1$ rather than $1$, and should divide by that fixed front-end factor to recover $r$. Everywhere below, $R$ denotes $r$ under the stated convention.

Box~1 draws the same construction in plain terms. The moment $c_1$ is the vector sum of the vote-vectors, and $\arg c_1$ is the compass direction that sum points, halved back to give the vertical $\phi$. The coherence $R$ is the sum's length as a fraction of the total --- how much the votes agree, $1$ for perfect agreement and $0$ when they cancel. We call this the \textbf{orientation order-parameter model} (PLUMB, after the plumb line). The name is literal: $R$ is an \emph{order parameter} in the sense of statistical physics, the alignment of a population of directors, and $\phi$ is the axis it points along; we call the pair $(R,\phi)$, equivalently $c_1$, the order parameter, and use the term loosely for $\phi$ alone. The factorisation in the box is a one-line calculation that makes the model discrete: substituting $E(\theta)=\sum_j A_j\kappa(\theta-\gamma_j)$ and changing variables $u=\theta-\gamma_j$,
\[
c_1=\sum_j A_j\!\int_0^\pi \kappa(\theta-\gamma_j)e^{i2\theta}\,d\theta=\Big(\!\int_0^\pi\kappa(u)e^{i2u}\,du\Big)\sum_j A_j e^{i2\gamma_j}=\hat\kappa_1\sum_j A_j e^{i2\gamma_j},
\]
so each inducer contributes a phasor $A_j e^{i2\gamma_j}$ --- a vote-vector in the plane, in the language of Box~1 --- scaled by the single constant $\hat\kappa_1=\int_0^\pi\kappa(u)e^{i2u}\,du$. The tuning profile $\kappa$ enters only through that scalar and drops out of $\phi$ entirely. For the isolated-line stimuli analysed here $E(\theta)$ is a sum of sticks, and $c_1$ reduces to this discrete phasor sum. For a continuous broadband field such as a natural scene, the same definition $c_1=\int_0^\pi E(\theta)e^{i2\theta}\,d\theta$ applies unchanged: the discrete form is the line-stimulus special case, not a restriction of the operator. This is the bridge to the image-statistics companion work (\S4). The \emph{operator} carries over to a broadband scene unchanged. The \emph{spatial} weighting such a cluttered field then needs is the aggregate model's stated limitation (\S6), not part of the present claim.

\textbf{Doubled angles represent unoriented axes.} Orientation is a \emph{director} --- an axis with no arrowhead: $\theta$ and $\theta+\pi$ are the same orientation. The lowest-order rotation-equivariant circular embedding invariant to this symmetry is the first moment of the \emph{doubled} angle $2\theta$ (twice the orientation --- the ``take twice the angle'' step of Box~1). Circular statistics calls this the ``axial'' mean (Mardia \& Jupp, 2000). Liquid-crystal physics calls the measure of how well a population of axes aligns the \textbf{nematic order parameter} (de Gennes \& Prost, 1993). The two-dimensional form used here, the mean of $\cos 2\theta$ measured from the director, is our reduction of it. Here $R=|c_1|/c_0$ is the scalar order parameter (the stimulus coherence, $0$--$1$) and $\phi$ the director, or mean axis. This physics formalism is standard. Doubled-angle means and vector-averaging are long-established for orientation perception and the tilt illusions. We are not aware of a prior account that formulates the induced visual vertical as this doubled-angle first moment. Averaging raw orientations is ill-posed (lines at $1^\circ$ and $179^\circ$ would average to $90^\circ$, orthogonal to both); the doubled-angle moment returns $0^\circ$, as it must. Higher harmonics and other circular loss functions remain possible. But $c_1$ is the lowest-order nonzero moment, and it is the readout tested here.

\textbf{Conditional uniqueness.} The doubled-angle first moment is not the only conceivable decoder. It \emph{is} the only additive axial summary with the covariance the model claims. Let $F$ assign a complex vote to a finite weighted line field. Assume four things: (i) component additivity, $F(X+Y)=F(X)+F(Y)$, where $X$ and $Y$ are added as orientation measures; (ii) salience homogeneity, $F(aX)=aF(X)$ for $a\geq0$; (iii) rotation covariance, $F(R_\alpha X)=e^{i2\alpha}F(X)$; and (iv) a nonzero vertical calibration, $F(A\delta_0)=KA$ with $K>0$. A line at $\gamma$ is a rotated vertical line, so covariance gives $F(A\delta_\gamma)=KAe^{i2\gamma}$. Additivity then leaves no freedom:
\[
F(X)=K\sum_j A_j e^{i2\gamma_j}=Kc_1/\hat\kappa_1.
\]
Continuity under weak convergence of finite orientation measures extends the same result to $F(E)=K\int_0^\pi E(\theta)e^{i2\theta}\,d\theta$. The substantive assumptions are therefore visible, and they divide in two. Director symmetry does part of the work unaided: because rotating a director field by $\pi$ returns the same field, only the \emph{even} rotational orders can carry a nonzero summary at all, so $c_1$ is the lowest nontrivial rung of the ladder $c_1,c_2,c_3,\ldots$ (for a summary that transforms under a rotation character, Appendix~\ref{app:unique}) rather than a harmonic order picked by hand. Within that rung, \emph{linear pooling plus positive homogeneity} pins $c_1$ uniquely. Dropping additivity admits winner-take-all and other nonlinear rules; keeping additivity but climbing the ladder produces the higher moments $c_k$ of \S5.4. Appendix~\ref{app:unique} proves both halves.

\section{The Li--Matin rules, derived}
From the boxed definition, with $\kappa$ dropping out:
\begin{itemize}
\item \textbf{R1 --- linear tracking.} One inducer at $\gamma$: $c_1\propto e^{i2\gamma}$, hence $\phi=\tfrac12(2\gamma)=\gamma$. Unit-slope tracking of a single line. \emph{(In vote language: one line casts one vote, halved back to its own tilt.)} Throughout, in the text and in the analysis code, orientation $\gamma$ and the readout $\phi$ are measured as \textbf{tilt from vertical} (vertical $=0^\circ$, horizontal $=90^\circ$), with a positive tilt the same rotational sense for every line, so a horizontal line tilted by $\delta$ sits at $90^\circ+\delta$. A symmetric pair about vertical therefore returns $\phi=0$.
\item \textbf{R2 --- symmetric cancellation.} Two equal inducers at $\pm\theta$: $c_1\propto e^{i2\theta}+e^{-i2\theta}=2\cos2\theta\in\mathbb R$, so for $|\theta|<45^\circ$ ($\cos2\theta>0$) $\arg c_1=0$ and $\phi=0$. Symmetric tilts cancel --- a non-trivial prediction, invisible from either line alone, and exactly what a linear/vector readout must do. \emph{(Worked example: two lines at $\pm30^\circ$ cast votes at $\pm60^\circ$; the two votes sum straight up the vertical, so $\phi=0$ --- Box~1, Figure~\ref{fig:2}, panel~4.)} (For $|\theta|>45^\circ$ the real part turns negative, $\arg c_1=\pi$, and the asserted axis is horizontal, $\phi=90^\circ$ --- the correct behaviour for a pair of near-horizontal lines. The mirror-pair data of \S5 are near-vertical and lie in the $|\theta|<45^\circ$ branch.)
\item \textbf{R3 --- weighted linear (vector) combination.} Two inducers at $\gamma_1,\gamma_2$ with weights $A_1,A_2$: $\phi=\tfrac12\arg(A_1e^{i2\gamma_1}+A_2e^{i2\gamma_2})$. This is \textbf{phasor addition in the doubled-angle domain}. It is the precise sense of Li and Matin's ``the two lines combine linearly'': linearity holds in the $2\theta$ representation proper to axial data. In the small-tilt limit $e^{i2\gamma}\!\approx\!1+i2\gamma$ it collapses to the familiar weighted average $\phi\approx(A_1\gamma_1+A_2\gamma_2)/(A_1+A_2)$. Ordinary linearity is thus recovered where their tilts were small, and the general law predicts departures at larger separations. The exact unequal-salience surface is especially diagnostic. Write the two orientations as $\gamma_{1,2}=m\pm d$, so $m$ is their mean and $d$ their half-separation, and let $\rho=(A_1-A_2)/(A_1+A_2)$ be the salience imbalance. Then
\[
\phi=m+\tfrac12\operatorname{atan2}\!\left(\rho\sin2d,\ \cos2d\right),\qquad
R^2=\cos^22d+\rho^2\sin^22d.
\]
Here $\operatorname{atan2}(y,x)$ is the quadrant-aware arctangent, so the formula keeps the axial angle on its correct branch. In words: the readout is the pair's mean orientation plus a pull toward the stronger line that grows with their separation, and the coherence falls with separation unless the imbalance $\rho$ props it up. For a geometrically symmetric pair ($m=0$, $d=\theta$) the imbalance susceptibility away from the null is $\left.\partial\phi/\partial\rho\right|_{\rho=0}=\tfrac12\tan2\theta$, which runs away as $\theta\to45^\circ$: the two votes have all but cancelled, so whichever line is stronger sets the axis on its own. Exactly at $45^\circ$ the asserted axis $\phi$ stays bounded but selects a branch --- the stronger line wins the $+45^\circ$ or the $-45^\circ$ axis. \textbf{That runaway belongs to $\phi$, not to a measured setting}, and the distinction is the same one \S5.6 draws for every other quantity here. A setting weighs $\phi$ against the body/gravity reference. Section~5.6 writes that weighting as a sum of two concentrations --- $\eta_1$ for the visual axis and $\beta$ for the upright body axis, whose ratio fixes the measured visual weight $k_V=\eta_1/(\beta+\eta_1)$ --- and derives the setting from it (Eq.~\ref{eq:cue-combination}). We quote the result here; it follows from that equation by the same differentiation of $\tfrac12\arg q_1$ that \S5.6 applies to the single line. Carried through the setting, the susceptibility $\tfrac12\tan2\theta$ becomes
\[
\left.\frac{\partial\widehat v}{\partial\rho}\right|_{\rho=0}=\frac{\eta_1\sin2\theta}{2\left(\beta+\eta_1\cos2\theta\right)},
\]
of which $\tfrac12\tan2\theta$ is the body-free ($\beta\to0$) limit. In words: the imbalance's pull on the setting is divided by the whole concentration holding that setting in place, visual \emph{and} body, and the body term never vanishes. At $\theta=45^\circ$ the visual term drops out of the denominator and the slope is the finite $\eta_1/2\beta$, so even a \emph{fully} one-sided pair --- which is just a single line at $45^\circ$ --- moves the setting by $\tfrac12\arctan(\eta_1/\beta)$ under the first harmonic alone: $4.0^\circ$ at $k_V=0.123$ and $5.0^\circ$ at $k_V=0.150$, not $45^\circ$. Branch selection is therefore a property of the stimulus-side axis alone, and \emph{no} observer displays it at the orthogonal pair: there $s_1=0$, so $q_1=\beta$, and the setting is continuous in the stimulus for every $\beta>0$ (\S5.6). This two-variable contrast-by-separation surface remains a sharper forward test than the null alone, but what it reads off near $45^\circ$ is, under the first harmonic alone, the visual-to-body ratio $\eta_1/\beta$, not a singularity.
\item \textbf{R4 --- coherence law (a new prediction, not a Li--Matin rule).} For a symmetric pair, $R=|2A\cos2\theta|/2A=|\cos2\theta|$ (using $\hat\kappa_1=\hat\kappa_0$, the convention fixed in \S2). The first-moment coherence falls from full agreement at $\theta=0$ to \textbf{zero at $\theta=45^\circ$}, where the orthogonal cross asserts no first-moment axis. It then rises back to full agreement as $\theta\to90^\circ$, where the pair re-aligns. Across the null the real moment $c_1\propto\cos2\theta$ changes sign, which for an axis is a change of phase: the asserted axis switches from vertical to horizontal, while the coherence itself is nonnegative and simply recovers. The coherence does not ``reverse.'' A measured reversal in perceived vertical needs one more ingredient --- an explicit response map that combines this visual evidence with the observer's body/gravity reference (\S5.6). Two honest caveats bound the novelty here. First, $|\cos2\theta|$ is the \emph{immediate} corollary of any two-component oriented-energy / structure-tensor sum: two unit phasors at $\pm2\theta$ sum to $2\cos2\theta$. The functional form is therefore no surprise to that literature. The contribution is to \emph{name} it as the modulus of the first moment and make the \textbf{whole curve}, not just the null, a zero-constant stimulus-side target. Second, that symmetric/$45^\circ$ configurations cancel is a classic observation --- the rod-and-frame illusion vanishes with the frame at $45^\circ$, where the frame's diagonal lies on the gravitational vertical and its other two axes stand equidistant from it (Beh, Wenderoth, \& Purcell, 1971; Beh \& Wenderoth, 1972). Those authors read the null as a major-axis effect; Hartley (1982) later traced orientation illusions of this family to the figure's axes of bilateral symmetry. \textbf{Two different $45^\circ$ nulls must not be conflated.} The classical null is a \emph{figure-tilt} null: rotating a fixed symmetric figure to $45^\circ$ brings a symmetry axis onto the vertical. There the independent variable is the whole figure's orientation, which is what Wenderoth and Beh's and Hartley's angular functions measure. R4's null is a \emph{separation} null: the pair's \textbf{mean is held at vertical} and only the half-separation $\theta$ is opened. The inducer's own axis therefore never leaves vertical, and the coherence falls to zero purely because the two phasors become antiparallel at $\pm45^\circ$. We could find no prior study that fixes a symmetric pair about vertical and sweeps the opening angle to map the $|\cos2\theta|$ coherence-vs-separation curve. That curve, with its null-strength tests (\S6), is the genuine forward target here. (The relevant caution from that literature is instead \emph{nonadditivity}: two-line tilt effects need not equal the sum of single-line effects --- Wenderoth \& Curthoys, 1974 --- which our leading-harmonic account captures only where $c_1$ dominates, \S5.4.)
\end{itemize}
Point (iii) --- retinal orientation drives it --- is automatic: $c_1$ is computed from the retinal orientation distribution, so the readout is in retinal coordinates by construction.

\begin{figure}[!htbp]\centering
\includegraphics[width=\linewidth]{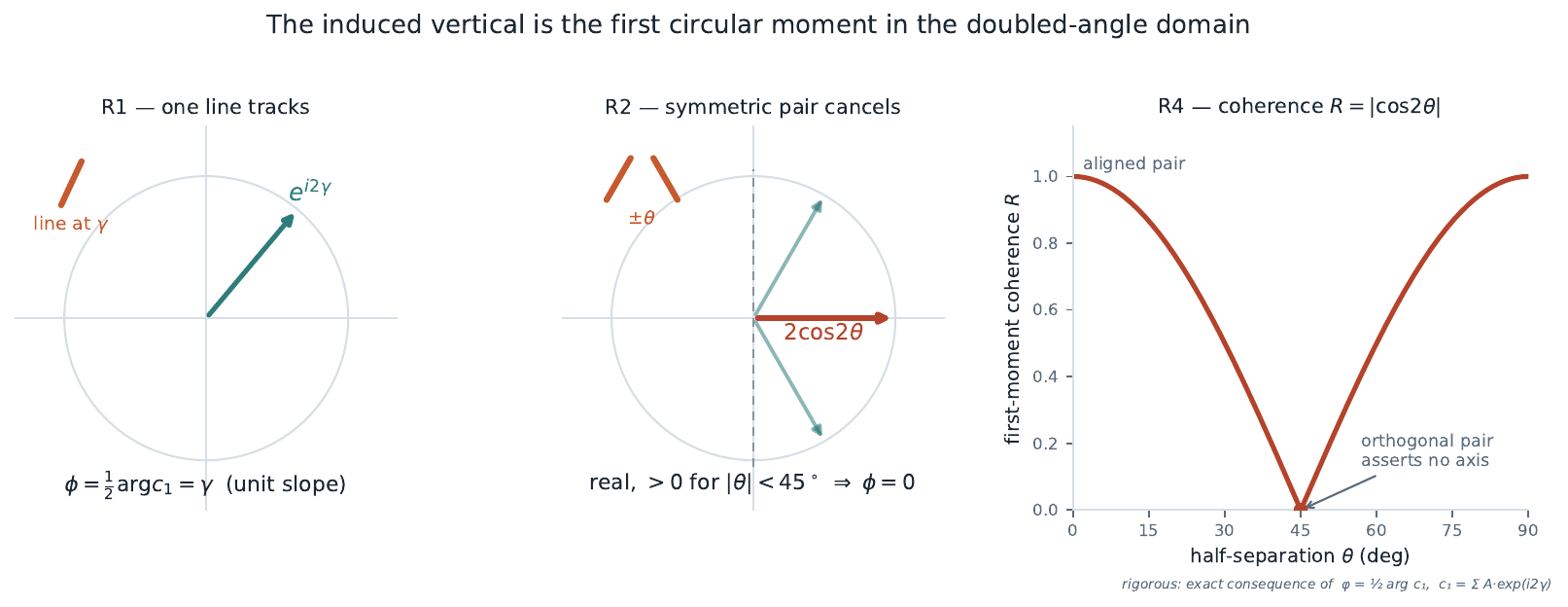}
\caption{The induced vertical as the first circular moment in the doubled-angle domain. R1: a single line at $\gamma$ maps to a phasor at $2\gamma$, so $\phi = \tfrac12\arg c_1 = \gamma$ (unit-slope tracking). R2: a symmetric $\pm\theta$ pair maps to conjugate phasors whose sum is purely real (zero imaginary part), so $\phi = 0$ for $|\theta|<45^\circ$ (the tilts cancel). R4: the first-moment coherence is $R = |\cos 2\theta|$, full at alignment and zero for an orthogonal ($45^\circ$) pair. Every panel is an exact consequence of the definition, not a schematic.}
\label{fig:3}
\end{figure}

\begin{figure}[!htbp]\centering
\includegraphics[width=\linewidth]{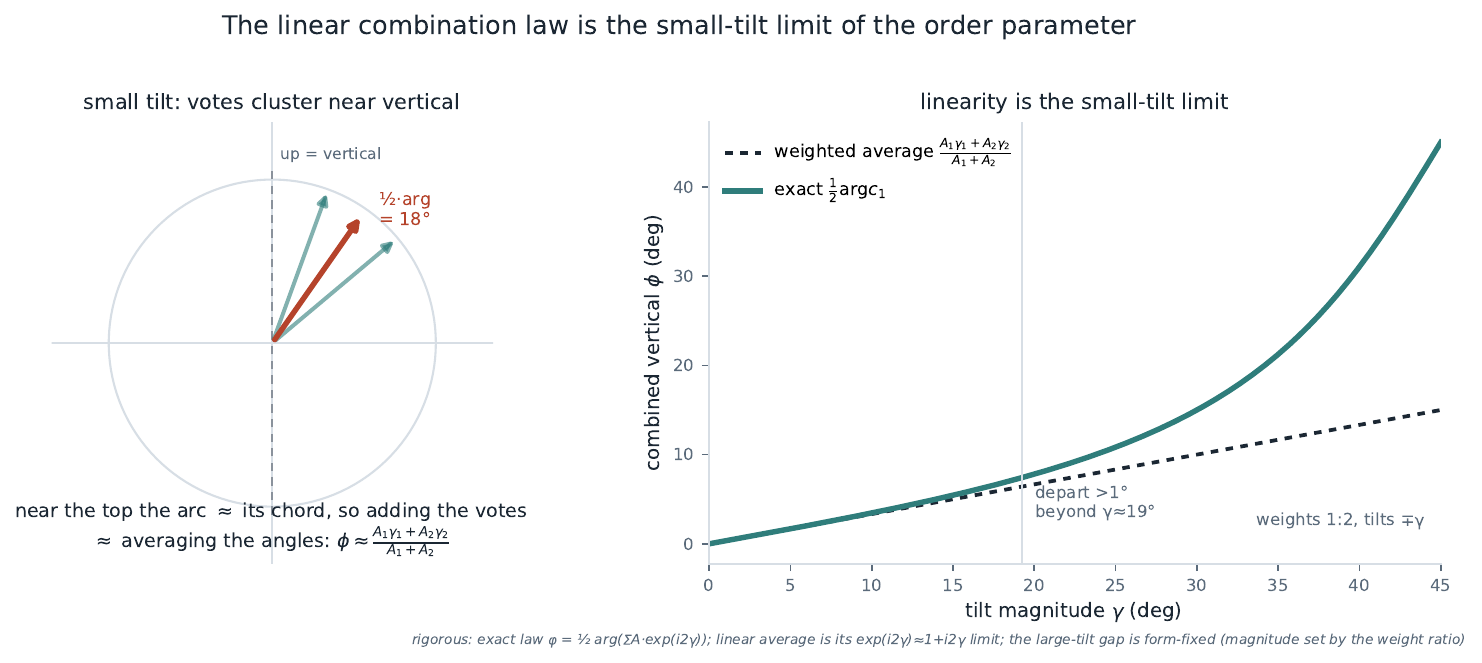}
\caption{The linear combination law (R3) is the small-tilt \emph{limit} of the order parameter. \emph{Left:} for small tilts the doubled-angle phasors sit near the top of the circle (arc $\approx$ chord), so adding the votes is the ordinary weighted average $\phi\approx(A_1\gamma_1+A_2\gamma_2)/(A_1+A_2)$. Li and Matin's tilts were small, so they observed this linearity. (For equal weights the readout is exactly the mean of the two angles at any tilt, once each angle is taken within $90^\circ$ of the other; exactly orthogonal lines cancel, $c_1=0$, and assert no axis. The small-tilt approximation does its work only when the weights differ.) \emph{Right:} for unequal weights ($1{:}2$, symmetric tilts $\mp\gamma$) the exact readout $\tfrac12\arg c_1$ tracks the weighted average at small tilt and \emph{departs} at larger separation. The gap's functional form is parameter-free in the angles, and the weight ratio sets its magnitude. The departure regime ($\gtrsim19^\circ$) lies beyond the $\pm15^\circ$ tilts of Li and Matin's (2005b) two-line data --- a forward prediction, not yet tested.}
\label{fig:smalltilt}
\end{figure}

\paragraph{A general influence and stability law.} The two-line susceptibility in R3 is one instance of a Jacobian that applies to every line field. Write $z=\sum_jA_je^{i2\gamma_j}=|z|e^{i2\phi}\ne0$. For small changes in salience and in orientation,
\begin{equation}
d\phi=\frac{1}{2|z|}\sum_j\sin 2(\gamma_j-\phi)\,dA_j
+\frac{1}{|z|}\sum_j A_j\cos2(\gamma_j-\phi)\,d\gamma_j.
\label{eq:influence}
\end{equation}
An inducer has maximal \emph{contrast} leverage when its doubled-angle vote lies transverse to the current resultant, and maximal \emph{orientation} leverage when it lies along it. Both leverages scale as $1/|z|=1/(A_\Sigma R)$, writing $A_\Sigma=\sum_jA_j$ for the total effective salience. Coherence is therefore a conditioning factor and not merely a descriptive length --- but a \emph{dimensionless} one only once a perturbation is measured relative to $A_\Sigma$. In absolute units the denominator is the product $A_\Sigma R$, so weak inducers and disagreeing inducers destabilise the axis by the same factor, and total salience is as load-bearing here as coherence. The same holds for finite perturbations. If a perturbation changes the vote by $\Delta z$ with $|\Delta z|<|z|$, plane geometry bounds the resulting shift of the axis:
\begin{equation}
\operatorname{dist}_{\pi}\!\left(\phi(z+\Delta z),\,\phi(z)\right)
\leq\tfrac12\arcsin\frac{|\Delta z|}{|z|}.
\label{eq:stability}
\end{equation}
Here $\operatorname{dist}_{\pi}$ is the shortest distance between axial angles modulo $\pi$. The bound is sharp, and for a perturbation of fixed size relative to $A_\Sigma$ it becomes nonuniform only as $R\to0$ --- exactly where the asserted axis stops being identifiable. Appendix~\ref{app:sensitivity} derives both formulas, and the probe experiment of \S6 is their behavioural version. Figure~\ref{fig:3} draws R1, R2 and R4 as exact consequences of the definition, and Figure~\ref{fig:smalltilt} shows R3's linear combination law emerging as the small-tilt limit of the same object.

\section{Three views of the order parameter}
\begin{figure}[!htbp]\centering
\includegraphics[width=\linewidth]{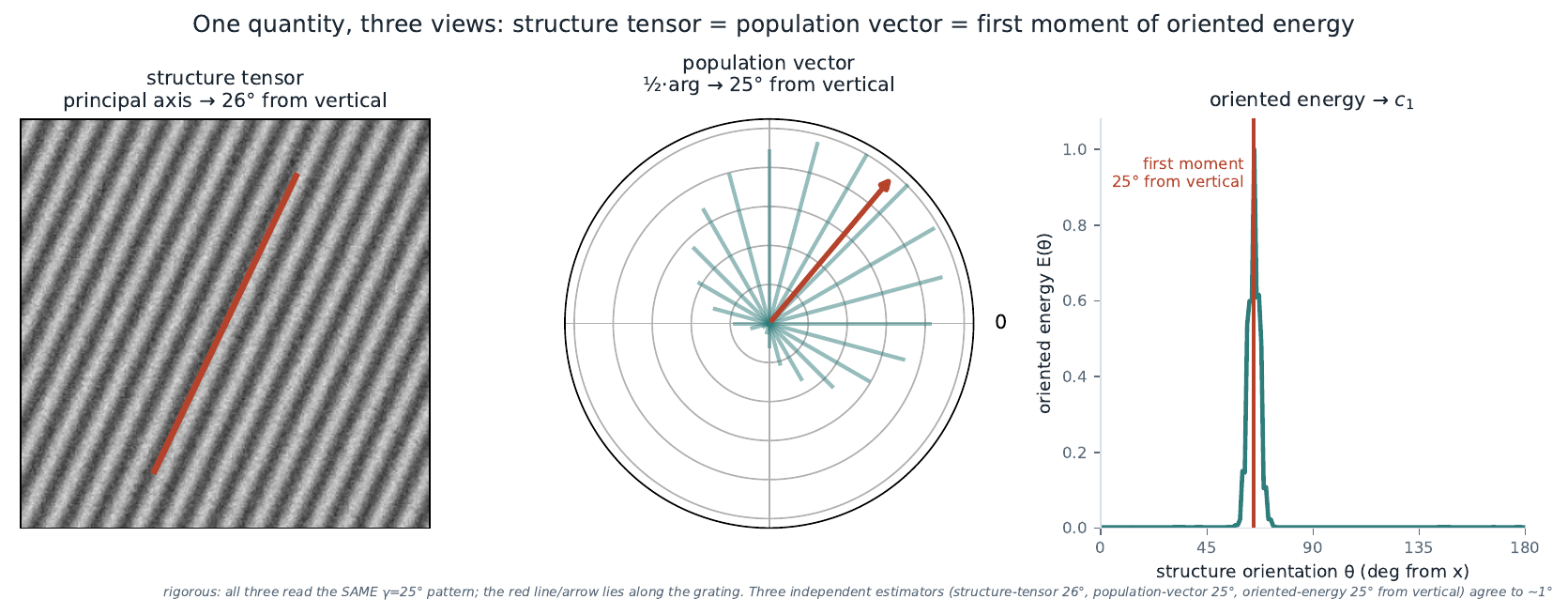}
\caption{One quantity, three views. All three panels are computed from the \emph{same} oriented pattern (a noisy grating at $\gamma=25^\circ$ from vertical) and return the same axis to within $1^\circ$: (left) the principal axis of the image \textbf{structure tensor}; (middle) the \textbf{population vector} $\tfrac12\arg c_1$ of a bank of cosine- (doubled-angle-) tuned orientation units; (right) the first moment of the \textbf{oriented-energy} profile $E(\theta)$. The order parameter $\phi=\tfrac12\arg c_1$ is these three objects at once (\S4). The right panel's abscissa is measured from the horizontal, so its peak near $65^\circ$ is $25^\circ$ from vertical.}
\label{fig:threeviews}
\end{figure}

The quantity $\phi=\tfrac12\arg c_1$ is not new mathematics; its identities are the point.
\begin{itemize}
\item \textbf{Structure tensor.} With $E$ the band-limited orientation energy of an image spectrum, $c_1$ is exactly the off-isotropic part of the spectral second-moment (structure-tensor) matrix, and $\phi$ is its principal eigenvector's orientation (Appendix~\ref{app:st}). One convention has to be stated, because the two literatures differ by a right angle: here $E(\theta)$ is indexed by \emph{line} (tangent) orientation, so the principal eigenvector lies \emph{along} the line, whereas a front end built on the conventional \emph{gradient} structure tensor returns a dominant eigenvector \emph{normal} to it, and the line-axis readout is then that matrix's other eigenvector. The two carry the same anisotropy and the same $R$; only the labelling of the axis differs. So the induced-VPV combination rule and the image \emph{structure tensor} (Bigün \& Granlund, 1987; Knutsson, 1989) are the same mathematical object, under the idealisations stated here. It is the very orientation descriptor applied to natural-scene and sport-environment spectra in the companion image-statistics work (Shavit \& Leopold, 2023).
\item \textbf{Population vector.} For a channel of orientation-tuned units with cosine (doubled-angle) tuning and firing rates $\propto E$, $c_1$ is a population vector and $\phi$ its preferred axis. The decoder is Georgopoulos, Schwartz, and Kettner's (1986). Theirs runs on the sphere of movement directions, where a direction and its reverse are opposite. Orientation needs the half circle, so the same vector average runs on the doubled angle. That axial form is what population-decoding models of orientation use (Schwartz, Sejnowski, \& Dayan, 2009; Tzvetanov, 2012); Webb, Ledgeway, and McGraw (2010) run the same vector average over a $180^\circ$ orientation population, written on the single angle. The readout thus has a standard neural-coding form, a vector average over an orientation population. Induced VPV can then arise from a low-level orientation readout, with no dedicated ``vertical'' computation to posit.
\item \textbf{Oriented energy.} $E(\theta)$ is an oriented-energy profile --- energy in the quadrature-pair sense of Adelson and Bergen (1985), read over static orientation rather than over their space-time orientation --- so the model is computable from any oriented-filter front end, not just the Fourier spectrum.
\end{itemize}
Under those idealisations the three are one mathematical quantity viewed three ways. That is an identity between descriptions, not evidence that the visual system carries out this particular computation. Figure~\ref{fig:threeviews} computes all three views from the same oriented pattern.

\textbf{Novelty and precedent.} The doubled-angle first moment as an orientation \emph{readout} is standard, and we claim no novelty for it. Perceived global orientation is well modelled as the vector average / circular mean of an orientation population (Webb, Ledgeway, \& McGraw, 2010; Dakin, 2001) --- with the qualification that Webb et al.\ find the vector average at short stimulus durations and a maximum-likelihood read-out at longer ones. The model that best predicts how humans \emph{average} orientation is precisely the centroid $\tfrac12\arg\sum_j p(\theta_j)\,e^{i2\theta_j}$, which carries \textbf{no free parameters} and beats peak, threshold-edge, and zero-crossing alternatives (Dakin \& Watt, 1997). PLUMB reads out the same operator: the \emph{classical} vector-average / population-vector decoder. It is deliberately not the information-optimal maximum-likelihood readout: Jazayeri and Movshon (2006) class the ``population vector'' and ``winner-takes-all'' decoders as suboptimal ``under most conditions of interest'' (p.~690), a contrast they draw in general rather than one that turns on how evenly the tuning is spread. Other parts of the literature use the same operation. Context-\emph{induced} perceived tilt has been modelled as a population read-out of context-shifted orientation responses (Schwartz, Sejnowski, \& Dayan, 2009). The doubled-angle ($2\gamma$) representation is the natural circular-feature framework for orientation (Tzvetanov, 2012). And the identical construction --- the argument of the first moment of doubled-angle gradient vectors --- is the classical structure-tensor / oriented-pattern flow of image analysis (Kass \& Witkin, 1987; Bigün \& Granlund, 1987).

The contribution here is the \textbf{transfer and its consequences}. Vector-sum accounts of the subjective vertical itself go back to Mittelstaedt (1983), whose setting is the resultant of a gravity vector and a body-fixed ``idiotropic'' vector; he later added a vector for the visual frame (Mittelstaedt, 1986, as cited in Vingerhoets et al., 2009). Vingerhoets, De Vrijer, Van Gisbergen, and Medendorp (2009) applied that model, and a Bayesian alternative, to a single peripheral line. They represented the line as a fourfold cue --- four frame vectors in Mittelstaedt's sum, or a von Mises frame likelihood in $\cos4r$ --- and fit its measured effect as a sum of twofold and fourfold harmonics. Our cue-combination bridge (\S5.6) is Mittelstaedt's resultant written in the doubled-angle plane, and our second-harmonic frame term re-derives what those authors model. The prominent current accounts of the \emph{induced visual vertical} include Bayesian subjective-visual-vertical (SVV) estimators (Alberts et al., 2016; Tian et al., 2025) and \emph{direction}-domain weighted vector sums of a few cue directions (Dyde, Jenkin, \& Harris, 2006); this list is a sample of the comparison frameworks rather than a census, and none of those listed is a first moment of the stimulus orientation \emph{distribution}. We claim that Li and Matin's combination rules R2 and R3, the new coherence law R4, and the VPV/VPEL sum/difference (\S5.5) follow from this order parameter in closed form; a fourfold term also yields R2, R3 and the sum/difference (\S5.3, \S5.5). R1, single-line tracking, follows from it too, but in Li and Matin's data a fourfold term carries almost all of that effect (\S5.3). The square-frame handoff to $c_2$ (\S5.4) is a re-derivation in this notation, not a new result. What is new there is the doubled-angle axial reading of the visual vote, the referral $s_k\bar g^{\,k-1}$ that makes every harmonic a natural parameter (\S5.6), and the no-go bracket on the frame peak for the phase-preserving family (\S5.6). The novelty is the application plus the derived combination structure, not the moment.

\section{Quantitative contact with the published data (Li \& Matin, 2005a, 2005b)}
The model is \textbf{parameter-free in its angles}, so it can be laid directly over the \emph{published} induced-VPV data. We do so here with two sets of published fitted coefficients. The first is from Li and Matin (2005a, \emph{Perception}), the rod-and-frame ``whole is less than the sum of its parts'' study that measured 1-, 2-, 3-, and 4-line combination. The second is from the companion orientation/length functions (Li \& Matin, 2005b, \emph{Vision Research}). No new data are collected. \texttt{analysis/fit\_vpv\_limatin.py} reproduces the comparison and figure below from the published slope values.

\subsection{R3 --- the combination law is the sharpest published test.}
\begin{figure}[!htbp]
\centering
\includegraphics[width=\linewidth]{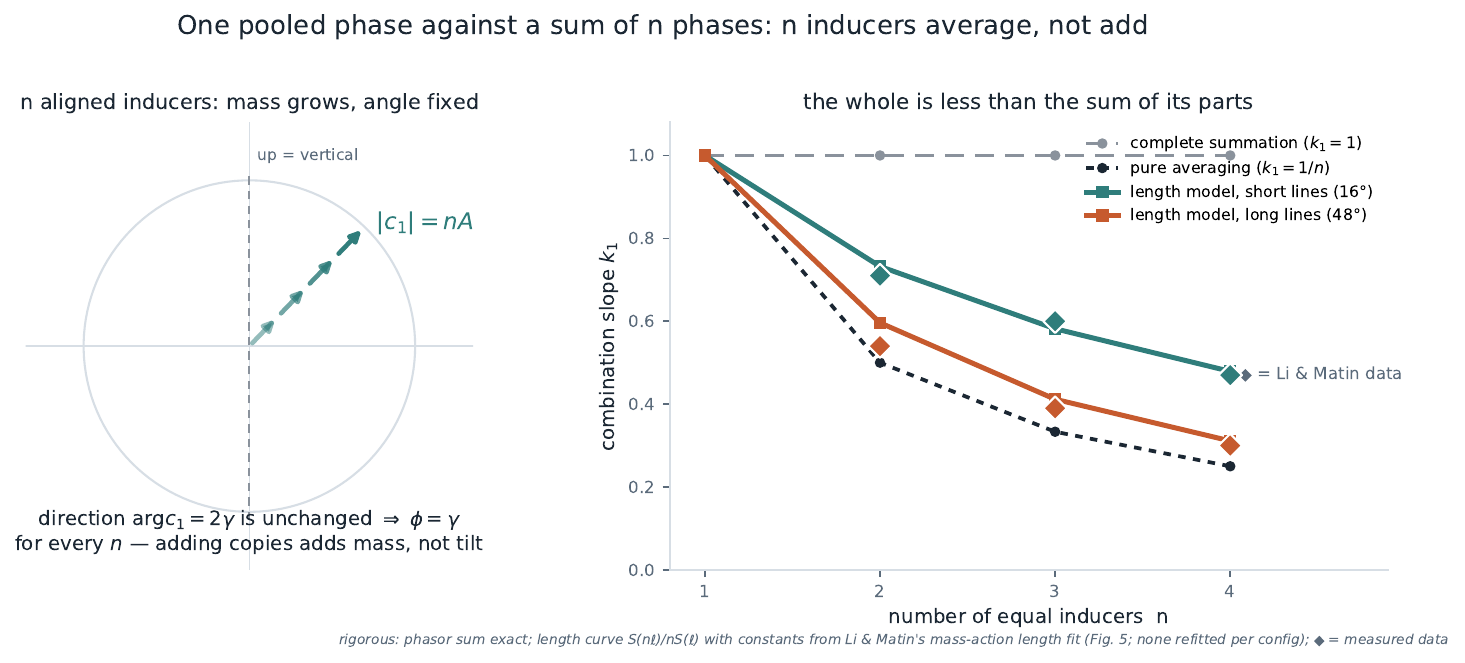}
\caption{The whole is less than the sum of its parts. \emph{Left:} $n$ aligned inducers add in \emph{mass} ($|c_1|=nA$ grows) but not in \emph{angle} ($\arg c_1=2\gamma$ is unchanged). The normalised readout $\phi=\tfrac12\arg c_1$ is therefore the same for every $n$: adding copies adds \emph{mass}, not tilt. \emph{Right:} the per-inducer combination slope $k_1$ therefore falls from complete summation ($k_1=1$) toward pure averaging ($1/n$). Li and Matin's length saturation fixes it in between, and the length-model curves reproduce their measured $k_1$ (diamonds, $n=2,3,4$: the parallel pair, three sides of the square and the full square, so not all aligned; the averaging comes from comparing one pooled phase with a sum of $n$ phases, not from the normalisation itself, \S5.1) with no per-configuration free parameters. The data refute $k_1=1$ but are \textbf{not} statistically separable from simple $1/n$ averaging at the long-line asymptote (\S5.1). The short-line points, which sit well above $1/n$, are the discriminating regime. The phasor sum is exact; the length curve is $S(n\ell)/nS(\ell)$ with constants from Li and Matin's mass-action length fit (their Fig.~5).}
\label{fig:whole}
\end{figure}

Li and Matin found that the multiline VPV is a \emph{linear} function of the \textbf{sum} of its constituent single-line VPVs, $V'_m = k_1\sum_i V_i + k_2$ (their eq.~1) --- exactly the linearity R3 asserts. The key is the slope $k_1$. A readout that summed influences without normalisation would give $k_1=1$ (``complete summation''). The first-moment \emph{phase} instead returns one pooled orientation: for small tilts, $\phi\simeq\sum_jA_j\gamma_j/\sum_jA_j$. Comparing that single pooled average against the \textbf{sum} of $n$ equal single-line phases gives the \textbf{averaging slope $k_1=1/n$}. Dividing by $c_0=\hat\kappa_0\sum_j A_j$ is a useful way to display the mean alongside its coherence, but it does not itself change the axis, because $\arg(c_1/c_0)=\arg c_1$ for $c_0>0$. The angular averaging and the scalar length gain below are therefore distinct parts of the prediction, and Figure~\ref{fig:whole} separates them.

\begin{figure}[!htbp]\centering
\includegraphics[width=\linewidth]{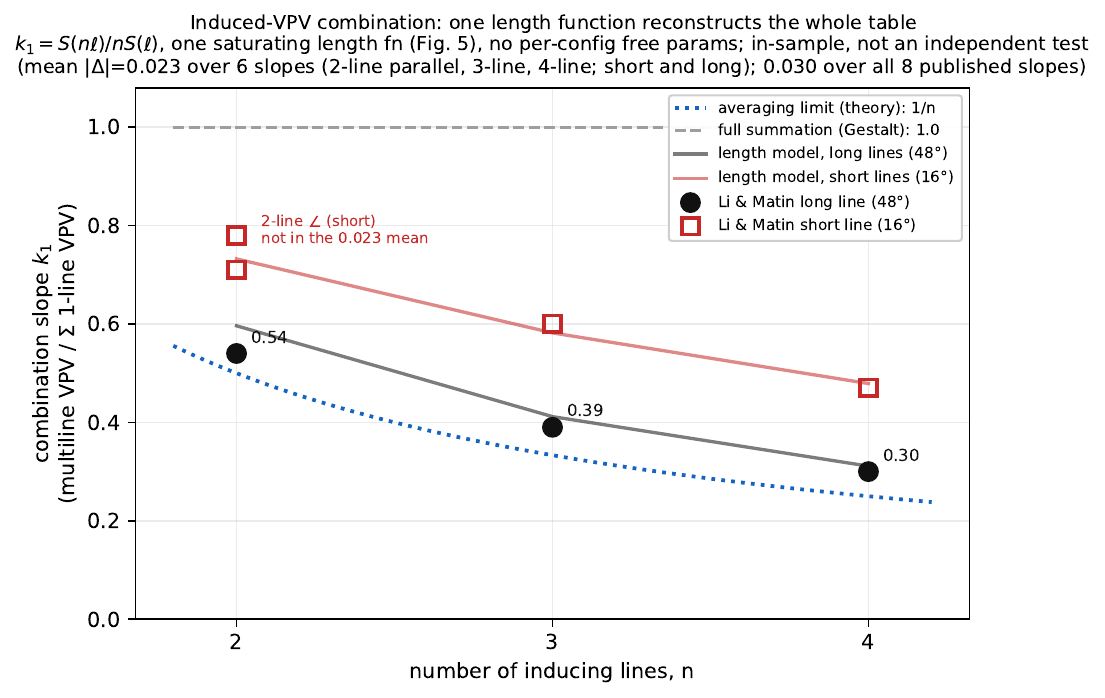}
\caption{(VPV-1). Induced-VPV combination coefficient $k_1$ for the 2-, 3-, and 4-line configurations versus the number of lines $n$. Li and Matin's measured slopes (long lines $48^\circ$, short $16^\circ$) are plotted against the length model $k_1=S(n\ell)/nS(\ell)$, whose constants are fixed by their own mass-action length fit (2005a, their Fig.~5; none refitted per configuration). The averaging limit $1/n$ and complete summation $1$ are shown for reference. Using the Fig.~5 curve as a descriptive gain law, this reconstructs the whole eight-slope table (mean absolute deviation 0.030); because both curves come from the same experiments, this is an internal-consistency check, not an independent test. The figure's own annotation reports a mean absolute deviation of 0.023 over six slopes (2-line parallel, 3-line and 4-line, each short and long); 0.030 is the figure over all eight published slopes, which add the two orthogonal 2-line fits. Neither set is the \emph{directional} one: of the eight, only the parallel 2-line and the 3-line have $c_1\neq0$, the square and the orthogonal pair both cancelling in the first moment (\S5.1).}
\label{fig:vpv1}
\end{figure}

In the \textbf{saturated (strong-inducer) limit} this averaging law is exact. Away from it, the same length dependence that governs R1 sets the \emph{degree} of normalisation: the single-line length-saturation function $S(\ell)=a+b\,(1-e^{-\ell/\ell_0})$ that Li and Matin fit to their own data (\S5.2). One function then reconstructs the whole table --- including where the short lines sit \emph{above} $1/n$. Because that function's constants come from the same experiments as the slopes it is being laid against, this is an internal-consistency check, not an independent test (Fig.~\ref{fig:vpv1}):

\begin{table}[!htbp]\centering
\small
\begin{tabular}{lrrrrr}
\toprule
Configuration & $n$ & $k_1$ (short, $16^\circ$) & $k_1$ (long, $48^\circ$) & length model $S(n\ell)/nS(\ell)$ & avg.\ limit $1/n$ \\
\midrule
2-line ($\parallel$ / $90^\circ$-angle) & 2 & 0.71 / 0.78 & 0.54 / 0.54 & 0.73 (short) / 0.60 (long) & 0.500 \\
3-line (three sides) & 3 & 0.60 & 0.39 & 0.58 / 0.41 & 0.333 \\
4-line (full square) & 4 & 0.47 & 0.30 & 0.48 / 0.31 & 0.250 \\
\bottomrule
\end{tabular}
\caption{Combination coefficient $k_1$ for the 2-, 3-, and 4-line configurations (built from Li and Matin's four single lines left, right, top and bottom of the test line; $\parallel$ is left plus right, the $90^\circ$ angle left plus top): measured least-squares slopes (short lines, $16^\circ$; long lines, $48^\circ$), the length-model reconstruction $S(n\ell)/nS(\ell)$ (constants fixed by Li and Matin's mass-action length fit, their Fig.~5 --- an in-sample calculation, not an independent test; see Fig.~\ref{fig:vpv1}), and the averaging limit $1/n$.}
\label{tab:limatin}
\end{table}

\begin{figure}[!htbp]\centering
\includegraphics[width=\linewidth]{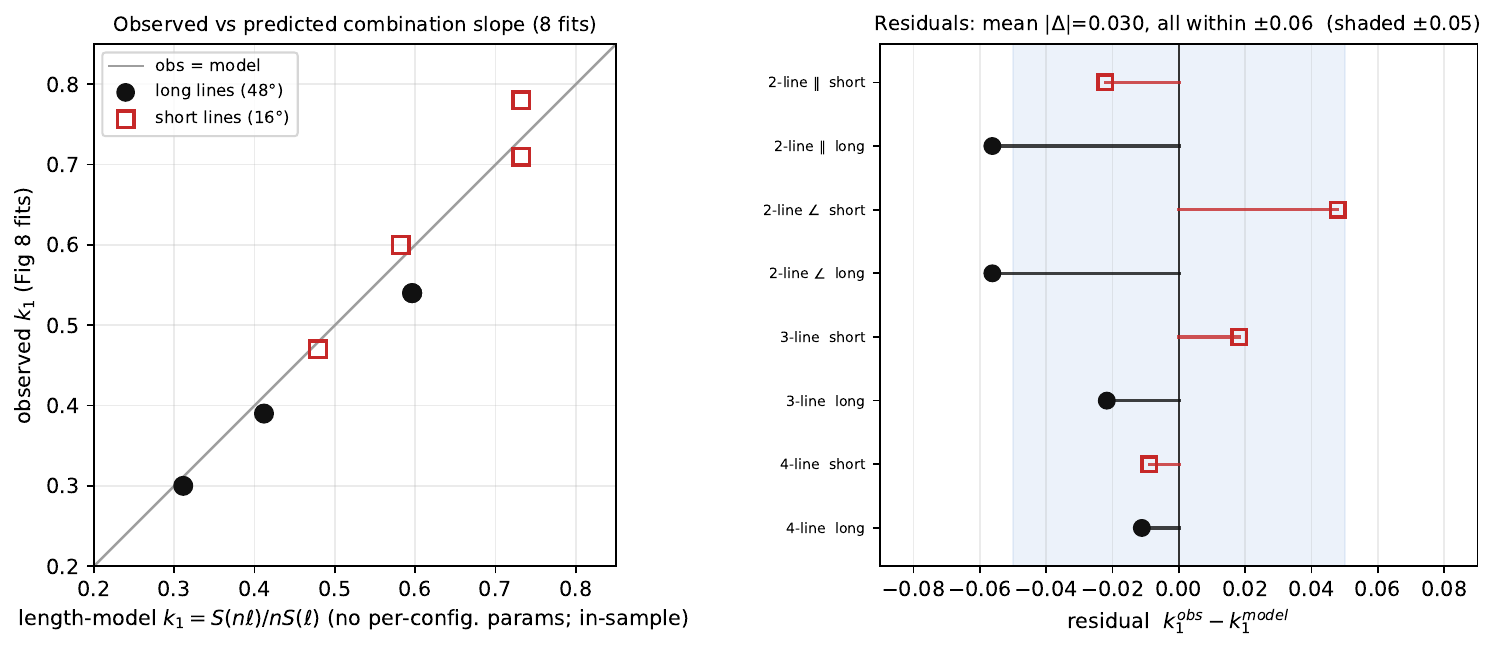}
\caption{(VPV-2). Goodness of fit for all eight least-squares slopes of Li and Matin (2005a). \emph{Left:} observed versus predicted $k_1$ on the identity line (obs $=$ model). \emph{Right:} residuals (observed $-$ predicted), all within $\pm0.06$ (mean $|\Delta|=0.030$). Long lines ($48^\circ$) are filled dark, short lines ($16^\circ$) open red.}
\label{fig:vpv2}
\end{figure}

\textbf{Limits of the published comparison.} \emph{Complete summation} ($k_1=1$) is refuted for every configuration --- no normalisation-free reading survives. The length model's constants $a,b,\ell_0$ are \textbf{fixed} by Li and Matin's own mass-action length fit (2005a, their Fig.~5, strength vs \emph{total} inducing length pooled over all configurations; none refitted per configuration here --- \S5.2 records the authors' own caveat on what that pooled constant measures). With those constants the length model matches all eight of the least-squares slopes printed in their Fig.~8 to within $\pm0.06$ --- mean absolute deviation \textbf{0.030}, versus \textbf{0.145} for the bare averaging limit $1/n$. But we do not overclaim a decisive win \emph{over averaging}. The improvement is \emph{concentrated in the short-line configurations}, where $1/n$ badly under-predicts, while the long lines already sit near the averaging asymptote. A two-sided exact-binomial sign test over the eight fits is therefore \textbf{not} significant (length model better on 6/8, $p=0.29$). Li and Matin themselves note the long-line 3-line value ``0.39 [is] not significantly more distant from 0.33 \ldots than is the 0.54 from 0.50 in the 2-line case.'' The honest reading is therefore this. The present data \textbf{refute complete summation and are closely matched by the one mass-action length function with no per-configuration free parameters, but cannot statistically separate it from simple $1/n$ averaging at the long-line asymptote}. The two predictions diverge, and a decisive test can be made, in the short-line regime (above) and in the $|\cos2\theta|$ coherence law (\S3, R4; \S6). The fitted intercepts $k_2$ (their Fig.~8: $-0.004$ to $-1.1^\circ$) are small, consistent with no additive offset ($\phi=0$ at $\sum_i V_i=0$). See \textbf{Figure~\ref{fig:vpv1}} (VPV-1: the combination coefficients from one length function) and \textbf{Figure~\ref{fig:vpv2}} (VPV-2: observed vs predicted for all eight fits on the identity line, residuals within $\pm0.06$).

\textbf{Two mechanisms, kept separate.} The coefficient $k_1=S(n\ell)/nS(\ell)$ (\S5.2) is a \emph{scalar} length / ``mass-action'' law: it depends only on \textbf{total inducing length}, not on orientation geometry. That is precisely why it also fits the \textbf{orthogonal} 2-line-angle and 4-line-square configurations. But for those the doubled-angle first moment is \emph{exactly zero for every configuration} ($c_1=e^{i2\gamma}+e^{i2(\gamma+90^\circ)}=0$; a square tilts to $c_1=0$ at every roll angle --- the cardinal-cross cancellation of \S5.4). Their nonzero induction is therefore Li and Matin's mass action, \textbf{not} the directional first-moment readout. The \emph{directional} content of the order parameter is separate: $n$ \textbf{coherent} inducers produce one scale-invariant pooled phase, hence the $1/n$ slope relative to the sum of $n$ phases. Li and Matin's single-line data (R1, \S5.2) and their 2-line-\textbf{parallel} runs bear on it. Their R2 data do not, because the mirror pair's null holds whatever the mix of harmonics (\S5.3). The single-line slopes behind R1 and the parallel pair are mostly fourfold, however: Li and Matin's near-horizontal lines give same-sign slopes, which put $c_1$'s share of a single line's effect at an estimated $2$--$4\%$ (\S5.3). No published data therefore separate $c_1$ from the fourfold term; \S6 lists the tests that could. The orthogonal configurations corroborate the scalar length law only. Conflating the two would overstate the case, so we keep them distinct.

\subsection{Single-line strength and multiline mass action share one length function --- no per-configuration free parameters.}
Over the near-cardinal range a single line induces VPV \emph{linearly} in its roll-tilt. For a near-vertical line this is R1, $\phi$ tracking the line; for a near-horizontal line the same-sign slope is the fourfold term's, not the first moment's (\S5.3). Its strength saturates as $S(\ell)=a+b\,(1-e^{-\ell/\ell_0})$, with $a=0.03$, $b=0.25$, $\ell_0=31.9^\circ$ (Li \& Matin, 2005a, their Fig.~5, where $\ell$ is \emph{total} inducing length) and small intercepts (mean $1.15^\circ$), consistent with a response that passes near zero when the inducer is untilted.

\textbf{One caveat on that constant, which its own authors state and we adopt.} Li and Matin themselves disown the mechanistic reading of $31.9^\circ$, in the text beneath that figure: the value ``is not a consequence of the length or of the eccentricity per se, but is largely a result of the fact that the plot in figure 5 combines results from two exponentials with different asymptotes, one for each of the two line lengths.'' Their short lines sat at $8.1^\circ$ eccentricity and their long lines at $26.4^\circ$, so pooling the two conditions inflates the fitted space constant; they show two exponentials of true constant $17.1^\circ$ but different asymptotes pool to an apparent $36^\circ$. The companion single-line measurement in the same year does give $17.1^\circ$ with asymptote $0.55$ (Li \& Matin, 2005b, their Fig.~4). The caveat has two consequences. The numerical result below is unaffected: $k_1=S(n\ell)/[n\,S(\ell)]$ is evaluated at a fixed eccentricity within each column of the table, so the eight slopes are matched by the curve Li and Matin published whatever generates its shape. But the \emph{mechanistic} gloss is weaker than ``mass action.'' Part of what $S$ tracks across the short-versus-long comparison is eccentricity, not total length, and the model's margin over $1/n$ averaging is concentrated in exactly that comparison (\S5.1). We therefore claim a fit to a published saturating function with no per-configuration free parameters, not a demonstration that induction is a pure function of total inducing length. Separating the two needs length varied at fixed eccentricity, which Li and Matin report they have measured but do not tabulate. This length-modulation of induction is itself a robust finding for both line and array inducers (Shavit, Li, \& Matin, 2008). If the published Fig.~5 curve is adopted as a descriptive gain law, it implies the combination coefficient: an $n$-line configuration has total length $n\ell$ and hence strength $S(n\ell)$, while the summed single-line reference is $n\,S(\ell)$, so
\[
k_1=\frac{S(n\ell)}{n\,S(\ell)}.
\]
Its saturated limit is the averaging law: for long, saturated $\ell$, $S(n\ell)\approx S(\ell)$ and $k_1\to1/n$. Complete summation ($k_1=1$) is the \emph{opposite}, no-saturation reference: the value if induction grew in strict proportion to length ($S\propto\ell$). Removing the offset does not produce it; with $a=0$ the same fit gives $k_1=0.80$ at $\ell=16^\circ$ and $0.61$ at $48^\circ$. Saturation alone keeps the sum short: $S$ is strictly concave with $S(0)=a\geq0$, which is sufficient for $S(n\ell)<n\,S(\ell)$ always, with or without a floor, so the predicted $k_1$ sits \textbf{strictly below 1 everywhere}. The nonzero floor ($a=0.03$) controls the short-line limit instead: $k_1\to1/n$ as $\ell\to0$ if and only if $a\neq0$, whereas $a=0$ sends $k_1\to1$. With the fitted floor, $k_1$ therefore rises from $1/n$ at very short lengths to an intermediate maximum ($\approx0.75$ for $n=2$, near $\ell\approx9^\circ$ --- below the $16^\circ$ shortest total length on their Fig.~5 abscissa, so that location is an extrapolation controlled by the fitted intercept $a$ rather than a measured feature), then falls back toward $1/n$ as the lines saturate. This \textbf{is} ``the whole is less than the sum of its parts,'' now quantitative. The saturating length function also places the short-line slopes (0.71--0.78 for $n=2$) between the $1/n$ asymptote and full summation. All of this is a statement about induction \emph{magnitude}, the scalar mass-action of \S5.1, and it reproduces the full eight-slope table with no per-configuration free parameters. The \emph{directional} combination --- which orientation the percept takes --- is the separate first-moment claim of R1--R3. The single-line slopes that calibrate $S(\ell)$ are mostly fourfold, since they include Li and Matin's near-horizontal lines with same-sign slopes; $S(\ell)$ is therefore a gain law for a mostly fourfold effect, not for $c_1$ alone (\S5.3). The length-saturation lives in the \emph{total unnormalised} inducer energy $C\propto\sum_j A_j$ (and hence in $S(\ell)$), not in the scale-invariant normalised ratio $R$ --- the direction/magnitude separation made precise in \S5.6.

\subsection{R2, the doubling signature, and what the full-circle line data rule out.}
\textbf{Symmetric cancellation.} R2 rests on Li and Matin's (2005b) equal-and-opposite pairs. Two lines at mirror-image tilts, within $\pm13.2^\circ$ of vertical, left VPV close to zero for short and long lines alike: the combined influence fell ``to zero for lines with equal-and-opposite orientations'' while each line alone shifted the setting. The fully symmetric square frame is the maximally cancelled case in the first moment, $c_1=0$. (Its residual rod-and-frame induction is the mass-action / higher-order effect of \S5.1 and \S5.4, not a first-moment effect.)

\textbf{The full-circle line data, and what they count against.} An earlier angle function is easily read as the same cancellation, and it is not. Li and Matin (1998a, 1998b, as cited in Li \& Matin, 2005a, \S1.2) rotated two $64^\circ$ long \emph{parallel} lines, $50^\circ$ apart, through twenty-four orientations covering the full circle. The induced VPV followed a sinusoid with four cycles per revolution: positive peaks at $22.5^\circ$ and $112.5^\circ$, zeros at $0^\circ$, $45^\circ$, $90^\circ$ and $135^\circ$, and negative peaks at $67.5^\circ$ and $157.5^\circ$. A parallel pair is not a symmetric configuration. Its two lines share one orientation, so its first moment is that of a single line, $c_1\propto2e^{i2\gamma}$, which never vanishes; the zero at $45^\circ$ is therefore not R2. The fourfold pattern holds for a single line as well, on two pieces of evidence. Li and Matin's (2005a) own single lines covered only the rising segments of their ``$4\alpha$'' function; their footnote~3 justifies fitting a complete sinusoid by appeal to the parallel-pair runs above, so it is not single-line evidence. The first piece is their near-horizontal lines, set out below. The second is Vingerhoets, De Vrijer, Van Gisbergen, and Medendorp (2009) swept a single $108^\circ$ frame line from $-90^\circ$ to $+90^\circ$ in $10^\circ$ steps and found a modulation of the subjective vertical with a period of about $90^\circ$. In their six upright observers it was small, about $2$--$6^\circ$ peak to peak (read off their Fig.~5), and it grew several-fold when the observers were rolled to $60^\circ$ or $120^\circ$.

These data count against the first moment at large tilts; the near-horizontal lines below count against it at small tilts as well. Coupled to the body reference (Eq.~\ref{eq:cue-combination}), the first moment gives a single line or parallel pair a bias with the sign of its tilt everywhere between $0^\circ$ and $90^\circ$. At $k_V=0.14$ it predicts $+2.9^\circ$ at $22.5^\circ$, $+4.6^\circ$ at $45^\circ$ and $+3.7^\circ$ at $67.5^\circ$, against a fitted sinusoid whose zero falls at $45^\circ$ and whose negative peak falls at $67.5^\circ$. (A single fourfold sinusoid has its zero at $45^\circ$ by construction; the measured fact is the reversal of sign beyond it.) A second harmonic can supply the reversal, but only by outweighing the first. With both terms of Eq.~\ref{eq:harmonic-completion} ($K=2$, upright observer) the bias of one line at tilt $\gamma$ has the sign of $\sin2\gamma\,(\eta_1+2\eta_2\cos2\gamma)$. Its interior zero therefore sits at $\cos2\gamma_0=-\eta_1/2\eta_2$, which lies beyond $45^\circ$ whenever $\eta_1>0$ and reaches $45^\circ$ only as $\eta_1/\eta_2\to0$; a negative bias at $67.5^\circ$ needs $\eta_2>\eta_1/\sqrt2$, a condition that is necessary and sufficient. In words: for Li and Matin's long parallel lines the fourfold term carries most of the effect over the full circle, and $c_1$ is at most a minor part of it. (Re-derived, with a control that fails, in \texttt{analysis/\allowbreak verify\_\allowbreak fourfold\_\allowbreak line\_\allowbreak 2026-10-15.py}.)

\textbf{Near-horizontal lines.} The small-tilt data separate the two harmonics too. Li and Matin (2005a) placed four single lines around the test line, two vertical (left, right) and two horizontal (top, bottom), each centred $26.4^\circ$ from fixation for the long lines, and tilted each over $\pm22.5^\circ$ about its own axis. All four functions rise with the same sign (their Fig.~3a, 3c), and the authors call the four slopes ``indistinguishable'', the horizontal ones ``slightly (but not significantly) smaller''. Read off their Fig.~4 (bar heights read by eye, so digitised rather than reported), the slopes are about $0.25$ and $0.25$ for the vertical and $0.24$ and $0.24$ for the horizontal long lines, and $0.13$ and $0.14$ against $0.12$ and $0.13$ for the short lines. The first moment forbids that. A horizontal line tilted by $\delta$ votes at $180^\circ+2\delta$, against the body reference, so under $c_1$ it moves the setting \emph{against} its tilt: at $k_V=0.14$ the horizontal-to-vertical slope ratio is $-1.39$. The measured ratio is about $+0.96$ (long) and $+0.93$ (short). The fourfold term votes alike for the two placements, since $4\times90^\circ=360^\circ$. With both terms, those ratios put the first moment's share of the near-vertical slope at about $2\%$ (long lines) and $4\%$ (short lines). The estimate errs high in one respect, because any intrinsic weakness of horizontal lines is charged to $c_1$ (\texttt{analysis/\allowbreak verify\_\allowbreak fourfold\_\allowbreak line\_\allowbreak 2026-10-15.py}, check~5). It does not cover sampling error. Li and Matin print these slopes without standard errors, so the share is a point estimate, and it moves quickly with the ratio: a horizontal-to-vertical ratio of $0.8$ would put it at $11.3\%$. The two equations behind those shares set the near-vertical slope $(\eta_1+2\eta_2)/(\beta+\eta_1+\eta_2)$ to $k_V=0.14$ (Shavit et al., 2013) and the near-horizontal slope $(2\eta_2-\eta_1)/(\beta-\eta_1+\eta_2)$ to the measured ratio times $k_V$; the share is $\eta_1/(\eta_1+2\eta_2)$. Solved exactly with the long-line ratio $0.96$, they give the two-harmonic weights that \S6 uses, $\eta_1/\beta=0.0035$ and $\eta_2/\beta=0.074$, point estimates like the share (\texttt{analysis/\allowbreak fig\_\allowbreak decisive\_\allowbreak test.py}). The solve combines two datasets and uses the long-line ratio only; the short-line ratio $0.93$ would give $\eta_1/\beta=0.0061$.

\textbf{Limits of the first-moment account.} (i) The first moment is not the dominant term of single-line induction, near the cardinal axes or beyond. Near vertical the two harmonics predict a bias of the same sign, since to first order the small-tilt slope is $(\eta_1+2\eta_2)/(\beta+\eta_1+\eta_2)$. Near-vertical tracking alone cannot separate them; Li and Matin's near-horizontal lines can, and they assign the effect to the fourfold term. (ii) R2 survives. Every moment of a mirror pair is real (\S6), so the pair's null holds whatever the mix of harmonics, and the cancellation data of Li and Matin (2005b) do not depend on the reading above. (iii) The combination coefficients of \S5.1 test a scalar length gain and are unaffected. (iv) The single-line turnover of \S5.6, a first-harmonic prediction, is already contradicted (\S6). What survives of the first-moment account is the doubled-angle combination structure, R2 and R3, and the algebra of \S\S2--4.

\textbf{VPV and VPH.} Li and Matin report the VPV and VPH (visually perceived horizontal) influences of a given inducer ``virtually indistinguishable'' --- their summary of the same parallel-pair design (Li \& Matin, 1998a, 1998b, as cited in Li \& Matin, 2005a), reached here only through the 2005a report. In the doubled-angle domain vertical ($\theta=0^\circ\!\to\!2\theta=0^\circ$) and horizontal ($\theta=90^\circ\!\to\!2\theta=180^\circ$) are antipodal, so a single $c_1$ readout drives both discriminations by the same amount. But any readout that rotates the perceived axes together predicts that equality, the fourfold one included. The observation is consistent with doubling and does not single it out. The \emph{single}-inducer version remains an open test (\S6).

\subsection{Scope of the fit: the square frame, harmonic survival, global versus local orientation, and Wenderoth and Beh.}
The core is parameter-free \emph{in its angles}, its combination \emph{magnitudes} riding on Li and Matin's three length constants. It reproduces the combination law (\S5.1) and symmetric cancellation (\S5.3). It reproduces linear tracking (\S5.2) as well, but a fourfold term carries almost all of that tracking (\S5.3), and \S5.3 also records the full-circle line data that the first moment cannot fit.

\textbf{The square frame and the second moment.} The classic rod-and-frame \emph{square} is the sharp test of scope. A square roll-tilted by $\theta$ has its four sides at just two orientations, $\theta$ and $\theta+90^\circ$. Opposite sides are parallel, so each orientation is carried by a collinear pair and enters the moments twice --- an overall factor absorbed by the $\propto$. Its first moment therefore vanishes \emph{identically},
\[
c_1\propto e^{i2\theta}+e^{i2(\theta+90^\circ)}=e^{i2\theta}\left(1+e^{i180^\circ}\right)=0,
\]
at \textbf{every} tilt --- yet a tilted square induces a large rod-and-frame effect. The four sides' votes cancel in orthogonal pairs, so the square is \emph{maximally sub-additive} in the first moment: its pooled signal is zero while its four lines' separate signals are not. Li and Matin's finding that the square is ``less than the sum of its parts'' (2005a) is therefore consistent with the first-moment account, not a counterexample. The square's residual induction is not a first-moment signal, and it cannot be the length term of \S5.1 either, which carries no orientation and so fixes neither the sign nor the tilt-dependence of a response. The square's orientation energy is instead \emph{promoted to the next harmonic}. Its second doubled-angle moment --- the fourth harmonic of the raw angle, matching the square's four-fold symmetry --- is
\[
c_2\propto e^{i4\theta}+e^{i4(\theta+90^\circ)}=e^{i4\theta}\left(1+e^{i360^\circ}\right)=2e^{i4\theta},
\]
which is nonzero and returns $\tfrac14\arg c_2=\theta$ (mod $90^\circ$). That $90^\circ$ ambiguity is the square's four-fold symmetry: $\theta$ and $\theta+90^\circ$ are the same frame, while a $45^\circ$ rotation gives the distinguishable \emph{diamond}. So the square's \textbf{axis} lives one harmonic up. Li and Matin's frame-tilt function nulls at both the upright square ($0^\circ$) and the $45^\circ$ diamond, is $90^\circ$-periodic, and reverses between --- the $c_2$ signature (\texttt{analysis/fit\_rodframe\_harmonic.py}). Alberts et al.\ (2016) measure the same null structure in their own rod-and-frame data.

\textbf{What $c_2$ does not carry: the envelope and its peak.} Because $|c_2|=2$ is constant in tilt, the second harmonic sets only the axis --- the nulls, the periodicity, the reversal --- and not the magnitude envelope. The rise to a peak and the fall come from coupling the harmonic to the body/cardinal reference, written in closed form in \S5.6 (Eq.~\ref{eq:square-readout}). The labour therefore divides in three: \S5.1's length curve can set a gain scale, $c_2$ the axis, and the coupling the envelope and sign. No two are rival mechanisms for the same part of the effect. The cardinal-anisotropy version of the envelope peaks near $24.6^\circ$, a pure number rather than a fitted one. Weighting the four edges by their cardinality, $w=1+a\cos2\gamma$, leaves the square a normalised first moment (\S5.6) $s_1=a\cos2\theta\,e^{i2\theta}$, so the curve $R\phi=a\,\theta\cos2\theta$ is stationary where $\cot2\theta=2\theta$, at $24.647^\circ$; the strength $a$ sets how large the effect is, never where it peaks (\texttt{analysis/\allowbreak verify\_\allowbreak rodframe\_\allowbreak peak\_\allowbreak provenance\_\allowbreak 2026-09-29.py}). The naive edge/diagonal-switch heuristic gives $22.5^\circ$. Either way the model overshoots the observed $15$--$20^\circ$ peak by $\sim5$--$9^\circ$, and \S5.6 shows the gap is structural: no member of the phase-preserving additive assimilative family peaks below $22.5^\circ$. We leave it as a stated limitation. Nor is $c_2=\sum_{\text{sides}}e^{i4\alpha}$ a configural exception to \S5.1: it is a linear sum over the four lines, with no whole-object term, consistent with Li and Matin's mass-action conclusion.

\textbf{Harmonic survival, and no switch at the handoff.} The square is one case of a general rule: for $M$ equal-weight orientations equally spaced (spacing $180^\circ/M$; a square's four sides give two orientations, each counted twice), the doubled-angle moments obey
\[
c_k=\sum_{j=0}^{M-1}e^{i2k\gamma_j}=0\quad\text{unless } M\mid k,
\]
so $c_M$ is the lowest-order moment symmetry allows, and for this equal-weight orbit it is nonzero. The orthogonal pair and the square ($M=2$) annul $c_1$ and reappear at $c_2$. The equilateral triplet at $0^\circ,60^\circ,120^\circ$ ($M=3$, \S6) annuls $c_1$ \emph{and} $c_2$ and survives only at $c_3$ --- a null no first- or second-harmonic (frame-like) mechanism can fill (pinned in \texttt{analysis/\allowbreak tests/\allowbreak test\_\allowbreak symmetry\_\allowbreak harmonics.py}). \textbf{We do not claim a \emph{phase}-mixing rule spanning the handoff} --- no $\text{bias}=w_1\phi_1+w_2\phi_2$ in which the visual system ``switches'' to $c_2$ when $c_1$ vanishes. Section~5.6 instead refers each harmonic back to the body/cardinal reference and adds it there as a concentration, with fixed gains and no switch (Eq.~\ref{eq:harmonic-completion}). The parameter-free claim of this paper is bounded to the angles $c_1$ predicts for the combination structure of the 1- and 2-line configurations, not to the weights; the harmonic completion adds one gain per harmonic. Within those configurations $c_1$ carries the combination structure, not the bulk of a single line's effect, which is fourfold (\S5.3).

\textbf{The single-line four-cycle function.} Li and Matin's \emph{single-line} results carry a four-cycle structure of their own, beside the square's rather than inside it. Replotting perceived vertical against 1-line orientation over the whole $360^\circ$ circle, they fit one sinusoid and describe ``$4\alpha$ periodicity'' (2005a, their \S4.1 and Fig.~7). The coefficient $\approx3.97$ is the \emph{angular frequency} inside that sinusoid; the fitted amplitudes are $4.95^\circ$ for the $48^\circ$ lines and $2.54^\circ$ for the $16^\circ$ lines. The function is not an artefact of the four 1-line placements, each covering one rising segment of the circular scale. The horizontal placements give same-sign slopes that the first moment forbids, and Vingerhoets et al.\ (2009) measured the falling segments for a single line (\S5.3); Li and Matin's footnote~3 rests the complete sinusoid on their parallel-pair runs. It is the fourfold line effect, which within the harmonic completion of \S5.6 is a second-harmonic term coupled to the body/cardinal reference. Their square-frame analysis (their \S4.2 and Fig.~8) reports no such coefficient, so nothing here imports a frame parameter. Vingerhoets et al.\ (2009) model the same fourfold single-line effect explicitly (\S4), so this is a re-derivation, not a new mechanism. (The cardinal reference has a geometric reading at the rod-and-frame $45^\circ$ null: a $45^\circ$-rotated square is a \emph{diamond} whose \textbf{diagonals} point cardinal, so the diagonal-versus-edge competition reported near $45^\circ$ is that same reference asserting itself.)

The idea of decomposing the inducing figure into orientation \emph{harmonics} is not new, and Wenderoth and Beh's (1977) data bear directly on ours. They measured the angular functions of the six two-sided figures formed by deleting two sides of a square frame. Parallel-side amputations, they found, induce tilt-illusion-like \emph{one-cycle} functions distinct from the \textbf{orthogonal-sided} ones. They drew the \textbf{global- vs local-orientation} distinction explicitly. Their proposal was that the tilt and rod-and-frame illusions arise from inducers ``composed of all or part of $n$ gratings \ldots intersecting at angles of $180^\circ/n$'' (the tilt illusion one cycle, the rod-and-frame two).

That distinction is not only formal. Segmenting an inducing line into a parallel array varies its \emph{global} (array) orientation and its \emph{local} (segment) orientation independently. In a long ($66^\circ$) array the induced VPEL then tracks the \textbf{global} orientation (slopes $0.45$--$0.52$), with almost no effect of the \textbf{local} segment orientation ($\approx0.02$). VPV shows the \emph{same} separation (global $0.17$--$0.24$, local $\approx0.01$): pooled over observers, local is about $4\%$ of global in \textbf{both} measures --- $4.1\%$ for VPEL and $3.7\%$ for VPV. (Fitting each observer separately and then averaging gives $3.9\%$ and $6.1\%$, \texttt{analysis/\allowbreak fit\_\allowbreak global\_\allowbreak local\_\allowbreak persubject.py}; the two estimators bracket the claim rather than agreeing to the decimal, and only the aggregate one is quoted above.) That fixes \emph{which} orientation signal the first moment reads --- the \textbf{global} (array-axis) orientation, not the local segment orientation (Shavit, Li, \& Matin, 2004; reported in full by Shavit, Li, and Matin, 2026; reanalysed here in \texttt{analysis/fit\_global\_local.py}). The global orientation is the coarse-scale input the position-blind $c_1$ takes as given; its extraction is the higher-level global stage of the neural model (\S5.5) and lies beyond the aggregate readout (\S5.6). The separation is established for one geometry, a $66^\circ$ array of segments at most $3^\circ$ long: in these data that geometry is sufficient for it, and neither its array length nor its segment length is shown to be necessary. Study~2 of the same report shortened the arrays to $6$--$18^\circ$ and lengthened the segments to $3$--$15^\circ$ together, and there segment orientation moves VPEL measurably, with slopes of $0.07$--$0.14$. Within Study~2 the segment slope rose with segment length ($0.07$, $0.14$ and $0.13$ for $3^\circ$, $9^\circ$ and $15^\circ$ segments, an interaction at $p=.035$ whose endpoint contrast is not significant at $n=8$) and showed no trend with array length ($0.10$, $0.11$ and $0.08$ for $6^\circ$, $12^\circ$ and $18^\circ$ arrays). So the local effect returns with longer segments, not demonstrably with shorter arrays (Shavit, Li, and Matin, 2026). That report's version~3 fits its own model to the same eight experiments, and other modelling work by this author uses them too, so their agreement with the present account is shared support, not independent evidence.

\begin{figure}[!htbp]\centering
\includegraphics[width=\linewidth]{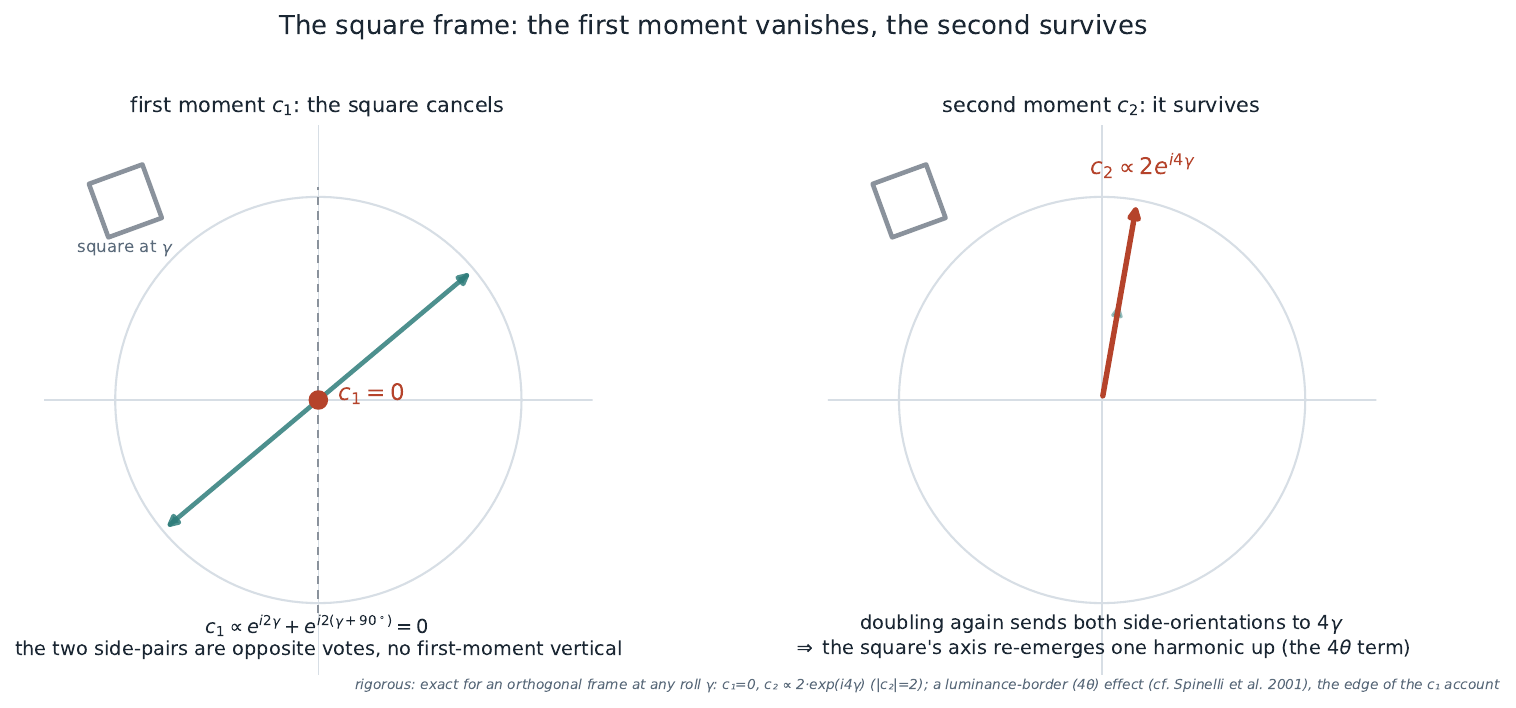}
\caption{The square frame: the first moment vanishes, the second survives. \emph{Left:} a square rolled by $\gamma$ has its four sides at $\gamma$ and $\gamma+90^\circ$, whose doubled-angle phasors are antiparallel. So $c_1=2e^{i2\gamma}+2e^{i2(\gamma+90^\circ)}=0$ at every tilt: no first-moment vertical. \emph{Right:} doubling again sends both side-orientations to $4\gamma$, so the second moment $c_2\propto e^{i4\gamma}$ survives and carries the square's axis one harmonic up ($\tfrac14\arg c_2=\gamma$, tracking the roll). Exact for an orthogonal frame at any roll --- the edge of the $c_1$ account. (The square's rod-and-frame induction is itself a luminance-border, not subjective-border, effect: Spinelli et al., 2001.)}
\label{fig:squareframe}
\end{figure}

\begin{figure}[!htbp]\centering
\includegraphics[width=\linewidth]{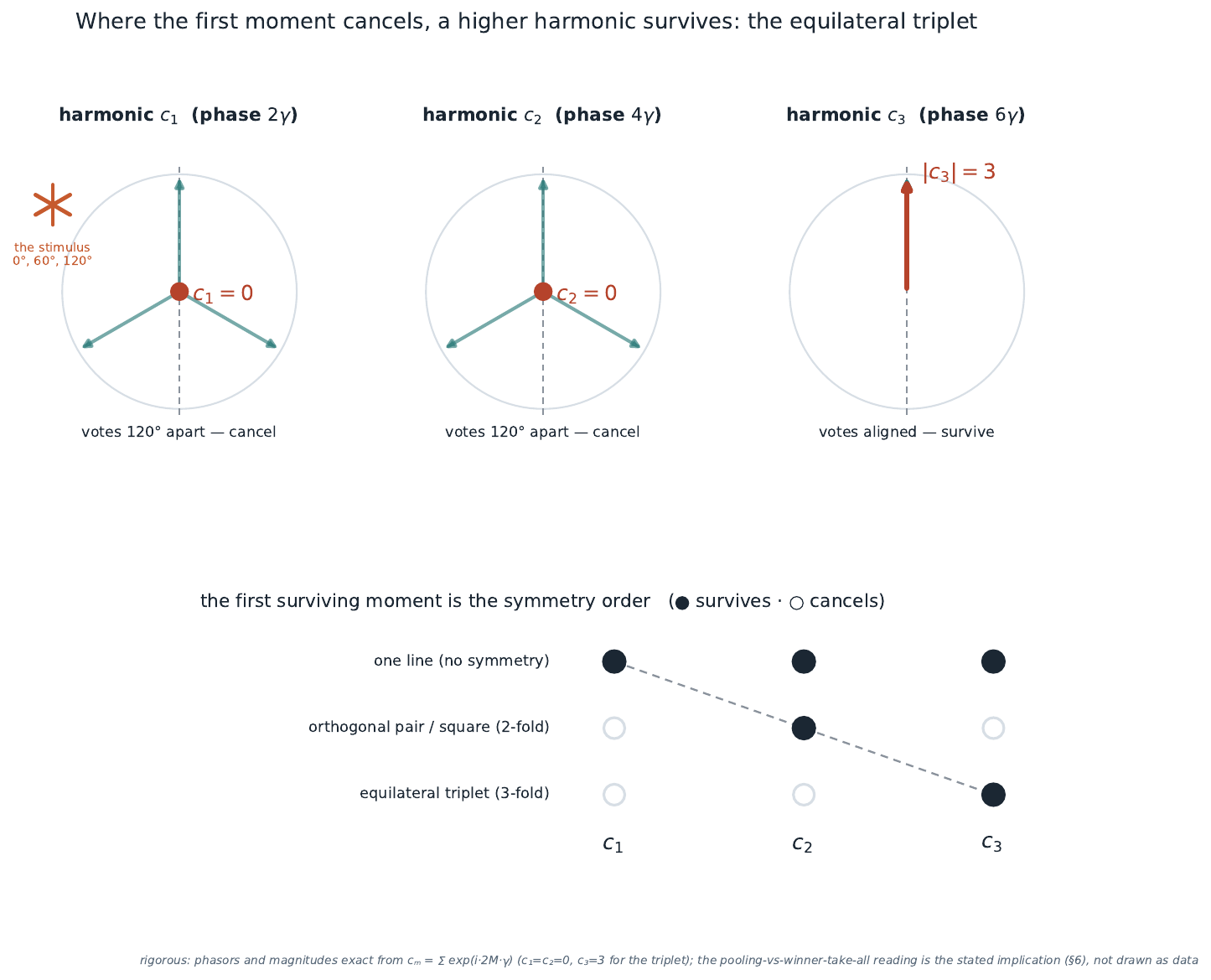}
\caption{The equilateral triplet: the first two moments cancel, and a higher harmonic survives. \emph{Top:} the three votes of a $0^\circ/60^\circ/120^\circ$ line triplet, shown in the first three circular-moment domains. In $c_1$ (phase $2\gamma$) the votes sit $120^\circ$ apart and cancel; doubling again ($c_2$, phase $4\gamma$) they are still $120^\circ$ apart and cancel; only at $c_3$ (phase $6\gamma$) do all three align, giving $|c_3|=3$. \emph{Bottom:} the first surviving moment of an $M$-fold symmetric set is the $M$-th --- a single line survives at $c_1$, an orthogonal pair or square at $c_2$, the triplet only at $c_3$. So the triplet is a null no first- or second-harmonic (line- or frame-like) mechanism can fill. A pooled first moment reads exactly zero, whereas a winner-take-all rule is three-way tied here; a slight imbalance then sends it to the strongest line while the pooled reading stays near zero (\S5.4, \S6).}
\label{fig:triplet}
\end{figure}

The harmonic split proper --- a line's $c_1$ versus the square's $c_2$ --- is the \emph{figure-symmetry} result above (Wenderoth and Beh's one- versus two-cycle functions), and it is a separate claim. The induced vertical from a line is the \textbf{leading ($n=1$) harmonic}, the first circular moment, while the square's structure lives in the $n=2$ harmonic ($c_2$, above). Two qualifications follow from their results. \textbf{(i)} Their angular functions vary the \emph{figure tilt}, rotating the whole inducer. They did \textbf{not} report the symmetric-pair \emph{coherence} as a function of half-separation $\theta$, which is the $|\cos2\theta|$ law of R4. That closed form, and the null-strength tests of \S6, therefore remain a genuine forward target rather than a re-description of theirs. \textbf{(ii)} They also reported \textbf{nonadditivity}: the summed functions of complementary amputations did not always equal the complete-frame function, with asymmetries no simple rule captured. Our phasor-additive combination is therefore the \emph{leading-harmonic} account, exact only where $c_1$ dominates, which single-line induction does not (\S5.3). The residual nonadditivity they document is precisely the higher-harmonic ($c_2$ and cardinal-reference) structure the first-moment core hands off to. To their qualitative proposal we add the closed-form readout $\phi=\tfrac12\arg c_1$ and its normalisation. From these, Li and Matin's combination coefficients follow with no per-configuration free parameters (the length constants are Li and Matin's own mass-action length fit, their Fig.~5). Figure~\ref{fig:squareframe} shows the square's first moment vanishing while its second survives. Figure~\ref{fig:triplet} shows the equilateral triplet, where the first and second moments both vanish.

\subsection{VPV and VPEL as symmetric and antisymmetric readouts of the per-hemifield signals.}
The 2013 dataset was built around a striking complementarity (Shavit, Li, \& Matin, 2013; presented earlier as Shavit, Li, Semanek \& Matin, 2004 and Shavit, Li, \& Matin, 2009). The \emph{same} roll-tilted frontoparallel lines drive both the perceived vertical (VPV) and the perceived eye level (VPEL), but with \textbf{reversed} integration. Two \textbf{parallel} lines shift VPV and leave VPEL unmoved, whereas two \textbf{bilaterally symmetric} (counter-rolled) lines shift VPEL and leave VPV unmoved --- the parallel-vs-counter-rolled dissociation of Matin and Li (1994), which Shavit, Li, and Matin (2009) restate and re-measure. And a single line's effect has the \emph{same} sign in the left and right fields for VPV but \emph{opposite} signs for VPEL. The order parameter reproduces all of this with \textbf{one added assumption} --- a per-hemifield decomposition of the orientation signal, read out two ways. Let each hemifield $h\in\{L,R\}$ contribute, for its near-vertical line, the orientation signal $s_h=\phi_h=\theta_h$ (half its doubled-angle phase, \S3/R1); an empty hemifield has no line and no phase, and contributes $s_h=0$ by convention (Fig.~\ref{fig:diamond}). The two norms are two orthogonal linear readouts of the pair:
\[
\text{VPV}\;\propto\; s_L+s_R \quad(\text{symmetric / sum}),\qquad
\text{VPEL}\;\propto\; s_R-s_L \quad(\text{antisymmetric / difference}).
\]
Here $s_h=\phi_h$ is the \emph{within-hemifield} normalised first-moment phase (\S5.1). Several inducers sharing one field are combined into one scale-invariant pooled phase \emph{before} this step, so a single line per field gives $s_h=\theta$. The fourfold term gives the same per-hemifield signal: a quarter of the phase of $c_2$ also returns $\theta_h$ for a single line tilted by less than $45^\circ$, and at small tilts both harmonics move the setting toward the line (\S5.3). The sum/difference split and every sign in Table~\ref{tab:readouts} therefore hold whatever the harmonic, so, like R2, they do not single out $c_1$. \textbf{This is an explicit structural assumption}: that the $c_0$ pooling is \emph{within-hemifield}, and that the two hemifields are then combined \emph{across} the midline without further renormalisation --- the $\pm$ readout of a whole-field comparator. We state it as an assumption, not a derivation, because nothing in \S2 forces normalisation to stop at the midline. We note a tension with \S5.1, where we account for Li and Matin's sub-additive multiline slopes by pooled normalisation although their lines span both fields (Table~\ref{tab:limatin} caption); the data here do not decide at which level that normalisation acts. The assumption's justification is the projective-geometry argument below (roll vs pitch) and, ultimately, the fit to the 2013 data. Under this assumption the pooled-phase $1/n$ law of \S5.1 governs \emph{within} a pool, the $\pm$ readout \emph{across} pools. In the 2013 data the cross-hemifield combination is \textbf{near-additive but not strictly so}; Li and Matin's cross-field pair is more sub-additive (the confirmatory re-analysis below). The two-line slopes sit at $k_1\approx0.84$--$0.90$ of the single-line sum/difference (the confirmatory re-analysis below), close to full additivity yet slightly discounted. Several lines in \emph{one} field are \textbf{predicted} to average far more strongly, toward $1/n$ (\S5.1). This within-hemifield prediction extrapolates from the two-level pooling and has no published test: the present four-stimulus set places at most one line per field, and Li and Matin's multiline configurations also span both fields. Because there is only one line per field, the table's structural predictions (which channel is active) stand, while the trial-level data below serve only to quantify the $\approx0.9$ cross-pool discount.

Geometrically, parallel lines ($s_L=s_R$) are a coherent net roll --- a pure sum --- so they drive verticality and vanish in the difference. Bilaterally symmetric lines ($s_L=-s_R$) form an up--down-asymmetric convergence --- a pure difference --- so they drive elevation and vanish in the sum. Every qualitative result of the 2013 study, its signs and nulls, then follows with no free parameters, as Table~\ref{tab:readouts} sets out.

\begin{table}[!htbp]\centering
\begin{tabular}{llcc}
\toprule
Stimulus & $(s_L,s_R)$ & VPV $\propto s_L{+}s_R$ & VPEL $\propto s_R{-}s_L$ \\
\midrule
1-left & $(\theta,0)$ & $+\theta$ & $-\theta$ \\
1-right & $(0,\theta)$ & $+\theta$ & $+\theta$ \\
2-parallel & $(\theta,\theta)$ & $2\theta$ & $0$ \\
2-symmetric & $(-\theta,\theta)$ & $0$ & $2\theta$ \\
\bottomrule
\end{tabular}
\caption{VPV and VPEL as the symmetric (sum) and antisymmetric (difference) readouts of the per-hemifield orientation signals $(s_L,s_R)$, for the four stimuli of Shavit, Li, and Matin (2013).}
\label{tab:readouts}
\end{table}

\begin{figure}[!htbp]\centering
\includegraphics[width=\linewidth]{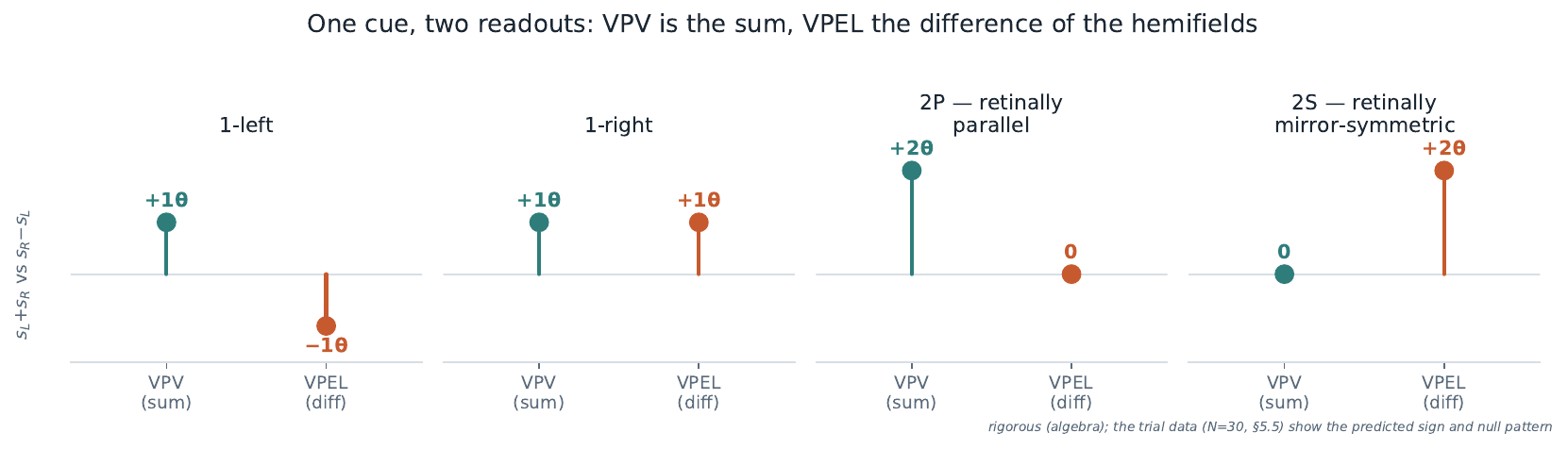}
\caption{One cue, two readouts. Across the four stimuli, VPV is the sum $s_L+s_R$ of the two hemifields' orientation signals and VPEL is the difference $s_R-s_L$. A retinally parallel pair (2P) drives VPV only; a retinally mirror-symmetric pair (2S) drives VPEL only; a single line has the same VPV sign but opposite VPEL sign in the two fields. The bar heights are exact algebra; the trial-level data (30 observers, \S5.5) show the predicted sign and null pattern.}
\label{fig:readouts}
\end{figure}

\begin{figure}[!htbp]\centering
\includegraphics[width=\linewidth]{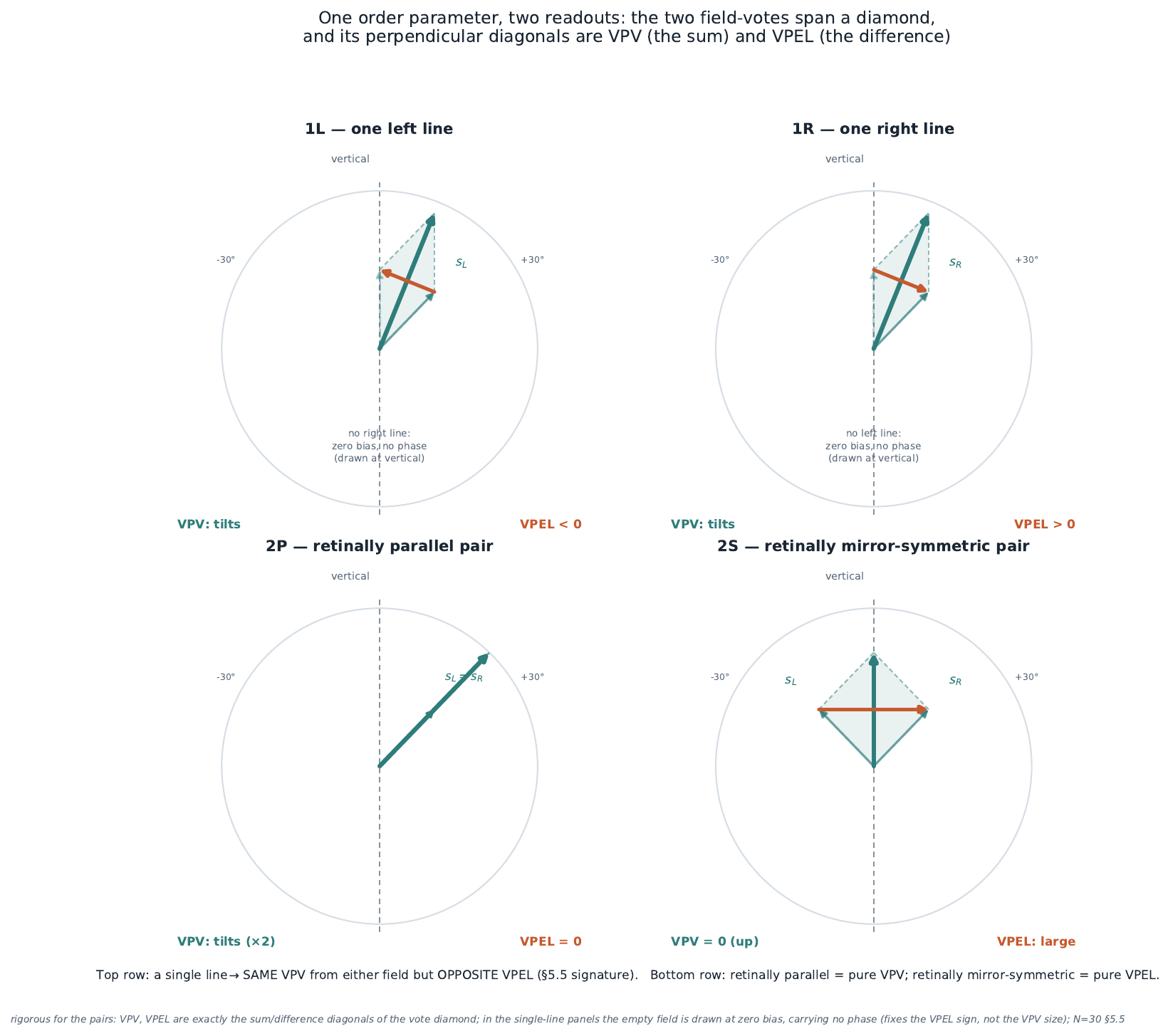}
\caption{One order parameter, two readouts on the doubled-angle circle. The two field-votes span a diamond whose perpendicular diagonals are the readouts. One diagonal is the sum, VPV --- the resultant's tilt off vertical. The perpendicular one is the difference, VPEL --- the left--right split between the two field-votes. A retinally parallel pair collapses the diamond onto the VPV axis (pure vertical); a retinally mirror-symmetric pair opens it onto the VPEL axis (pure eye level); a single line lands off both, with the same VPV but opposite VPEL in the two fields. An empty field is drawn at zero bias: it contributes no orientation evidence and is assigned no phase ($c_1^h=0$, so $\arg c_1^h$ does not exist), so the drawn arrow fixes the VPEL sign but is not an axis read off that field.}
\label{fig:diamond}
\end{figure}

The table matches their Fig.~3 exactly (VPV same-signed across fields, VPEL opposite; 2P affects only VPV, 2S only VPEL). Figure~\ref{fig:readouts} plots the four stimuli against both readouts, and Figure~\ref{fig:diamond} puts the same pair of readouts on the doubled-angle circle, where they are the two perpendicular diagonals of the field-vote diamond.

\textbf{The projective geometry of self-motion motivates the sum and the difference.} The two readouts are not an arbitrary orthogonal pair; they are the retinal signatures of the two ways the world's verticals move on the eye. Under head \textbf{roll} the whole field rotates rigidly, so verticals in the left and right hemifields tilt the \emph{same} way ($s_L=s_R$): a pure \textbf{sum}, which the vertical system reads. Under \textbf{pitch} --- the line of sight, or a surface, tilting in depth --- perspective projection sends verticals toward a vanishing point. Left- and right-field verticals then tilt in \emph{opposite} directions ($s_L=-s_R$): a pure \textbf{difference}, which the elevation/eye-level system reads. This is precisely Matin and Li's observation that the projections of the two sides of a frame are \emph{bilaterally parallel for roll but bilaterally symmetric for pitch} (Matin \& Li, 1994). It casts the sum/difference architecture as a linearised, low-dimensional shadow of their \textbf{Great Circle Model}: the spherical geometry that separates roll from pitch is what separates VPV from VPEL in the plane. The 2P and 2S stimuli are thus lab-isolated roll and pitch cues, and their clean single-channel effects (2P $\to$ VPV only, 2S $\to$ VPEL only) are what that geometry demands. Projective optics therefore \emph{motivates} the sum/difference basis independently of the fitted gains. It does not by itself fix this decoder: the readout stage, the use of the same $c_1$ summary, the stopping of normalisation at the midline, and the cross-field gain are the structural assumption stated above, not consequences of the geometry.

The rule is therefore \textbf{retinal}, not world-referenced. ``Parallel drives vertical'' holds for \emph{retinal} parallelism --- the frontoparallel 2P pair, $s_L=s_R$. Two lines held \textbf{parallel in depth} instead project to bilaterally \emph{symmetric} retinal orientations ($s_L=-s_R$) and so drive VPEL, not VPV. That is exactly Matin and Li's (1994) plane-of-origin reversal, and the reason the $s_h$ are defined on retinal orientation throughout. The ``single, shared visual orientation process that feeds to separate integration processes'' that Shavit, Li, and Matin (2013) infer \emph{is} one first-moment orientation signal read through the sum and through the difference. This \textbf{recasts} Matin and Li's (2001) \textbf{four-channel neural model} of VPEL --- two channels per side of the median plane, one reading clockwise and one counter-clockwise retinal orientation. On their account, a higher-level stage builds the \emph{global} orientation from V1's \emph{local} (segment) orientation responses. That global-orientation readout is what $\tfrac12\arg c_1$ computes (the global-over-local result of \S5.4), and their four channels are the per-hemifield sum and difference above. The reformulation is ours: Matin and Li do not derive $\tfrac12\arg c_1$ in closed form. Within a \emph{single} field, several inducers are assumed to combine first by the sub-additive averaging of \S5.1, a property of the normalised readout. A within-field combination coefficient would therefore enter each hemifield's $s_h$ before the sum/difference is taken, and the one-line-per-field entries of the table above are unaffected. Neither Li and Matin's $k_1$ nor Matin and Li's (1999) two-line VPEL averaging measures that within-field step: both come from lines on opposite sides of the midline. That equivalence is exact within this idealization, not rhetorical. A population of cosine-tuned orientation channels, pooled as a population vector, returns $\tfrac12\arg c_1$ \emph{to machine precision} on every stimulus that asserts an axis at all --- single lines, parallel pairs, and symmetric pairs away from $\pm45^\circ$. On the orthogonal pair, the $0/60/120^\circ$ triplet and the square frame the two agree in the only respect that exists: both pooled votes vanish, so neither asserts an axis and neither has a phase to compare (\texttt{analysis/neural\_channel\_model.py}). For an equal-gain, uniformly-tiled, linearly-pooled channel bank --- the idealisation derived in a supplementary neural-equivalence note --- the pooling model and the order parameter are the \emph{same mathematical object} on this stimulus family, and no member of it can tell them apart. This is a structural recasting, not a claim to reproduce the full non-linear four-channel 2001 parameterisation. The \emph{perturbed} triplet and the symmetric-pair null of \S6 \emph{would} separate a pooled readout, in either form, from a \emph{non-linear} one: once a slight imbalance breaks the triplet's three-way tie, a winner-take-all rule jumps to the strongest line while a pooled first moment vanishes linearly with the imbalance. Those experiments thus test pooling versus non-linearity, not our formulation versus Matin and Li's.

\textbf{Confirmatory re-analysis of the trial-level data (30 observers).} A first-hand re-analysis of the trial-by-trial settings recovers the sign and null structure of the sum/difference architecture: \textbf{30 observers, 4{,}800 individual settings} (8 conditions $\times$ 5 orientations $\times$ 4 repetitions per observer). These are the author's own data from the already-published study of \textbf{Shavit, Li, and Matin (2013)}, the individual-differences study designed to test whether the two norms share a common processing stage. It is a secondary re-analysis with no new data collected, and \texttt{analysis/\allowbreak fit\_\allowbreak common\_\allowbreak mechanism.py} reproduces every slope and gain below, while \texttt{analysis/\allowbreak p39c\_\allowbreak b367\_\allowbreak null\_\allowbreak ci\_\allowbreak 2026-10-10.py} computes their 95\% intervals (seeded bootstrap for the gains, so the last digit of an interval bound can move with the seed). Each observer's setting was regressed on inducer orientation over the five frontoparallel tilts ($\pm15^\circ,\pm7.5^\circ,0^\circ$). The two \textbf{nulls the model requires} are indistinguishable from zero: VPEL under parallel lines (mean slope $-0.006$, 95\% CI $[-0.027,+0.015]$) and VPV under symmetric lines ($+0.003$, 95\% CI $[-0.013,+0.018]$). Both intervals lie within $\pm0.03$, under a quarter of the smallest active slope ($0.12$). The single lines carry its \textbf{sign structure}: VPV tracks the line with the \emph{same} sign from either field (1L $+0.12$, 1R $+0.15$), whereas VPEL flips sign between fields (1L $-0.12$, 1R $+0.14$). The qualitative signs and nulls are, in fairness, what the sum/difference assumption was constructed to capture, so recovering them is an internal check rather than an independent one. Above we assume that normalisation acts within a pool and stops at the midline, despite the tension with \S5.1 noted there, so the cross-field combination is \emph{compatible with} that two-level picture rather than predicted by it. Its point estimates are sub-additive in each active channel, as the picture requires, though bootstrap 95\% intervals over observers reach full summation. Here the cross-field ratio, written $k_1$ by analogy with Li and Matin's slope, is $k_1=\text{VPV}(2P)/[\text{VPV}(1L){+}\text{VPV}(1R)]=0.90$ (95\% CI $[0.78,1.03]$) and $k_1=\text{VPEL}(2S)/[\text{VPEL}(1R){-}\text{VPEL}(1L)]=0.84$ ($[0.66,1.06]$) --- both point estimates below full summation, and well \emph{above} Li and Matin's long-line two-line value $k_1\approx0.54$ (\S5.1; these 2013 inducers are long lines too). That value is not a same-field contrast: Li and Matin's parallel pair also places one line on each side of the test line (their Fig.~2), so the two datasets estimate a cross-field gain for long lines and disagree ($0.54$ against $\approx0.9$). The reason for the difference is open, and neither dataset places several lines in one field. Across the 30 observers the two nulls are not significant ($|t|<0.7$) while all six active slopes are ($|t|=4.1$--$7.7$, all $p<3\times10^{-4}$). So the sum/difference structure $\text{VPV}\propto s_L+s_R$ and $\text{VPEL}\propto s_R-s_L$ holds in sign and null pattern, with every active slope reliably nonzero. Each proportionality carries its own visual weight $k_V$ (\S5.6), so the two channels need not share a scale.

\textbf{Limits of the trial-level re-analysis.} The sign and null structure above is a recovery of the pattern the sum/difference assumption was built to capture, not an independent confirmation of it; what it confirms is that the assumption is consistent with trial-level data it was not fitted to. The \emph{sub-additive magnitude} is a different kind of claim again: given the linear sum/difference architecture, $k_1$ is a comparator gain read from the data, not a value the architecture derives (an independent re-derivation confirms this for these within-paper roll data). That re-derivation also shows the gain cannot be localised to the cross-field stage specifically: a pair response combining an unattenuated cross-field sum with an equal-sized \emph{per-line} discount is observationally identical, on the pair response, the marginal slope, and the 2:1 VPV/VPEL ratio, to the same discount applied only at the comparator. The present data support the sign/null structure and sub-additive point estimates, but do not locate the attenuation at the cross-field stage rather than earlier.

\textbf{The precision cross-over --- confounded, and left open.} We had flagged the \emph{direction} of a trial-to-trial variance effect as genuinely open: does a null inducer \emph{raise} or \emph{lower} the scatter? It could leave the setting \emph{less} constrained, raising variance --- the reliability$\to$precision reading --- or return the percept to a well-anchored gravity reference, \emph{lowering} it. The \emph{raw} within-cell trial SD is lower under each setting's null stimulus (VPV: $1.17$ under the null $2S$ vs $1.76$ under the active $2P$; VPEL: $1.05$ vs $1.41$), which is superficially the anchoring direction. (We pool it as $\sqrt{\text{mean variance}}$, not by averaging per-cell SDs, which is downward-biased; averaging the SDs instead would return $1.01$/$1.37$ and $0.94$/$1.19$, the same ordering at smaller values.) But this is largely a \textbf{magnitude confound.} The active two-line settings are far larger at the extreme tilts (up to $\sim6^\circ$ vs $\sim1$--$3^\circ$), and under signal-dependent (Weber-type) noise larger settings scatter more. Controlling for it at the $0^\circ$ cell, where settings are of comparable size across conditions, leaves a difference that is \textbf{small for VPV} (SD $1.10$ vs $1.23$ at matched magnitude, a $12\%$ gap against the $51\%$ gap before control) and \textbf{absent for VPEL} ($1.04$ vs $0.99$, and in the \emph{opposite} direction to the uncontrolled comparison). So this variance effect does \textbf{not} cleanly separate the two accounts. We leave its direction genuinely open, and note that a magnitude-matched or coefficient-of-variation design is the way to settle it. (The eccentricity-dependence of VPEL is left to separate work, out of scope here.)

\subsection{From the order parameter to a measured bias --- the visual weight.}
The readout $(\phi,R)$ is an orientation \emph{summary}, not yet a bias in degrees. The measured setting is that summary weighted against the observer's body/gravity reference. Matin and Fox's own linear model connects the two (Matin \& Fox, 1989). A setting takes a fraction $k_V$ of the visual estimate and $k_B=1-k_V$ of the body-referenced (upright) estimate, so for near-vertical inducers
\begin{equation}
\text{bias}=k_V\,\phi,\qquad k_V\in[0,1],\quad k_V+k_B=1,
\label{eq:matinfox}
\end{equation}
with $k_V$ the measured slope of the setting-vs-orientation function. \textbf{Direction and magnitude are separate stimulus-side quantities, and must not be conflated.} The axis $\phi=\tfrac12\arg c_1$ comes from the \emph{normalised} moment, and $R=|c_1|/c_0$ is a scale-invariant ratio. For a single line it is the fixed coherence $1$ (\S2's $\hat\kappa_1=\hat\kappa_0$ convention), \emph{independent of length}; for a symmetric pair it is $|\cos2\theta|$ (\S3/R4). The \emph{magnitude} comes instead from the \textbf{total, unnormalised} inducer energy $C=\hat\kappa_0\sum_j A_j$, which grows and saturates with length as $S(\ell)$ (\S5.2). So the visual weight is, approximately, $k_V=g(C)\,R$ (the exact form follows Eq.~\ref{eq:cue-combination}). The energy $C$ sets \emph{how much} through the saturating $S$, and the coherence $R$ discounts incoherent configurations: unity for aligned inducers, zero for an orthogonal pair. $R$ itself does \textbf{not} grow with length --- it is a normalised ratio. The two factors then split the combination law of \S5.1 cleanly. The \textbf{$1/n$ asymptote is the angular averaging of the pooled phase} --- one pooled orientation set against the sum of $n$ single-line phases. That phase is scale-invariant, so the asymptote does not come from the $c_0$ denominator itself (\S5.1). The \textbf{deviations of $k_1$ above $1/n$} are the $C$-driven saturation $g(C)/S(\ell)$. Both factors of $g(C)\,R$ are load-bearing. Writing the bridge explicitly keeps the two quantities the model delivers --- an axis $\phi$ and a visual weight $k_V=g(C)\,R$ --- distinct from the single number they combine into.

\textbf{The linear bridge is local, and completing it produces the large-tilt prediction.} Equation~\ref{eq:matinfox} is a small-tilt statement, and it misbehaves when read beyond one. Taken outside the near-cardinal chart, the linear rule treats an axial variable as Euclidean and ignores its branch structure, and it extrapolates straight through the configurations where $c_1$ vanishes. Both are artefacts of writing a circular quantity as a scalar. The repair keeps the same two cues and adds them in the doubled-angle plane. Remove the constant tuning factor with $s_1=(\hat\kappa_0/\hat\kappa_1)\,c_1/c_0$, and represent the body/gravity axis $b$ by the unit phasor $g=e^{i2b}$ (not the gain $g(C)$ above). Each cue is then a vector in the doubled-angle plane: its direction is the axis the cue favours, and its length --- the cue's \emph{concentration}, a precision-like strength --- is how firmly it favours it. For a von Mises distribution (the circular analogue of a Gaussian: peaked at one angle and falling off symmetrically around it) that vector is the distribution's \emph{natural parameter}, and independent cues combine by adding these vectors. Let $\beta>0$ be the body reference's concentration, and let $\eta_1(C)\geq0$ be the saturating visual concentration the energy $C$ drives. Adding the two axial natural parameters gives the doubled-angle form of Mittelstaedt's (1983) construction, in which the subjective vertical is the resultant of a gravity vector and a body-fixed idiotropic vector (\S4):
\begin{equation}
q_1=\beta g+\eta_1(C)\,s_1,\qquad
p(v\mid E)\propto\exp\!\left\{\operatorname{Re}\!\left(q_1e^{-i2v}\right)\right\},\qquad
\widehat v=\tfrac12\arg q_1 .
\label{eq:cue-combination}
\end{equation}
In words: the body cue and the visual cue add as vectors, and the setting $\widehat v$ is the direction of their sum, halved back to an orientation. That density is a von Mises in the doubled angle, and its mode is $\tfrac12\arg q_1$. With no body cue ($\beta=0$) it returns the order parameter's own $\phi$ unchanged. For an upright observer ($b=0$) and one line at small tilt $\gamma$ it gives
\[
\widehat v=\frac{\eta_1}{\beta+\eta_1}\,\gamma+O(\gamma^3),\qquad
k_V=\frac{\eta_1}{\beta+\eta_1},\qquad k_B=\frac{\beta}{\beta+\eta_1},
\]
so Matin and Fox's $k_V+k_B=1$ rule is the small-angle limit of circular cue combination, with a cubic rather than quadratic first correction. Everything upstream survives untouched. The length function $S(\ell)$ still calibrates the measured coherent-line weight, so the multiline coefficient remains $S(n\ell)/[n\,S(\ell)]$ and every number in \S5.1--\S5.2 stands. We must, however, revise the standing of $k_V=g(C)\,R$: at coherence $r$ the exact weight is $\eta_1r/(\beta+\eta_1r)$, which is proportional to $r$ only while the visual cue is weak. The two forms differ by at most a factor $1+\eta_1/\beta$, which is $1.16$ at the measured single-line weight. So the $R$-proportional law is an approximation with a stated error, not an identity. The same expansion covers an array and its own segments. With the array at tilt $\theta_A$ and concentration $\eta_A$, and its segments near the vertical at tilt $\theta_L$ and concentration $\eta_L$, $\tfrac12\arg\bigl(\beta+\eta_Ae^{i2\theta_A}+\eta_Le^{i2\theta_L}\bigr)=(\eta_A\theta_A+\eta_L\theta_L)/(\beta+\eta_A+\eta_L)+O(\theta^3)$, which is exactly the drive-weighted committee rule that Shavit, Li, and Matin (2026, version~3, Appendix~A.1) fit to eight line-array experiments. (That version of the preprint is submitted to PsyArXiv and pending moderation, so it is not yet public; the posted version~2 contains neither the committee rule nor its fit.) Two conditions are needed for that match. Each of their drives must be read as a concentration here, an identification rather than a result, and the segments must lie near the vertical; near-horizontal segments enter with the opposite sign, giving $\beta+\eta_A-\eta_L$ in the denominator. The two rules part at third order. That report fits its drives in units of the eye-level body drive: the perceived-vertical body drive is $4.9$ (version~3, Appendix~A.7) and the visual drive of its $66^\circ$ array is $0.93$ (version~3, \S4.2). With $\beta=4.9$ and $\eta_A=0.93$, the segments' smaller drive set aside, the readout predicts a setting $0.15^\circ$ closer to vertical than the committee rule's at a $20^\circ$ tilt, $0.52^\circ$ closer at $30^\circ$, and $1.81^\circ$ closer at $45^\circ$. Over the $\pm10^\circ$ those experiments tested the gap is at most $0.02^\circ$, between $0.02$ and $0.08$ standard errors of a Study~1 perceived-vertical condition mean (median $0.04$), so their data cannot separate the two rules; tilts of $30^\circ$ or more could.

\textbf{The turnover near the oblique, derived rather than asserted.} For a single line at roll tilt $\gamma$ the completion is explicit, and so is its slope:
\begin{equation}
\widehat v(\gamma)=\tfrac12\operatorname{atan2}\!\left(\eta_1\sin2\gamma,\ \beta+\eta_1\cos2\gamma\right),
\qquad
\frac{d\widehat v}{d\gamma}=\frac{\eta_1\left(\eta_1+\beta\cos2\gamma\right)}{\beta^2+\eta_1^2+2\beta\eta_1\cos2\gamma}.
\label{eq:single-line-turnover}
\end{equation}
The denominator is positive, so the slope vanishes exactly where $\cos2\gamma_\ast=-\eta_1/\beta$ and nowhere else. The single-line function therefore has at most one turning point, it has one if and only if $0<\eta_1<\beta$, and that point sits at
\[
\gamma_\ast=\tfrac12\arccos\!\left(-\eta_1/\beta\right)\in(45^\circ,90^\circ),
\qquad
\widehat v(\gamma_\ast)=\tfrac12\arcsin\!\left(\eta_1/\beta\right),
\]
approaching the oblique from above as the visual cue weakens. Past $\gamma_\ast$ the bias falls, reaching exactly zero at $\gamma=90^\circ$, where a horizontal line leaves the setting upright. The condition and the location are fixed by the one weight the experiment already measures, because $\eta_1/\beta=k_V/k_B$. The turnover exists precisely when $0<k_V<\tfrac12$, and then $\cos2\gamma_\ast=-k_V/k_B$. Shavit et al.\ (2013) measured $k_V=0.123$ and $0.150$ in the two single-line conditions, which place $\gamma_\ast$ at $49.0^\circ$ and $50.1^\circ$ with peak biases of $4.0^\circ$ and $5.1^\circ$. Across the $\pm15^\circ$ window they tested, the local slope falls only to $0.91$--$0.92\,k_V$, which is why the measured function looks straight there.

Three qualifications limit the turnover prediction. First, the scalar bridge could not have delivered it. For a single line both $R$ and $C$ are independent of tilt, so $\text{bias}=k_V\,\phi$ is exactly linear in $\gamma$. It has no turning point at all, and the prediction follows from Eq.~\ref{eq:cue-combination} or from nothing. Second, the maximum is shallow. The predicted bias stays within $0.1^\circ$ of its peak from $44^\circ$ to $55^\circ$. The experiment therefore reads a plateau and then a decline to zero at horizontal, not a sharp peak whose position can be pinned. Third, $\gamma_\ast$ is a first-harmonic result, and a second-harmonic coupling $\eta_2\geq0$ changes it in two ways. Stationarity stops being a single condition and becomes a quadratic in $y=\cos2\gamma$, namely $4\beta\eta_2y^2+\eta_1(\beta+3\eta_2)y+\eta_1^2+2\eta_2^2-2\beta\eta_2=0$: read plainly, the tilt at which the curve turns over now solves a quadratic, so there can be none, one or two such tilts instead of always exactly one. We hold the model's slope at vertical, $(\eta_1+2\eta_2)/(\beta+\eta_1+\eta_2)$, at the pooled $k_V\approx0.14$ of the two single-line conditions by re-solving $\eta_1$ for each $\eta_2$. The leading turning point then moves downward as $\eta_2$ grows, from the $49^\circ$--$50^\circ$ of the first harmonic alone to $39.9^\circ$ at $\eta_2/\beta=0.02$ and $28.7^\circ$ at $\eta_2/\beta=0.05$ ($\eta_1/\beta=0.055$), where a second turning point sits at $74.3^\circ$. The same calibration caps the second harmonic: with $\eta_1\geq0$, $k_V=0.14$ allows $\eta_2/\beta$ only up to $0.075$, because the second harmonic alone already gives a slope of $2\eta_2/(\beta+\eta_2)$. So the second harmonic leaves the turnover in place but moves it a long way. \textbf{For $0<\eta_1<\beta$, a turning point exists if and only if $\eta_2<\beta$} --- necessary as well as sufficient, because the quadratic's discriminant factors as $(\beta-\eta_2)(\beta\eta_1^2-9\eta_1^2\eta_2+32\beta\eta_2^2)$ and the second factor is positive for every $\eta_2\geq0$ once $\eta_1<\beta$, so only the first factor can change its sign. In words: the turnover survives exactly as long as the second harmonic stays weaker than the body reference. At the boundary $\eta_2=\beta$ the two turning points do not merge into a smooth bend; they run into a point where the three cue terms cancel. The quadratic becomes the perfect square $(2\beta y+\eta_1)^2$, but the resultant itself factors as $\beta+\eta_1e^{i2\gamma}+\beta e^{i4\gamma}=e^{i2\gamma}(\eta_1+2\beta\cos2\gamma)$, so the double root at $\gamma=\tfrac12\arccos(-\eta_1/2\beta)$ is where that resultant vanishes. In words: at that one tilt the summed vote has zero length, so the setting is undefined there and jumps by $90^\circ$ across it, while on either side its slope is exactly $1$. Above $\beta$ no stationary point survives at all. The turnover's \emph{existence} is therefore robust to any second harmonic the body reference dominates, but not to an arbitrary one, and its location is not robust even there. Re-derived symbolically and numerically in \texttt{analysis/\allowbreak verify\_\allowbreak x7\_\allowbreak turnover.py}, with the threshold settled by quantifier elimination in \texttt{analysis/\allowbreak verify\_\allowbreak turnover\_\allowbreak harmonic2\_\allowbreak 2026-09-29.wls}.

The mechanism is the body/gravity reference itself, acting as an upright-concentrated prior on a circular variable. That is a cardinal-prior effect of the kind Girshick, Landy, and Simoncelli (2011) formalise. The model does not separate a body-referenced cardinal prior from a visual-cardinal one, because both enter as $\beta g$ with $g$ upright. At a fixed stimulus strength, writing the cardinal anisotropy into the visual gain instead would add nothing. A real, mirror-symmetric gain $\eta_1(\gamma)=a_0+a_1\cos2\gamma$ multiplies the line's vote $e^{i2\gamma}$, and the product is identically $(a_0+a_1\cos2\gamma)\,e^{i2\gamma}=\tfrac{a_1}{2}+a_0e^{i2\gamma}+\tfrac{a_1}{2}e^{i4\gamma}$. Read term by term, the gain shifts $\beta$ by $\tfrac{a_1}{2}$, keeps a first harmonic of weight $a_0$, and adds a second-harmonic term. If the anisotropy scales with the visual energy $C$, that shift changes with strength, so the equivalence holds one strength at a time. The turnover therefore needs no free mechanism beyond the two cues already in Eq.~\ref{eq:cue-combination}. Data outside the $\pm15^\circ$ window already exist, and they run against it. Li and Matin's parallel pair (1998a, 1998b, as cited in Li \& Matin, 2005a) and Vingerhoets et al.'s (2009) single line both reverse past $45^\circ$, where the first-harmonic curve is still near its peak (\S5.3). The first-harmonic turnover is therefore contradicted; a turnover can survive only in a curve dominated by the second harmonic, whose location the first-harmonic formula does not give.

\begin{figure}[!htbp]\centering
\includegraphics[width=\linewidth]{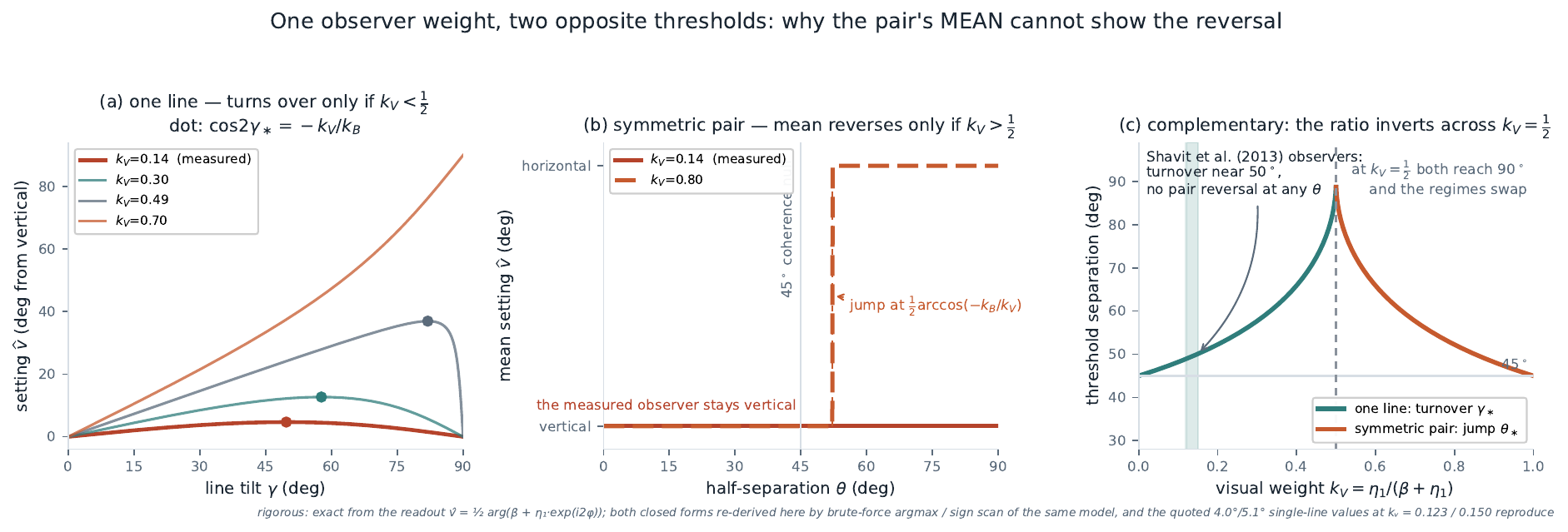}
\caption{One observer weight, two opposite thresholds. The visual weight $k_V=\eta_1/(\beta+\eta_1)$ is the only observer-side number in the cue-combination readout $\widehat v=\tfrac12\arg(\beta+\eta_1e^{i2\phi})$, and it decides both of the model's qualitative signatures --- in opposite directions, with the ratio inside the arccosine \emph{inverted} between them. \emph{(a)} One roll-tilted line: the bias turns over at $\cos2\gamma_\ast=-k_V/k_B$, and only for a body-dominant observer, $k_V<\tfrac12$; the turnover is marked on each curve that has one. \emph{(b)} A symmetric pair with its mean held at vertical: the mean setting is pinned exactly vertical --- the pair's first moment is real, so the sum of the two cues is real --- until the visual cue outweighs the body reference, and then it \emph{jumps} to horizontal at $\tfrac12\arccos(-k_B/k_V)$, which requires $k_V>\tfrac12$. It is a discontinuity, not a smooth crossing. \emph{(c)} The two thresholds as one map: complementary, meeting at $90^\circ$ where $k_V=\tfrac12$, and neither ever below $45^\circ$. Shavit et al.'s (2013) observers sit at $k_V\approx0.14$ (shaded), so the first-harmonic single-line turnover is predicted near $50^\circ$ while the pair's mean never reverses at any separation. Panel (a) draws the first harmonic alone; published full-circle line data reverse past $45^\circ$ and contradict it there (\S5.3). That is why the decisive test of Fig.~\ref{fig:decisive} is read off how the null resists perturbation rather than off the mean setting.}
\label{fig:thresholds}
\end{figure}

The same two concentrations carry between-observer variation, and they sharpen it into a test. Frame dependence differs widely across people. Here that variation is the single ratio $\eta_1/\beta=k_V/k_B$, acting on an \emph{unchanged} order parameter rather than on a different readout. A body-dominant observer ($k_V<\tfrac12$) shows the turnover. A strongly frame-dependent one ($k_V>\tfrac12$) shows no turning point at all, because the setting then follows the line monotonically to horizontal --- visual capture. Those are two qualitatively different curves, not two slopes, so a single-line function measured out to $90^\circ$ sorts observers by which side of $k_V=\tfrac12$ they fall on. The measured $k_V\approx0.14$ places the 2013 observers well inside the turnover regime, a regime in which the first-harmonic curve is already contradicted by the full-circle line data (\S5.3). The two thresholds are complementary --- set by the same measured weight, but with the ratio inside the arccosine inverted between them --- and Figure~\ref{fig:thresholds} puts both on one map.

\textbf{The same reference completes every harmonic, and supplies the square's envelope.} The second moment $c_2$ fixes the square frame's \emph{axis}, but its tilt-invariant magnitude $|c_2|$ cannot produce the rise-and-fall envelope (\S5.4). The obstacle is a mismatch of type, not of size. The phase of $c_2$ turns four times per full rotation of the scene, whereas the upright reference $g=e^{i2b}$ turns twice, so the two cannot simply be added. Equation~\ref{eq:cue-combination} already contains the repair: refer the harmonic back to the reference before adding it there. Write the normalised doubled-angle moments of the effective-salience distribution as
\[
s_k=\frac{\sum_jA_je^{i2k\gamma_j}}{\sum_jA_j},\qquad |s_k|\leq1 ,
\]
so that $s_k$ is $c_k$ with the tuning factor and the total salience divided out, keeping only the $k$-th harmonic's phase and coherence. In particular $s_1$ is the $s_1$ of Eq.~\ref{eq:cue-combination}, and $s_2$ is the square's surviving moment of \S5.4. Rotating the whole scene by $\alpha$ sends $s_k\mapsto e^{i2k\alpha}s_k$ while $g\mapsto e^{i2\alpha}g$. The product $s_k\bar g^{\,k-1}$ therefore transforms as $e^{i2\alpha}$ for \emph{every} $k$: dividing out $k-1$ copies of the reference removes the surplus turning, and what is left is an axial natural parameter the reference can accept --- every vote then turns at the same rate as the body axis. Adding one such term per harmonic gives the completion
\begin{equation}
q_K=\beta g+\sum_{k=1}^{K}\eta_k(C)\,s_k\bar g^{\,k-1},\qquad
p(v\mid E)\propto\exp\!\left\{\operatorname{Re}\!\left(q_Ke^{-i2v}\right)\right\},\qquad
\widehat v=\tfrac12\arg q_K .
\label{eq:harmonic-completion}
\end{equation}
In words: each harmonic casts a vote on the vertical, that vote is first expressed in the reference's own coordinates, and the votes are then summed as concentrations rather than as angles. At $K=1$ this is Eq.~\ref{eq:cue-combination} unchanged, so nothing above is disturbed. Real gains $\eta_k$ preserve the mirror symmetry, and their signs carry downstream assimilation or repulsion without making any stimulus weight $A_j$ negative (\S6). Other decoders could also span the handoff of \S5.4. Equation~\ref{eq:harmonic-completion} is a rule that is \emph{complex}-linear in each moment and covariant under rotation, which is the pre-specified mixing rule \S5.4 declined to assert on its own. Complex-linearity is what makes it a natural-parameter sum, and it is a genuine restriction: real-linearity alone would also admit a conjugate term $\bar s_k g^{k+1}$ at the same harmonic, whose consequences the no-go bound below has to reckon with.

For a square tilted by $\theta$ the moments are $s_1=0$ and $s_2=e^{i4\theta}$, so the body reference competes with a single $c_2$ vote and the setting is the direction of their sum. Keeping the first two harmonics gives the frame response in closed form,
\begin{equation}
\widehat v_{\square}(\theta)=\tfrac12\operatorname{atan2}\!\left(\eta_2\sin4\theta,\ \beta+\eta_2\cos4\theta\right).
\label{eq:square-readout}
\end{equation}
In words: the frame pulls the setting toward its nearest edge with a strength that rises and falls four times per turn, against a body reference that holds it upright. The harmonic completion yields the signed, tilt-dependent frame response that \S5.4 could not write down. It is $90^\circ$-periodic, as the square's four-fold symmetry requires. It nulls at the upright square and again at the $45^\circ$ diamond, provided the body reference outweighs the frame, $\beta>\eta_2$. For a weak frame it reduces to $(\eta_2/2\beta)\sin4\theta+O((\eta_2/\beta)^2)$, so near upright the frame-tilt function is a pure fourth-harmonic sinusoid whose amplitude is the frame-to-body concentration ratio. The nulls, the $90^\circ$ periodicity and the reversal were already the $c_2$ signature of \S5.4; what Eq.~\ref{eq:square-readout} adds is the envelope between them.

\textbf{A no-go bound, and the limitation it makes explicit.} Differentiating Eq.~\ref{eq:square-readout} leaves a strictly positive denominator and a numerator that factors exactly as $2\eta_2(\beta\cos4\theta+\eta_2)$. Every stationary point of the frame response therefore satisfies
\[
\cos4\theta_\ast=-\eta_2/\beta ,
\]
and for assimilative coupling $0<\eta_2<\beta$ that condition confines the positive peak to
\[
22.5^\circ<\theta_\ast<45^\circ .
\]
The bracket is a necessary consequence of the form, not a fitted range: no choice of gain escapes it, and the peak runs monotonically across the bracket as $\eta_2/\beta$ runs from $0$ to $1$. The measured rod-and-frame function peaks near $15$--$20^\circ$ (\S5.4), which lies \emph{below} the bracket's floor. So the \emph{phase-preserving} additive assimilative family --- the one Eq.~\ref{eq:harmonic-completion} writes, in which a single real gain $\eta_2$ multiplies the whole harmonic vote --- is excluded as a complete account of the frame envelope, by proof rather than by a failed fit. \textbf{Symmetry permits a broader family than the phase-preserving form.} Rotation covariance and reflection equivariance permit the two quadratures of the harmonic to carry \emph{separate} real gains, $q_\square=\beta+A\cos4\theta+iB\sin4\theta$ with $A\neq B$ --- equivalently $\beta g+\tfrac{A+B}{2}s_2\bar g+\tfrac{A-B}{2}\bar s_2g^{3}$, which is reflection-equivariant for any real $A,B$. The stationary condition becomes $\cos4\theta_\ast=-A/\beta$, set by $A$ alone, while the sign of the response near upright is set by $B$ alone. The two therefore \emph{decouple}, and the escape the previous paragraph closes reopens: $A=-\beta/2$ with $B>0$ puts the peak at exactly $15^\circ$, keeps the response assimilative at small tilt, and preserves both the upright and the $45^\circ$ null, with $|q_\square|$ bounded away from zero throughout. Running $A/\beta$ from $-0.15$ to $-0.5$ sweeps the peak across the observed $15$--$20^\circ$ window. The phase-preserving form keeps, and this wider family gives up, the cue-combination reading: $\eta_2s_2\bar g$ adds axial natural parameters, so Eq.~\ref{eq:cue-combination}'s product-of-von-Mises interpretation survives, whereas an $A\neq B$ deformation keeps such a reading only by adding a second cue, $\tfrac{A-B}{2}\bar s_2g^{3}$, whose favoured axis turns \emph{against} the frame. The algebra permits that factor; it does not show that any independent signal supplies it. The bound is therefore a statement about the single-gain, phase-preserving harmonic vote, not about additivity or cue combination as such.

Neither of the two alternative gain regimes resolves the discrepancy. A repulsive coupling ($\eta_2<0$) does move the extremum below $22.5^\circ$, to $18.1^\circ$ at $\eta_2/\beta=-0.3$, but it also reverses the sign of the whole function. It then predicts a rod set \emph{away} from the frame at every small tilt, which is the opposite of the measured assimilation. A frame stronger than the body reference ($\eta_2>\beta$) instead destroys the $45^\circ$ null: the diamond captures the setting to horizontal, and the observed null there rules it out. Both regimes are re-derived in \texttt{analysis/verify\_v4\_revision\_math.py}.

The bound places the overshoot of \S5.4 rather than excusing it. The cardinal-weighted curve that peaks at $24.6^\circ$ sits \emph{inside} the bracket, so it is a consistent member of this family rather than a broken fit. No adjustment of the first moment can pull it down. The equal-channel neural model's $22.5^\circ$ --- the same number the edge/diagonal-switch heuristic of \S5.4 returns --- sits at the bracket's floor, which is why both harmonic accounts overshoot by the same $\sim5$--$9^\circ$. The observed peak lies outside the reach of the \emph{phase-preserving} family, and the candidates now come in two grades. The cheapest is the two-quadrature completion above, which stays first-order, additive and symmetry-admissible and reaches $15^\circ$ --- but it replaces the single phase-preserving harmonic vote with a second, counter-rotating cue that no measured signal yet supplies. Beyond it sit a tilt-dependent gain, a cardinal prior that is not von Mises, or a target-relative spatial interaction of the kind \S6 already requires for the sign reversal with cortical distance. We record the gap as a stated limitation of the model, not as a residual to be tuned away, and we do not adopt the two-quadrature form here: naming the escape route is not the same as taking it, and which of these the data actually select is an open question.

\section{Predictions and scope}
\textbf{Predictions.} (a) The first-moment coherence of a symmetric $\pm\theta$ pair follows $|\cos2\theta|$, vanishing near $45^\circ$. It is a stimulus-side target, parameter-free in its angles, whose perceptual expression is set by the visual-versus-body weighting of \S5.6. (b) For two inducers of unequal salience, the induced vertical follows the exact contrast-by-separation surface of R3. Varying the contrast imbalance therefore tests the curvature of that surface, and at $45^\circ$, under the first harmonic alone, it measures the visual-to-body ratio $\eta_1/\beta$ --- which is what a measured setting shows there in place of the stimulus-side axis's branch selection (\S3, \S5.6). The branch selection itself never reaches a measured setting. At the orthogonal pair $s_1=0$, so $q_1=\beta g$ is nonzero for every body concentration $\beta>0$ and $\widehat v$ stays continuous; the jump is recovered only in the body-free limit $\beta\to0$. A strongly frame-dependent observer ($k_V>\tfrac12$) does acquire a discontinuity, but at a \emph{different} stimulus: the separation $\tfrac12\arccos(-k_B/k_V)$, strictly beyond $45^\circ$, where the pair asserts a \emph{horizontal} axis of coherence $k_B/k_V$ and the visual term cancels the body reference exactly (\S5.6). That is the mean reversal, not the null. The threshold $k_V=\tfrac12$ divides the single-line turnover of \S5.6. (c) More than two inducers combine by the same phasor sum. Three equally spaced lines ($0^\circ,60^\circ,120^\circ$), for instance, cancel --- and by the harmonic-survival rule (\S5.4) they annul $c_1$ \emph{and} $c_2$, so the null survives even a frame-like second-harmonic mechanism. That makes the triplet a sharp discriminator --- but only once the tie is broken. At exactly equal salience a hard winner-take-all rule has \emph{three} tied winners and no specified tie-breaker, so it predicts no particular line, and its mean over trials can look misleadingly symmetric. \textbf{Breaking the tie separates the two rules cleanly.} Give all three lines weight $A$ and add $\varepsilon>0$ to the $60^\circ$ line: the equal baseline still cancels, leaving $s_1=[\varepsilon/(3A+\varepsilon)]e^{i120^\circ}$. In words, the surviving vote is the imbalance alone, pointing along the line that received it. Additive pooling then returns a setting that vanishes \emph{linearly} with that imbalance --- for the first harmonic alone at $k_V=0.14$, $0.13^\circ$ for a $10\%$ imbalance and $0.013^\circ$ for a $1\%$ one --- while winner-take-all switches to the now-strongest line and holds a fixed $4.4^\circ$ for every $\varepsilon>0$. The discriminator is therefore \emph{continuity at the symmetry point}, a ramp against a step: at a $1\%$ imbalance the two rules differ by more than two orders of magnitude, which is a far easier measurement than agreeing on a null. With the two-harmonic weights solved in \S5.3 both readings reverse sign and shrink, to about $-0.06^\circ$ and $-0.006^\circ$ against $-1.8^\circ$, and the ratio stays above two orders of magnitude ($313$ against $325$). (d) Two stimuli matched in $(c_0,c_1)$ are first-moment \emph{metamers}: the core predicts the same percept for both, even when their higher moments differ.

\paragraph{A moment-matched metamer test.} The metamer claim gives a cleaner test of the $c_1$-to-$c_2$ handoff than comparing unrelated figures does. Match total energy, then compare an equal pair at $\pm\theta$ against a vertical/horizontal mixture whose normalised weights are $\cos^2\theta$ and $\sin^2\theta$. Writing $s_k$ for the normalised $k$-th doubled-angle moment, the two stimuli have identical first moments,
\[
s_1^{\rm pair}=\cos2\theta=\cos^2\theta-\sin^2\theta=s_1^{\rm cardinal},
\]
but different second moments,
\[
s_2^{\rm pair}=\cos4\theta,\qquad s_2^{\rm cardinal}=1.
\]
They therefore separate by $1-\cos4\theta$, which is strictly positive for every $\theta\in(0^\circ,90^\circ)$ and vanishes only at the endpoints, where the ``pair'' degenerates to a single axis. The separation is largest at $\theta=45^\circ$, which is the angle to use. At $\theta=30^\circ$, for instance, both stimuli have $s_1=1/2$ while $s_2$ is $-1/2$ for the pair and $1$ for the mixture. Any decoder that reads only $c_1$ --- the core account of this paper --- must predict the same percept for the two. A decoder with a second-harmonic channel has the information to separate them --- but in this cardinal construction it expresses that information in the setting's \emph{concentration}, not in its mean. Both stimuli are mirror-symmetric about vertical, so every $s_k$ is real and the completed readout of \S5.6 stays real. A real sum leaves the mean setting vertical only while it stays positive. With the second harmonic of Eq.~\ref{eq:harmonic-completion} the two sums are $\beta+\eta_1\cos2\theta+\eta_2\cos4\theta$ for the pair and $\beta+\eta_1\cos2\theta+\eta_2$ for the mixture, and $\beta>\eta_1+\eta_2$ is sufficient to keep both positive at every $\theta$. Below that bound the pair's sum can turn negative while the mixture's does not --- at $\beta=1$, $\eta_1=0.5$, $\eta_2=0.99$ it is negative from about $45.6^\circ$ to $51.7^\circ$, most deeply at $48.6^\circ$ --- and the pair's mean then jumps to horizontal. Within the bound the mean setting is vertical for both while $|q|$ differs. The cardinal comparison is therefore a precision and conditioning test within that bound, and turning it into a mean-difference test needs the construction rotated off the cardinal axes. The comparison estimates whether such a channel contributes at all, while holding the entire first-moment prediction fixed.

\textbf{Test record.} The model makes twelve predictions, and the table sorts them by the kind of standing each has. \emph{Proven} marks an algebraic identity of the model itself: true by derivation, pinned by the test suite, and not an empirical result. \emph{Confirmed empirically} marks a prediction that published data already bears out. For a regularity used to construct the model, the table labels its recovery an internal check, not an independent confirmation. \emph{Open test} marks a forward prediction no one has yet run. \emph{Partial} marks a case where model and data part ways in one feature, and the entry says where the model falls short. \emph{Contradicted} marks a prediction that published data already reject. The two labels must not be run together: a derivation cannot be evidence for itself. The archived rows are computed in \texttt{analysis/test\_all\_predictions.py}, and the uniqueness, influence-and-stability, and metamer identities added here are re-derived independently in \texttt{analysis/verify\_v4\_revision\_math.py}.

{\small
\begin{longtable}{p{0.42\linewidth} p{0.13\linewidth} p{0.35\linewidth}}
\hline
\textbf{Prediction, in plain terms} & \textbf{Standing} & \textbf{What the evidence shows} \\
\hline
\endfirsthead
\hline
\textbf{Prediction, in plain terms} & \textbf{Standing} & \textbf{What the evidence shows} \\
\hline
\endhead
A single tilted line makes vertical \emph{appear} tilted the same way (rule R1). & partial & Li and Matin (2005b) and Shavit et al.\ (2013) both measured this tracking; it is linear over the $\pm15^\circ$ tested. Near vertical a fourfold term predicts the same sign. Li and Matin's (2005a) near-horizontal lines shift VPV the same way, which $c_1$ forbids, and put $c_1$'s share at an estimated $2$--$4\%$; past $45^\circ$ the effect reverses (\S5.3). \\
Two mirror-image lines about vertical leave the vertical setting unmoved (rule R2). & confirmed empirically & Li and Matin (2005b): equal-and-opposite pairs within $\pm13.2^\circ$ null the VPV shift; that such a pair instead shifts perceived \emph{eye level} is the VPEL result of Shavit, Li, and Matin (2013). \\
Many lines combine \emph{sub-additively}, between complete summation and $1/n$ averaging. & confirmed empirically & Li and Matin (2005b), an experiment separate from the 2005a slopes of \S5.1: the two-line VPV function reaches 0.66--0.91 of the sum of the separate effects (long vs short lines; cf.\ the \emph{Perception} multiline fits of \S5.1). The short-line value is close to complete summation; Li and Matin read the long-line value of 0.66 as near averaging. A recovery: these data exclude complete summation, but for long lines they do not separate the account from simple averaging (\S5.1). The wider class of nonlinear selection rules is separated only by the perturbed-triplet test below. \\
With a long array of short segments, the percept follows the \emph{whole array's} orientation, not the individual segments. & confirmed empirically & Shavit et al.\ (2026), 8 observers: in Study~1's $66^\circ$ arrays of segments at most $3^\circ$ long the local segments add only ${\sim}4$--$6\%$; in Study~2 ($6$--$18^\circ$ arrays, $3$--$15^\circ$ segments) segment orientation reaches VPEL slopes of $0.07$--$0.14$, rising with segment length rather than with shorter arrays. \\
Equally spaced lines cancel --- $M$ equally spaced lines annul $c_1,\dots,c_{M-1}$, and the first surviving moment is $c_M$ (harmonic survival). & proven & An algebraic identity of the model, re-derived and pinned in \texttt{analysis/\allowbreak tests/\allowbreak test\_\allowbreak symmetry\_\allowbreak harmonics.py}. It is not an empirical confirmation. \\
A square frame gives \emph{no} first-moment ($c_1$) signal; its bias lives one harmonic up (a $90^\circ$-periodic term). & partial & The rod-and-frame bias nulls at $0^\circ$ and $45^\circ$ and reverses between, as predicted (Alberts et al., 2016). The second harmonic sets only the axis ($|c_2|$ is constant in tilt); the envelope comes from coupling it to the body reference (Eq.~\ref{eq:square-readout}) and peaks near $24.6^\circ$ versus the observed $15$--$20^\circ$. That ${\sim}5$--$9^\circ$ overshoot is structural within the phase-preserving family: no completion that adds axial natural parameters peaks below $22.5^\circ$, though letting the harmonic's two quadratures carry separate gains does reach $15^\circ$ (\S5.6). \\
An equal-gain, uniformly tiled, cosine-tuned, linearly pooled bank of orientation channels computes \emph{exactly} $\tfrac12\arg c_1$, so the neural model and this one coincide. & proven & The two agree to machine precision on every stimulus that asserts an axis, and vanish together on the cancelling ones (\S5.5). That is an identity under the stated idealisations, not a measurement. \\
A symmetric $\pm\theta$ pair in one hemifield has first-moment coherence $\lvert\cos2\theta\rvert$ --- full at $0^\circ$, zero at $45^\circ$, recovering beyond --- while its signed vertical component $\cos2\theta$ changes sign there. & open test & The decisive experiment (Figure~\ref{fig:decisive}); with the two-harmonic weights solved in \S5.3 its body-tilt response nulls near $22.5^\circ$ instead, so it also separates the two harmonics. The same shape already appears in the tilt illusion and its motion analog. The setting an observer actually makes depends on the visual-versus-body weighting of \S5.6. \\
Three lines at $0^\circ,60^\circ,120^\circ$ cancel, and a slight imbalance then parts pooling from winner-take-all. & open test & At equal salience winner-take-all is three-way tied and separates nothing, so the test is the \emph{perturbed} triplet (\S6): a ramp against a step, two orders of magnitude apart at a $1\%$ imbalance. Not yet run. \\
First-moment metamers --- an oblique mirror pair and a contrast-matched vertical/horizontal mixture with the same $(c_0,c_1)$ --- give the same core prediction but differ in $c_2$. & open test & Isolates first- from second-harmonic decoding while holding the whole first-moment summary fixed; not yet run. \\
A single line's induced bias turns over near the oblique and returns to zero at horizontal, but only for an observer whose visual weight is below one half (first harmonic alone). & contradicted & Derived from the cue-combination bridge (Eq.~\ref{eq:cue-combination}): $\cos2\gamma_\ast=-k_V/k_B$, so the measured $k_V\approx0.14$ puts the turn near $50^\circ$ with a ${\sim}4.7^\circ$ peak and a positive bias out to $90^\circ$. Vingerhoets et al.\ (2009) ran the test, a single line from $-90^\circ$ to $+90^\circ$, and found a $90^\circ$-periodic effect; Li and Matin's parallel pair (1998, as cited in 2005a) reverses past $45^\circ$. Only a second harmonic with $\eta_2>\eta_1/\sqrt2$ reproduces the reversal (\S5.3). \\
One inducer shifts perceived vertical and perceived horizontal by the \emph{same} amount (doubling). & open test & Li and Matin call the two ``virtually indistinguishable'' (Li \& Matin, 1998a, 1998b, as cited in Li \& Matin, 2005a), from a design that measured both under the same \emph{2-line} stimulus; we identified no \emph{single}-inducer within-design comparison, and did not read the 1998 abstracts themselves. A fourfold readout predicts the same equality (\S5.3). \\
\hline
\end{longtable}
}

\emph{In short: every published regularity the model was built to recover is recovered, and the quantitative features checked against it are compatible with it; there are two documented misses. The rod-and-frame peak is about $5^\circ$ off, and the single-line data are fourfold: near-horizontal lines shift VPV the same way as near-vertical ones, and the effect reverses past $45^\circ$, both of which the first moment forbids (\S5.3). The genuinely discriminating evidence for the combination structure is still forward. Section~5.6 bounds the rod-and-frame miss rather than merely flagging it: the observed peak lies outside the reach of the phase-preserving additive family, the one that adding axial natural parameters forces, though a two-quadrature completion of the same order does reach it at the cost of that reading. The open tests now form a hierarchy: the $45^\circ$ pair separates a first-moment-only readout from the two-harmonic one, the triplet annuls both $c_1$ and $c_2$, and the moment-matched metamers isolate $c_2$ while holding $(c_0,c_1)$ fixed. Under the stated idealisations, the pooled neural readout is mathematically the same object as $\tfrac12\arg c_1$ (\S5.5). Those experiments therefore decide between a pooling rule and a non-linear one, not between our version and Matin and Li's.}

\begin{figure}[!htbp]\centering
\includegraphics[width=0.80\linewidth]{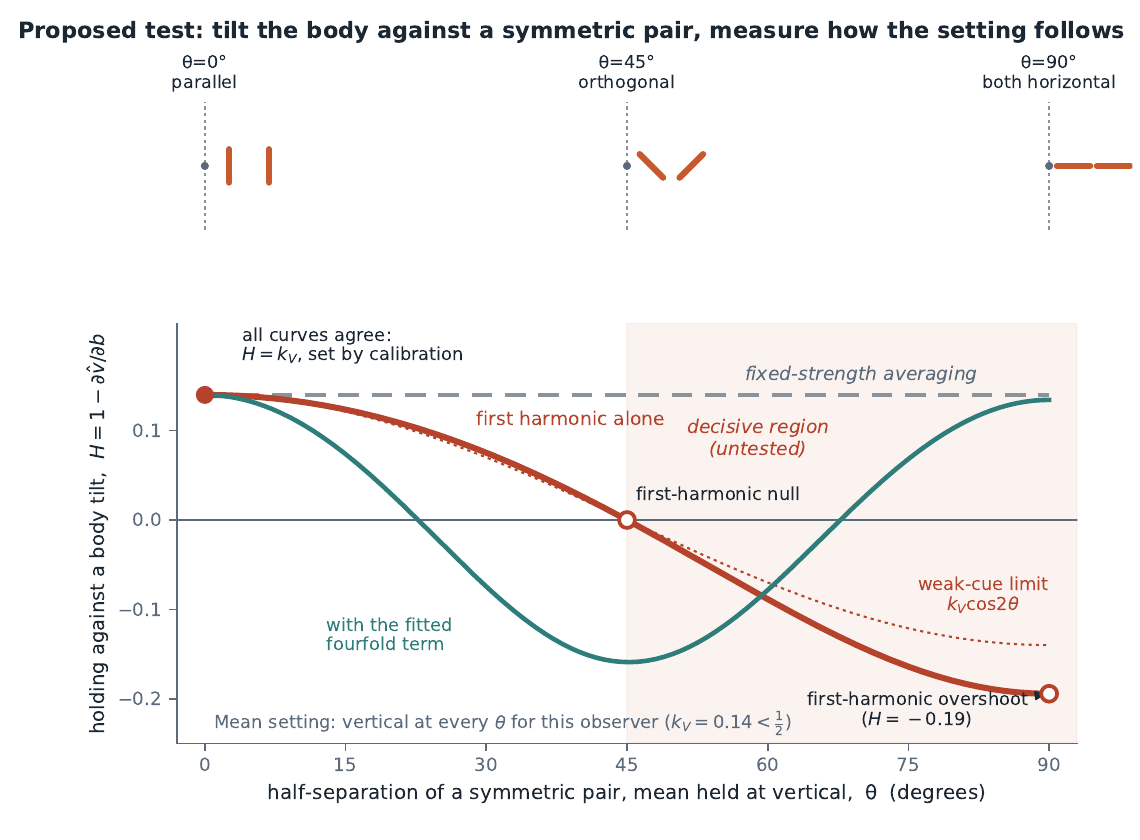}
\caption{The decisive forward test (not yet run). Open a symmetric roll-tilted pair, both lines in one hemifield so that they pool (``One experiment, both references'', \S6), from $\theta=0^\circ$ (both vertical) through $\theta=45^\circ$ (orthogonal, $90^\circ$ apart) to $\theta=90^\circ$ (both horizontal), holding the pair's mean at vertical. The pair's signed first moment $\cos2\theta$ is a stimulus-side quantity, and it never moves the mean setting of an observer with $k_V\approx0.14$ (Fig.~\ref{fig:thresholds}b). The plotted quantity is one that observer's settings do reveal: roll the body by a small angle $b$ while the pair stays fixed on the retina, and measure how far the setting follows. The holding coefficient $H=1-\partial\widehat v/\partial b$ is the share of that body tilt the pair resists. Red: the first harmonic alone, $H=\eta_1\cos2\theta/(\beta+\eta_1\cos2\theta)$ (paradigm (ii) of ``Measuring the strength of a null'', \S6), drawn at $k_V=0.14$, so $\eta_1/\beta=k_V/k_B$. It equals $k_V$ at $\theta=0^\circ$, vanishes at $\theta=45^\circ$ where orthogonal inducers assert no first-moment axis (R4), and turns negative beyond: the pair's horizontal visual axis pushes the setting past the body tilt, an overshoot reaching $H=-0.19$ at $90^\circ$. Red dotted: the weak-cue limit $k_V\cos2\theta$, the stimulus-side cosine scaled by the visual weight. Grey dashed: an averaging account whose pooled axis stays vertical and whose visual strength is held fixed as the pair opens, so $H=k_V$ at every $\theta$. Averaging fixes the pooled axis but not its strength, so the flat line rests on that added assumption. The first-harmonic and averaging accounts (red, grey) are calibrated to agree at $\theta=0^\circ$ (filled marker), where they cannot discriminate, and diverge in the shaded region: the first-harmonic null at $45^\circ$ and the overshoot beyond (open markers). The prediction assumes the retinal image stays fixed while the body reference rolls; if the image rolls with the body, the measured slope is not $\partial\widehat v/\partial b$. A reversal of the mean itself needs a visual cue that outweighs the body reference ($k_V>\tfrac12$; the crossing then sits at $\tfrac12\arccos(-k_B/k_V)$, strictly beyond $45^\circ$), a condition Shavit et al.'s (2013) observers do not meet. The pair's own second harmonic is parked at a rod-and-frame null at every $\theta$ and adds no mean bias (``Does the orthogonal pair hide a frame?'', \S6), but it does change $H$. Teal (panel label ``fitted fourfold term''): both harmonics (Eq.~\ref{eq:harmonic-completion}, $K=2$) with the weights solved in \S5.3 ($\eta_1/\beta=0.0035$, $\eta_2/\beta=0.074$), assuming those single-line weights carry over to a pooled pair. $H$ then nulls near $22.5^\circ$, falls to about $-0.16$ at $45^\circ$ and returns to about $+0.13$ at $90^\circ$, so the same test separates the two harmonics. Under these weights the first moment changes $H$ by at most $0.003$, so the fourfold term sets its sign, and the teal curve rejoins averaging near $90^\circ$.}
\label{fig:decisive}
\end{figure}

\textbf{One experiment, both references.} The $|\cos2\theta|$ test above pools its two lines into a single signal ($c_1=e^{i2\theta}+e^{-i2\theta}=2\cos2\theta$), so it is read most directly as perceived vertical. Its signed vertical component nulls at $45^\circ$ and reverses, but the mean setting of an observer with $k_V\approx0.14$ does not reverse (\S5.6), so the reversal could show only through the perturbation paradigms below. With the weights solved in \S5.3 the sign of the body-tilt response is set by the fourfold term, and the first moment changes $H$ by at most $0.003$ (Fig.~\ref{fig:decisive}, teal), so paradigm (ii) separates the two harmonics rather than isolating $c_1$. Placed within a \emph{single} hemifield that same pooled pair drives \emph{both} readouts identically: with only one field active $s_L=0$, so VPV $=s_R$ and VPEL $=s_R$. The eye-level version of the perturbation paradigms below is not derived here. The same stimulus is run as a vertical setting and, in a \emph{separate} session, as an eye-level setting, following Shavit, Li, and Matin (2013, 2026), who measured the two norms separately. One stimulus thus reaches both references (Fig.~\ref{fig:1}), though only the vertical setting has a derived perturbation test. (A \emph{bilateral} symmetric pair --- one line per field, $s_R=\theta$, $s_L=-\theta$ --- is instead the \S5.5 case: it holds VPV flat at zero and drives only VPEL, with no reversal.) So the stimulus reaches eye level as well as vertical, and its placement selects which: a single-field pair drives both readouts, a split-field pair eye level alone.

\textbf{The null-and-reversal shape is not unprecedented --- adjacent paradigms show the same geometry.} For a symmetric pair, the first-harmonic curve of Fig.~\ref{fig:decisive} predicts the signature \emph{shape} of orientation induction in general --- its angular form, not its sign. The tilt illusion repels a test line from a nearby inducer, passes through a null near $45$--$50^\circ$, and reverses to \emph{attraction} beyond --- the classical direct and indirect effects. Gibson and Radner (1937) mapped that function, and named those two effects, for the \emph{successive} after-effect: at $45^\circ$ the effect ``has disappeared for one subject and is almost negligible'' for the other, and ``beyond $45^\circ$, an after-effect in the opposite direction begins to show itself'' (their Fig.~3, p.~466). For the \emph{simultaneous} illusion the companion paper measures induced tilt falling from $2.01^\circ$ at a $10^\circ$ inducing grille to $0.08^\circ$ at $45^\circ$ (Gibson, 1937, Table~I, p.~556), but it stops at the null; the crossover past it is Wenderoth and Beh's (1977), whose direct effects ``cross to negative indirect effects at about $+50^\circ$'' (p.~66). Its sign is opposite to ours: the induced vertical is \emph{assimilative}, the percept moving toward the inducer (R1), whereas the tilt illusion repels. We therefore take from it only the doubled-angle null-and-reversal geometry, not the direction of the effect. Its motion-direction analogue nulls near $90^\circ$, about twice that angle (Marshak \& Sekuler, 1979) --- the scaling a direction variable gives where orientation needs the doubling. We take that as a qualitative parallel and do not check their angular function point by point against the doubled-angle prediction. The rod-and-frame frame-tilt function is $90^\circ$-periodic, nulling at both the upright square ($0^\circ$) and the $45^\circ$ diamond, with its measured peak near $15$--$20^\circ$. That is the $c_2$ axis signature the same formalism assigns to a four-edged frame (Alberts et al., 2016); the magnitude peak is the \S5.4 cardinal-prior term, not $c_2$ itself. And the population readout that generates the tilt illusion is the vector average our $\phi=\tfrac12\arg c_1$ names (Schwartz, Sejnowski, \& Dayan, 2009). The untested part is the \emph{specific} instance --- a symmetric line pair read out as perceived vertical, its null at $45^\circ$ --- not the shape.

\textbf{Two boundary conditions the same literature imposes.} First, the null is \emph{near} $45^\circ$, not exactly: the measured zero-crossing sits at $45$--$50^\circ$ --- $45^\circ$ in Gibson and Radner's (1937) after-effect function, ``about $+50^\circ$'' in Wenderoth and Beh's (1977) simultaneous one --- and efficient-coding repulsion from the cardinal axes warps the readout (Wei \& Stocker, 2015, built on the cardinal-peaked orientation prior measured by Girshick, Landy, \& Simoncelli, 2011, whose own measured bias runs the other way, toward the cardinals). The model's null is exact only in the doubled-angle mathematics ($\cos90^\circ=0$); the \emph{measured} null is an approximate, observer-specific version of it. Second, induction changes \emph{sign} with the inducer's cortical distance from the target --- assimilation when the two are close, repulsion at intermediate spacing, tracing half a Mexican hat (Mareschal, Morgan, \& Solomon, 2010). That sign change cannot be put into $A_j$. A negative energy weight would break $E\geq0$, break $R\in[0,1]$, and break the positive-semidefinite structure-tensor identity of Appendix~\ref{app:st}. And in axial geometry $-e^{i2\gamma}=e^{i2(\gamma+90^\circ)}$ encodes an orthogonal director, not the bias $-\gamma$. The sign must instead live downstream, in a target-relative spatial coupling that multiplies the already-formed visual summary. The parameter-free claim of this paper is for the \emph{angles} --- the phasor phases $2\gamma_j$ --- not for the weights that scale them, and that spatial coupling is a separate front-end term.

\textbf{Measuring the strength of a null.} Because a symmetric $\pm\theta$ pair sets $\phi=0$, its $|\cos2\theta|$ coherence is invisible in the mean setting and must be read from how the null \emph{resists perturbation}. Three paradigms make it measurable. Each is derived first for the first harmonic alone ($K=1$); the two-harmonic case with the weights solved in \S5.3 ($K=2$, $\eta_1/\beta=0.0035$, $\eta_2/\beta=0.074$) follows each, and under those weights the fourfold term, not $c_1$, carries what (i) and (ii) measure. \textbf{(i) Probe:} add a weak unilateral probe line to the pair. The pair anchors vertical with a visual coherence $\propto|\cos2\theta|$, so the probe's leverage on the setting grows as that anchor weakens. Writing the probe as weight $\varepsilon$ at $\delta$ against anchors of weight $A$ at $\pm\theta$, the first-harmonic completion $q_1$ of \S5.6 makes that leverage explicit:
\[
\left.\frac{\partial\widehat v}{\partial\varepsilon}\right|_0=\frac{\eta_1\sin2\delta}{4A\,(\beta+\eta_1\cos2\theta)}.
\]
Only in the body-free limit $\beta=0$ does this scale as $1/|\cos2\theta|$; that limit is a non-uniform local approximation, not the exact response. \textbf{At the visual null itself nothing can diverge.} Since $\cos90^\circ=0$, the visual term drops out of the denominator at $\theta=45^\circ$ and the body concentration $\beta$ is left holding the setting alone, so the susceptibility there is $\eta_1\sin2\delta/(4A\beta)$ --- finite for \emph{any} $\beta>0$, with the visual weight $\eta_1$ surviving in the numerator only. The condition $\beta>\eta_1$ gives something stronger: boundedness across the \emph{whole} separation sweep out to $\theta=90^\circ$, for which it is necessary as well as sufficient. The denominator $\beta+\eta_1\cos2\theta$ falls monotonically to its minimum $\beta-\eta_1$ at $\theta=90^\circ$, and whenever $\eta_1\geq\beta$ it vanishes at the real angle $\tfrac12\arccos(-\beta/\eta_1)$. \textbf{That angle lies above $45^\circ$ and at or below $90^\circ$} --- the arc-cosine of a negative argument exceeds $90^\circ$, so halving it cannot land at or below the null, and it reaches $90^\circ$ exactly in the boundary case $\eta_1=\beta$ --- and it is precisely the angle at which the balanced pair's own mean setting swings from vertical to horizontal, because $q_1=\beta+\eta_1\cos2\theta$ is real for a mirror pair and changes sign exactly there. The divergence the body/gravity reference removes therefore sits \emph{beyond} the visual null, never at it, and it is the same event as the axis flip rather than a separate pathology. For the first harmonic alone, at the measured $k_V\approx0.14$ read as a first-harmonic weight (i.e.\ $\eta_1/\beta\approx0.16$), the smallest denominator is still $84\%$ of $\beta$, and the leverage rises by the factor $1/k_B$ --- about $16\%$ --- from $\theta=0^\circ$ to the null, and by $(\beta+\eta_1)/(\beta-\eta_1)$, about $39\%$, across the full sweep. The experiment measures how steeply that bounded leverage rises, and under the first harmonic alone that steepness identifies the visual-to-body weighting $\eta_1/\beta$ rather than a singularity. With the solved weights the leverage still rises by about $16\%$ to $45^\circ$, but by only about $0.7\%$ across the full sweep, against $39\%$ for the first harmonic alone. The rise to $45^\circ$ cannot separate the harmonics: it equals $(\beta+\eta_1+\eta_2)/(\beta-\eta_2)$, which for any weights that reproduce the measured slope $k_V$ is exactly $1/(1-k_V)$, so it measures $k_V$ alone (the fourfold term alone, at the solved $\eta_2$, gives $15.9\%$). Only the full-sweep rise depends on the mix, so paradigm (i) separates the harmonics through its $90^\circ$ endpoint, which together with $k_V$ fixes both $\eta_1/\beta$ and $\eta_2/\beta$ if the readout has exactly these two harmonics. Because a $16\%$ rise against a flat baseline is a precision measurement rather than a qualitative contrast, paradigm (i)'s power against realistic setting variability is an open question, and paradigm (ii) below carries the discrimination. \textbf{(ii) Body tilt:} pit the pair against a vestibular (body-roll) tilt $b$, with the pair held fixed on the retina. Differentiating the phase of $q_1=\beta e^{i2b}+\eta_1\cos2\theta$ gives the share of the tilt the pair resists,
\[
H(\theta)=1-\left.\frac{\partial\widehat v}{\partial b}\right|_{b=0}=\frac{\eta_1\cos2\theta}{\beta+\eta_1\cos2\theta},
\]
valid across the whole sweep when $\beta>\eta_1$. In words: $H$ equals $k_V$ at $\theta=0^\circ$, vanishes at $45^\circ$, and is negative beyond, where the pair's horizontal visual axis pushes the setting \emph{past} the body tilt --- by $19\%$ at $90^\circ$ for $k_V\approx0.14$. Its magnitude scales as $|\cos2\theta|$ while the visual cue is weak, but the sign carries the discrimination: a fixed-strength averaging account predicts $H=k_V$ at every $\theta$, and no scalar weighting of a vertical visual estimate against the body cue with nonnegative weights can make $H$ negative. This signed curve is the one drawn in red in Fig.~\ref{fig:decisive}. With the weights solved in \S5.3, Eq.~\ref{eq:harmonic-completion} at $K=2$ gives instead $H=(\eta_1\cos2\theta+2\eta_2\cos4\theta)/(\beta+\eta_1\cos2\theta+\eta_2\cos4\theta)$, which nulls near $22.5^\circ$, reaches about $-0.16$ at $45^\circ$ and returns to about $+0.13$ at $90^\circ$ (teal curve). Where $H$ first nulls therefore separates the two harmonics, as its sign separates the account from averaging. \textbf{(iii) Variance (magnitude-controlled):} \emph{if} setting precision tracks reliability, the trial-to-trial VPV variance rises as the visual coherence $R=|\cos2\theta|$ falls. It need not diverge at the null, because the body/gravity reference remains available there and still constrains the setting. The \emph{direction} of the dependence is in any case genuinely open (\S5.5), so a magnitude-matched design tests the reliability-to-precision mapping itself rather than assuming it. For the first harmonic alone, all three convert the $|\cos2\theta|$ law --- parameter-free in the angles it predicts --- into a directly falsifiable measurement of a cancelled state. Under those weights, (i) and (ii) read mainly the fourfold term through Eq.~\ref{eq:harmonic-completion}, so they test the two-harmonic readout rather than $c_1$ alone.

\textbf{Does the orthogonal pair hide a frame?} The pair also carries a second harmonic, so the first-moment null must be checked against that term. Its second doubled-angle moment is $c_2=2\cos4\theta$, whose magnitude is \emph{maximal} exactly where $c_1$ vanishes: at $\theta=45^\circ$, $c_1=0$ but $c_2=-2$. Geometrically the $\pm45^\circ$ pair carries the same two orientations as a diagonal square frame --- an X has a diamond's edge orientations --- and therefore the same second-harmonic phase as the stimulus that drives the rod-and-frame effect (\S5.4). The two differ in spatial layout, which this aggregate model cannot see (the limitation stated below); what follows uses their orientation content only. A frame-sensitive mechanism could add bias at the predicted $45^\circ$ first-moment null. It does not, for a definite reason. Because the two lines are mirror images about vertical, $c_2$ is always \emph{real}, so the pair's frame axis $\tfrac14\arg c_2$ is pinned to $0^\circ$ (upright) for $\theta<22.5^\circ$ and to $45^\circ$ (diagonal) for $22.5^\circ<\theta<67.5^\circ$ --- never an oblique angle between. Both $0^\circ$ and $45^\circ$ are \emph{nulls} of the rod-and-frame function, which vanishes for an upright or a diagonal frame and peaks near $15$--$20^\circ$ (\S5.6). The pair's second harmonic is thus \emph{parked} at a frame orientation that adds no mean tilt, at every $\theta$; the first-moment null survives in the mean, which stays vertical while the completed sum stays positive ($\beta>\eta_1+\eta_2$ suffices).

The second harmonic contributes a variance signature, and the literature bounds the claims it can support. \textbf{A $45^\circ$ frame is not a variance maximum.} Alberts et al.\ (2016) find setting variability rising with frame orientation but ``leveling off again at $\pm45^\circ$'', with a peak near $\pm30^\circ$ that their own model does not capture; and Beh, Wenderoth, and Purcell (1971) report a variance of $0.761$ at a $45^\circ$ frame against $3.432$ at $75^\circ$ and $3.334$ at $15^\circ$, which places $45^\circ$ among the smallest of the seven tilts they measured, beside the $0^\circ$ and $90^\circ$ nulls. Their figures are between-observer variances of the condition means --- the only variance that paper reports --- so they bound the claim in the measure available rather than in within-observer trial-to-trial scatter. We therefore predict no divergence of setting variance at the null, and paradigm (iii) does not need one: it reads the magnitude-controlled \emph{change} in scatter as the coherence falls, whose direction \S5.5 leaves genuinely open, while paradigm (ii) reads the perturbation response. Those same sources do support the claim this paragraph actually needs --- that a $45^\circ$ square frame is a null of the rod-and-frame function.

This protection is specific to the \emph{mirror-symmetric} pair. An asymmetric pair --- two lines at unrelated tilts and unequal salience --- has an \emph{oblique} $c_2$, a frame axis at neither $0^\circ$ nor $45^\circ$, so there a frame-sensitive mechanism \emph{would} add a bias. For that case the cleanest pooling-versus-winner-take-all discriminator is instead the equilateral triplet, whose $c_2$ vanishes outright (\S5.4) and so leaves no second-harmonic loophole at all. The completion of \S5.6 settles the higher harmonics in general, not just $c_2$. Every moment of an equal mirror pair is real, $s_k=\cos2k\theta$, so with an upright reference and real gains $q_K$ in Eq.~\ref{eq:harmonic-completion} is real at every $K$. A retained harmonic can therefore switch the setting between vertical and horizontal, or change its concentration, but it cannot manufacture an oblique mean.

\textbf{Scope.} This is a \emph{stimulus-side} account of the orientation cue, not a claim about the neural locus at which vertical is computed; it says the \emph{combination structure} of the induced percept is inherited from a first-moment orientation readout. It \emph{complements}, rather than competes with, Bayesian optimal-integration accounts of the rod-and-frame effect and subjective vertical (Alberts, de Brouwer, Selen, \& Medendorp, 2016). Those accounts fuse the frame with head and gravity cues by reliability weighting, specifying \emph{how} orientation cues are weighted. The present account instead specifies the stimulus-side orientation \emph{summary} --- the first moment of the orientation distribution --- that such weightings act on. The strongest of them already represents the square frame \emph{circularly}. Alberts et al.\ model it as a sum of four von Mises distributions at the cardinal $90^\circ$ intervals, with a vertical-versus-horizontal concentration asymmetry that equalises at $45^\circ$ (isotropy, so the bias nulls). That is an arrival, independent of this paper, at the square's fourth-harmonic ($c_2$) structure and cardinal reference of \S5.4, and evidence that a circular description of the frame is the right currency. It builds on Vingerhoets et al.\ (2009), who had already treated a \emph{single} line as a fourfold cue, in Mittelstaedt's vector sum and as a $\cos4r$ von Mises likelihood (\S4). The remaining distinction is therefore not Euclidean-versus-circular but \emph{fitted-versus-closed-form}: those models fit per-axis concentrations and infer head-in-space by MAP, whereas the present first moment $\tfrac12\arg c_1$ fixes the \emph{angles} of the 1- and 2-line \textbf{combination structure} with no per-configuration parameter, while Li and Matin's length function sets the coefficients' magnitudes (\S5.1). The most recent sensory-prior-estimation models still enter the stimulus as scalar cue tilts (Tian et al., 2025), leaving the orientation \emph{summary} PLUMB supplies unspecified --- so the two levels compose rather than compete. For the same reason it is compatible with Bayesian cue-combination views of perceived eye level (Orendorff et al., 2016), supplying the likelihood term's orientation summary, and with evidence for a shared orientation-sensitive process behind VPEL and VPV (Shavit, Li, \& Matin, 2013).

\textbf{A stated limitation: the model is spatially aggregate.} The weights $A_j$ are scalars, so $c_1$ collapses all \emph{retinal-position} information --- it is blind to where in the field a given orientation sits. The \S5.5 hemifield split is the coarsest possible spatial structure (a binary left/right partition), not a continuous kernel. Yet VPV and VPEL are strongly eccentricity-dependent (Shavit, Li, \& Matin, 2013), so a full account needs a \emph{spatially weighted} first moment $c_1(\mathbf{x})$ with an eccentricity kernel --- which the present, position-agnostic order parameter does not provide. That extension (the continuous spatial weighting) is left to separate work and is out of scope here; likewise gravity/vestibular integration and adaptation dynamics.

\textbf{Complex and natural scenes.} The definition carries over from lines to any field: replace the sum with $c_1=\int_0^\pi E(\theta)\,e^{i2\theta}\,d\theta$ over the scene's oriented-energy profile $E(\theta)$ (\S2), the same mathematical object as the image structure tensor (\S4). The pooling already scales past single lines --- the global-first-moment result (\S5.4) has the percept follow a multi-element array's \emph{global} orientation, not its local parts.

But an everyday carpentered scene is \emph{cardinal-dominated}. In the doubled-angle domain its verticals and horizontals are antipodal, so to the extent their energies balance they cancel, $c_1$ is small, and the first moment predicts little net tilt. The residual signal there lives in the second harmonic $c_2$ --- the frame/cardinal term (\S5.4, \S5.6) --- not the first. The first moment is thus most predictive for a scene with a coherent \emph{oblique} bias (a hillside, a tilted horizon), and quietest for a balanced rectilinear one.

The closed form does not itself supply the oriented-energy map $E(\theta)$. The model leaves the edge-weighting front end unspecified: contrast, spatial frequency, eccentricity (the spatial limitation above), depth, and semantic weight (a horizon versus a stray edge). The parameter-free claim is for the \emph{angles} --- the phasor phases $2\gamma_j$ --- not for the weights.

PLUMB therefore accounts for the orientation-induced \emph{component} of how a complex field biases the vertical. It hands a fuller layout-and-gravity account --- Bayesian integration (Alberts et al., 2016), with vestibular and visual-polarity cues --- the orientation \emph{summary} that account consumes. It does not by itself predict the whole everyday upright, and a direct natural-scene test remains forward work.

\section{Conclusion}
Two of the empirical rules Li and Matin established for induced visual vertical --- linear combination and symmetric cancellation --- are the closed-form behaviour of the first circular moment of stimulus orientation in the doubled-angle domain, though not uniquely so (next paragraph). The third, linear tracking of a single line, follows from it as well, but in their data a fourfold term carries almost all of it (\S5.3). That moment is the structure-tensor principal axis, or equivalently the orientation population vector. We make the logical status of the doubling precise. Up to a positive scale, the moment $c_1$ is the unique additive, salience-homogeneous, spin-2 rotation-covariant axial summary that reads a vertical line as vertical, and its coherence $R$ sets the conditioning of that axis: a perturbation measured as a fraction of the total effective salience shifts $\phi$ inversely with $R$, while one of fixed absolute size is divided by the product $A_\Sigma R$. For a summary that transforms under a rotation character, director symmetry alone restricts the rotational order to the \emph{even} spins, so $c_1$ is the lowest nontrivial rung rather than a chosen harmonic; linear pooling and positive homogeneity then fix it uniquely within that rung. The account fits their combination table with no per-configuration free parameters beyond Li and Matin's own mass-action length constants (their Fig.~5), and it adds a testable coherence law ($|\cos2\theta|$).

The existing data support a narrower result. R2 and R3 are what a doubled-angle first-moment readout produces, but the fourfold term produces them as well, and no published data separate the two. A single line's effect is fourfold: near-horizontal lines shift VPV the same way as near-vertical ones, and the effect reverses past $45^\circ$, both of which the first moment forbids (Li \& Matin, 2005a; Vingerhoets et al., 2009). In Li and Matin's data the first moment carries an estimated $2$--$4\%$ of a single line's effect, a point estimate without standard errors (\S5.3). The present data refute complete summation but do not separate the model from simple averaging for long lines. And the square frame marks the edge where the first moment cancels and the second, $c_2$, carries the axis. Coupling that harmonic to the body reference (\S5.6) supplies the frame's signed response, and it also proves the limit of doing so: no \emph{phase-preserving} additive assimilative completion --- the form that adding axial natural parameters forces --- places the frame peak below $22.5^\circ$, whereas the data place it at $15$--$20^\circ$. Letting the harmonic's two quadratures carry separate gains does reach $15^\circ$ while keeping both nulls and the assimilative sign, so the residual is specific to the cue-combination reading rather than to additivity (\S5.6). That gap is a real residual of the model as interpreted here, and we leave it standing.

Read symmetrically and antisymmetrically across the visual hemifields, the per-hemifield orientation signals yield perceived vertical and perceived eye level as one orientation process with two integration rules (Shavit, Li, \& Matin, 2013). The fourfold term gives the same signals, so this result, too, does not single out $c_1$ (\S5.5). That split mirrors the projective geometry of roll and pitch.

A trial-to-trial variance signature we had proposed is confounded by how large the settings are (larger settings scatter more under Weber-type noise), so we leave that question open.

Because the readout \emph{is} low-level orientation pooling written in closed form (\S5.5), it is not a rival to higher-level grouping but a \textbf{baseline} for it. The baseline shows how much of Li and Matin's apparently cognitive multi-line weighting falls out of scalar length saturation alone. The baseline thereby marks off what a genuine Gestalt effect --- such as the rod-and-frame peak asymmetry --- must account for \emph{beyond} the first moment. Recasting a psychophysical combination rule as an order parameter both explains its form and ties it to the standard image-analysis and population-coding descriptors of orientation.

\appendix
\section{The structure-tensor identity}\label{app:st}
We show that $\phi=\tfrac12\arg c_1$ and $R=|c_1|/c_0$ are the principal-axis orientation and the eigen-anisotropy of the orientation structure tensor. Under these assumptions the induced-VPV readout (\S2) and the classical image structure tensor are therefore the same mathematical object --- the claim used in \S4.

Let $\mathbf u(\theta)=(\cos\theta,\sin\theta)^{\!\top}$ and form the second-moment (structure) matrix of the orientation energy $E(\theta)$,
\[
M=\int_0^\pi E(\theta)\,\mathbf u(\theta)\mathbf u(\theta)^{\!\top}\,d\theta=\begin{pmatrix} M_{uu} & M_{uv}\\[2pt] M_{uv} & M_{vv}\end{pmatrix}.
\]
With $\cos^2\theta=\tfrac12(1+\cos2\theta)$, $\sin^2\theta=\tfrac12(1-\cos2\theta)$, $\cos\theta\sin\theta=\tfrac12\sin2\theta$, and the moments $c_0=\int_0^\pi E\,d\theta$, $\operatorname{Re}c_1=\int_0^\pi E\cos2\theta\,d\theta$, $\operatorname{Im}c_1=\int_0^\pi E\sin2\theta\,d\theta$, the entries give
\[
M_{uu}-M_{vv}=\operatorname{Re}c_1,\qquad 2M_{uv}=\operatorname{Im}c_1,\qquad M_{uu}+M_{vv}=c_0.
\]
So $\operatorname{tr}M=c_0$ and the traceless (off-isotropic) part of $M$ is
\[
M-\tfrac12 c_0\,I=\tfrac12\begin{pmatrix}\operatorname{Re}c_1 & \operatorname{Im}c_1\\[2pt] \operatorname{Im}c_1 & -\operatorname{Re}c_1\end{pmatrix},
\]
a symmetric traceless matrix whose leading eigenvector points along orientation $\tfrac12\arg c_1$ and whose eigenvalues are $\pm\tfrac12|c_1|$. Hence the principal axis of $M$ is $\phi=\tfrac12\arg c_1$ and its normalised eigen-anisotropy is
\[
\frac{\lambda_1-\lambda_2}{\lambda_1+\lambda_2}=\frac{|c_1|}{c_0}=R.
\]
The induced vertical $\phi$ is therefore the structure-tensor principal axis and $R$ its coherence. (Equivalently, for a population of orientation-tuned units with cosine tuning and firing rates $\propto E$, $c_1=\sum_j(\text{rate}_j)\,e^{i2\gamma_j}$ is the population vector and $\phi$ its preferred axis, so the two read-outs coincide.) Only $\{c_0,c_1\}$ enter $M$: the doubled-angle moment $c_2$ of \S5.4 (the square-frame axis) is invisible to the structure tensor, which is why the orthogonal square gives $M\propto I$ (isotropic at second order) yet a well-defined $\tfrac14\arg c_2$.

\section{Uniqueness of the additive axial summary}\label{app:unique}
Let a finite line field be the positive measure $X=\sum_jA_j\delta_{\gamma_j}$ on orientations modulo $\pi$, and let $F(X)\in\mathbb C$ be its vote. Suppose
\begin{align*}
F(X+Y)&=F(X)+F(Y), & F(aX)&=aF(X)\quad(a\geq0),\\
F(R_\alpha X)&=e^{i2\alpha}F(X), & F(\delta_0)&=K>0.
\end{align*}
The first line says the summary pools effective nonnegative saliences linearly; $X+Y$ is addition of measures, so two coincident lines add their weights. The homogeneity line says that multiplying every salience by $a\geq0$ multiplies the summary by $a$, so the readout has no built-in scale of its own. The covariance line says the output rotates as an axial (spin-2) quantity, and fixes its units. Since $\delta_\gamma=R_\gamma\delta_0$, covariance already fixes every atom,
\[
F(A\delta_\gamma)=AKe^{i2\gamma},
\]
and additivity then leaves no freedom at all:
\[
F(X)=K\sum_jA_je^{i2\gamma_j}.
\]
Axial invariance comes for free, because $e^{i2(\gamma+\pi)}=e^{i2\gamma}$. If $F$ is also continuous under weak convergence of finite measures, atomic quadrature measures converge weakly to an integrable orientation field and give $F(E)=K\int_0^\pi E(\theta)e^{i2\theta}\,d\theta$.

\textbf{Director symmetry selects even spins, beginning with spin 2.} Written as above, the covariance line can look as though it simply assumes the doubling. It does less than that. Suppose only that a nonzero summary transforms under \emph{some} continuous one-dimensional character of rotations, $F(R_\alpha X)=\chi(\alpha)F(X)$. Every such character of $SO(2)$ has the form $\chi(\alpha)=e^{im\alpha}$ with $m$ an integer, since a continuous homomorphism of a compact group into $\mathbb C^\times$ lands in the unit circle and must return to $1$ after a full turn. Orientation is a director, so rotating the field by $\pi$ returns the same field, $R_\pi X=X$, and therefore $e^{im\pi}F(X)=F(X)$. For $F(X)\neq0$ this forces $e^{im\pi}=1$, so \textbf{$m$ is even}. The odd orders are not merely disfavoured: there $\chi(\pi)=-1$, so $F(X)=-F(X)$ and the summary is identically zero. Director symmetry thus permits exactly the ladder $m=2,4,6,\ldots$, which is exactly $c_1,c_2,c_3,\ldots$ --- reading spin $2k$ in place of spin $2$ replaces $e^{i2\gamma}$ by $e^{i2k\gamma}$ in the atom, and additivity then returns $c_k$. The negative even orders add nothing new: on a nonnegative weight field the spin-$(-2)$ summary is the complex conjugate of the spin-$2$ one, the same axis with the doubled-angle plane traversed the other way. So $c_1$ is \textbf{the lowest nontrivial rung of a ladder that director symmetry itself selects}, not a representation chosen by hand --- and that is also why $c_2$ and $c_3$ are legitimate extensions rather than ad-hoc repairs (\S5.4, \S6).

The uniqueness theorem depends on its stated assumptions, which have two distinct roles. Director symmetry gives the even ladder and its lowest rung; additivity and positive homogeneity give uniqueness \emph{within} that rung. Abandoning additivity admits nonlinear competitors such as winner-take-all, and that, not the choice of harmonic order, is what the experiments of \S6 are built to test. Verified in \texttt{analysis/\allowbreak verify\_\allowbreak even\_\allowbreak spin\_\allowbreak ladder\_\allowbreak 2026-09-29.wls}.

\section{Differential influence and finite stability}\label{app:sensitivity}
Let $z=\sum_jA_je^{i2\gamma_j}=|z|e^{i2\phi}\ne0$. The standard differential of a complex argument is
\[
d\arg z=\operatorname{Im}\frac{dz}{z}
=\frac{1}{|z|}\operatorname{Im}\!\left(e^{-i2\phi}dz\right).
\]
Perturbing the saliences and the orientations gives
\[
dz=\sum_je^{i2\gamma_j}dA_j
+2i\sum_jA_je^{i2\gamma_j}d\gamma_j,
\]
and halving $d\arg z$ yields Eq.~\ref{eq:influence}. Taking moduli gives the scalar form
\[
|d\phi|\leq\frac{|dz|}{2|z|}=\frac{|dz|}{2A_\Sigma R},
\]
since $|z|=A_\Sigma R$ exactly, for \emph{any} tuning, with the tuning factor already stripped from $z$. The denominator is $A_\Sigma R$ and not $c_0R$: because $c_0=\hat\kappa_0A_\Sigma$, the product $c_0R$ equals $\hat\kappa_0|z|$, which is $|z|$ only when the tuning happens to carry unit mass --- and \S2's convention fixes the ratio $\hat\kappa_1/\hat\kappa_0$, never the overall scale $\hat\kappa_0$. In words: a perturbation measured as a \emph{fraction} of the total salience shifts the axis by an amount governed by $1/R$, so coherence alone sets the dimensionless conditioning; a perturbation of fixed \emph{absolute} size is divided by $A_\Sigma R$, so a faint stimulus is as ill-conditioned as an incoherent one.

The finite bound follows from plane geometry. Rotate coordinates so that $z=|z|>0$. Every perturbed endpoint $z+\Delta z$ with $|\Delta z|\leq\varepsilon<|z|$ lies in the closed disk of radius $\varepsilon$ centred at $z$. The rays from the origin that make the largest angle with $z$ and still meet that disk are its tangents, at angle $\arcsin(\varepsilon/|z|)$ from $z$. Halving the doubled-angle displacement proves Eq.~\ref{eq:stability}, and the tangent case shows the bound is attained. If $\varepsilon\geq|z|$ the disk reaches the origin, and no nontrivial uniform phase bound exists --- which is exactly the loss of identifiability at $R=0$.

\section*{Code and data availability}
\begin{sloppypar}
Every script that produces a number or falsification test in this paper accompanies this submission as an \textbf{ancillary file} (the figures are in the paper, so a script that only draws one is not included), together with the verification suite, the re-analysed trial-level settings, and the supplementary neural-equivalence derivation cited in \S5.5. The OSF deposit (\href{https://doi.org/10.17605/OSF.IO/6R7UE}{DOI: 10.17605/OSF.IO/6R7UE}) holds the code as of 30 September 2026; scripts added or changed since, including those behind the \S5.3 fourfold share, the \S5.5 intervals and the \S6 values computed with the \S5.3 weights, are for now only in these ancillary files. The file \texttt{analysis/README.md} maps the analysis scripts to the specific manuscript claim, figure, or test each one backs, and the symbolic re-derivation scripts state their purpose in their own headers; running \texttt{python3 -m pytest analysis/tests -q} reproduces the pinned numbers, and \texttt{python3 analysis/test\_all\_predictions.py} prints the falsification checklist. The uniqueness theorem of \S2, the influence and stability bounds of \S3, the harmonic completion of \S5.6 with its square readout and no-go bound, and the metamer identities of \S6 are all re-derived from scratch, symbolically and numerically, by \texttt{python3 analysis/verify\_v4\_revision\_math.py}. The cue-combination bridge and the single-line turnover of \S5.6 are re-derived the same way by \texttt{python3 analysis/verify\_x7\_turnover.py}. The trial-level settings re-analysed in \S5.5 are the author's own data from the already-published study of Shavit, Li, and Matin (2013); no new data were collected. The \S5.4 global-versus-local settings belong to the companion paper of Shavit, Li, and Matin (2026) and are deposited with it, not here; the re-analysis scripts ship in this supplement, and the four checks that recompute from those settings skip when the files are absent.
\textbf{Trust boundary.} Every check in this supplement was written within this project, with the AI assistants named below. The Python and Wolfram Mathematica re-derivations share no code, but they share the model's equations as written here. Every estimate of the first moment's share (\S5.3) starts from the same by-eye readings of Li and Matin's Fig.~4, so those checks are not independent of that digitisation.
\end{sloppypar}

\section*{Use of AI tools}
AI tools assisted under my direction. Claude (Anthropic) wrote analysis and verification code to my specification and gave drafting and editorial support. ChatGPT (OpenAI) and Gemini (Google) were used for adversarial review and independent mathematical checking, Perplexity for literature search and citation verification, and Wolfram Mathematica for checking derivations.

\section*{Acknowledgements}
I thank C. S. for helpful comments on the manuscript. The accompanying code reproduces the results.

\vspace*{\fill}
\begin{center}\footnotesize\emph{In awe of the world's deep wonder\\
And life's essential light\\
Grateful to pursue the beauty\\
As the waking mind gives sight}\end{center}
\vspace*{\fill}

\end{document}